\pdfoutput=1
\documentclass[prd,showpacs,letterpaper,10pt,aps,notitlepage,superscriptaddress,nofootinbib,preprintnumbers]{revtex4-1}

\usepackage{amsfonts,amsmath,graphicx}
\usepackage{slashed}
\usepackage{placeins}

\newcommand{\wm}{\phantom{-}}
\newcommand{\bs}[1]{\ensuremath{{\boldsymbol{#1}}}}

\begin{document}

\title{$\Lambda_b \to \Lambda\: \ell^+ \ell^-$ form factors and differential branching fraction from lattice QCD}

\author{William Detmold}
\affiliation{Center for Theoretical Physics, Massachusetts Institute of Technology, Cambridge, MA 02139, USA}
\author{C.-J.~David Lin}
\affiliation{Institute of Physics, National Chiao-Tung University, Hsinchu 300, Taiwan}
\affiliation{Physics Division, National Centre for Theoretical Sciences, Hsinchu 300, Taiwan}
\author{Stefan Meinel}
\email{smeinel@mit.edu}
\affiliation{Center for Theoretical Physics, Massachusetts Institute of Technology, Cambridge, MA 02139, USA}
\author{Matthew Wingate}
\affiliation{DAMTP, University of Cambridge, Wilberforce Road, Cambridge CB3 0WA, UK}


\date{December 19, 2012}

\preprint{MIT-CTP 4430}

\begin{abstract}
We present the first lattice QCD determination of the $\Lambda_b \to \Lambda$ transition form factors
that govern the rare baryonic decays $\Lambda_b \to \Lambda\: \ell^+ \ell^-$ at leading order
in heavy-quark effective theory. Our calculations are performed with 2+1 flavors of domain-wall
fermions, at two lattice spacings and with pion masses down to 227 MeV. Three-point functions
with a wide range of source-sink separations are used to extract the ground-state contributions.
The form factors are extrapolated to the physical values of the light-quark masses and to the continuum limit.
We use our results to calculate the differential
branching fractions for $\Lambda_b \to \Lambda\: \ell^+ \ell^-$ with $\ell=e,\mu,\tau$ within the
standard model. We find agreement with a recent CDF measurement of the $\Lambda_b \to \Lambda\: \mu^+ \mu^-$
differential branching fraction.
\end{abstract}

\maketitle

\FloatBarrier
\section{Introduction}
\FloatBarrier

The flavor-changing neutral-current transitions $b \to s\: \gamma$ and $b \to s\: \ell^+\ell^-$ can provide
strong constraints on models of new physics. The effective weak Hamiltonian describing decays of this type has the form
\cite{Grinstein:1988me,Grinstein:1990tj,Misiak:1992bc,Buras:1993xp,Buras:1994dj,Buchalla:1995vs,Chetyrkin:1996vx,Bobeth:1999mk}
\begin{equation}
\mathcal{H}_{\rm eff} = -\frac{4 G_F}{\sqrt{2}}V_{tb}V_{ts}^*  \sum_{i=1,...,10,S,P} (C_i {O_i} + C_i' {O_i'}) , \label{eq:Heff}
\end{equation}
where the operators that directly give ``short-distance'' contributions to these decay amplitudes are
\begin{align}
\nonumber O_7    =& \frac{e}{16\pi^2} m_b\: \bar{s} \sigma^{\mu\nu} {P_R} b \: F_{\mu\nu}^{(\rm e.m.)}, &  O_7' =&
\frac{e}{16\pi^2} m_b \: \bar{s} \sigma^{\mu\nu} {P_L} b \: F_{\mu\nu}^{(\rm e.m.)}, \\
\nonumber O_9    =& \frac{e^2}{16\pi^2} \bar{s} \gamma^\mu {P_L} b\: \bar{l} \gamma_\mu l, & O_9' =&
\frac{e^2}{16\pi^2} \bar{s} \gamma^\mu {P_R} b\: \bar{l} \gamma_\mu l, \\
\nonumber O_{10} =& \frac{e^2}{16\pi^2} \bar{s} \gamma^\mu {P_L} b\: \bar{l} \gamma_\mu \gamma_5 l, & O_{10}' =&
\frac{e^2}{16\pi^2} \bar{s} \gamma^\mu {P_R} b\: \bar{l} \gamma_\mu \gamma_5 l,\\
\nonumber O_S =& \frac{e^2}{16\pi^2} m_b \: \bar{s} {P_R} b\: \bar{l} l, & O_S' =&
\frac{e^2}{16\pi^2} m_b \: \bar{s} {P_L} b\: \bar{l}  l, \\
O_P =& \frac{e^2}{16\pi^2} m_b \: \bar{s} {P_R} b\: \bar{l} \gamma_5 l, & O_P' =&
\frac{e^2}{16\pi^2} m_b \: \bar{s} {P_L} b\: \bar{l} \gamma_5 l, \label{eq:HeffSD}
\end{align}
and the $C_i^{(\prime)}$ are Wilson coefficients. In the standard model, the Wilson coefficients of the scalar and
pseudoscalar operators $O_{S,P}$, as well as those of the opposite-chirality operators $O_i'$, are highly suppressed \cite{Altmannshofer:2012az}.
Experimental measurements of $b \to s$ decay rates, angular distributions, and related observables provide
constraints on various functions of the Wilson coefficients $C_i$ and $C_i'$ \cite{Altmannshofer:2012az}. In this way,
these measurements restrict models of new physics and their allowed parameters. Most of the existing studies have focused on mesonic decays
such as $B \to K^* \gamma$ and $B \to K^{(*)} \ell^+\ell^-$, for which experiments have reached a high level of precision.
To get the most complete set of constraints on new physics, it is important to consider many different observables.
To this end it is useful to analyze also \emph{baryonic} $b \to s$ decays such as $\Lambda_b \to \Lambda \: \gamma$ and
$\Lambda_b \to \Lambda \: \ell^+ \ell^-$. The decay $\Lambda_b \to \Lambda \: \mu^+ \mu^-$ has recently been observed
using the Tevatron \cite{Aaltonen:2011qs}, and is being measured at LHCb.

One important aspect that distinguishes $\Lambda_b$ decays from $B$ meson decays is the spin of the $\Lambda_b$ baryon,
which in principle provides an additional handle on the fundamental interactions. When produced through $Z$ bosons at
$e^+ e^-$ colliders, $b$ quarks have a strong longitudinal polarization, and the $\Lambda_b$ baryons keep most of that
polarization \cite{Falk:1993rf, Bonvicini:1994mr, Diaconu:1995mp, Buskulic:1995mf, Abbiendi:1998uz, Abreu:1999gf}.
At the Tevatron and the LHC, the $\Lambda_b$ baryons produced in proton-(anti)proton collisions are expected to have some
degree of transverse polarization \cite{Dharmaratna:1996xd, Ajaltouni:2004zu, Hiller:2007ur}, which can be measured accurately
using the method proposed in Ref.~\cite{Hrivnac:1994jx}. As first mentioned in Ref.~\cite{Gremm:1995nx} and later
studied in detail in Ref.~\cite{Hiller:2001zj}, the $\Lambda_b$ polarization can be exploited to test the
``helicity structure'' of $\mathcal{H}_{\rm eff}$, that is, to disentangle the contributions from the Wilson coefficients $C_i$ and $C_i'$.
In practice, this entails measuring an asymmetry in the angular distribution between the $\Lambda_b$ spin and the momentum
of a particle in the final state \cite{Hiller:2001zj}. Even for unpolarized $\Lambda_b$ baryons, the spin of the
final-state $\Lambda$ baryon can also be exploited to test the helicity structure of $\mathcal{H}_{\rm eff}$, as discussed
for $\Lambda_b \to \Lambda \: \gamma$ in Refs.~\cite{Mannel:1997xy, Chua:1998dx, Huang:1998ek, Hiller:2001zj, Wang:2008sm, Mannel:2011xg}
and for $\Lambda_b \to \Lambda \: \ell^+ \ell^-$ in Refs.~\cite{Huang:1998ek, Chen:2001ki, Chen:2002rg, Wang:2008sm, Aslam:2008hp}.
To this end, an angular analysis needs to be performed for the secondary weak decay $\Lambda \to p\: \pi^-$. Lepton
asymmetries for $\Lambda_b \to \Lambda \: \ell^+ \ell^-$ have also been considered \cite{Chen:2001sj, Chen:2001zc, Wang:2008sm, Aslam:2008hp}.

In order to use these $\Lambda_b$ decays to search for new physics, the matrix elements
$\langle\Lambda \gamma | \mathcal{H}_{\rm eff} | \Lambda_b \rangle$ or
$\langle\Lambda \ell^+\ell^- | \mathcal{H}_{\rm eff} | \Lambda_b \rangle$ must be determined. For the operators
in Eq.~(\ref{eq:HeffSD}), this then requires the computation of the hadronic matrix elements
$\langle \Lambda | \:\bar{s} \Gamma b\: | \Lambda_b \rangle$, which are expressed in terms of ten QCD form factors.
When using heavy-quark effective theory (HQET) for the $b$ quark, the number of independent $\Lambda_b \to \Lambda$
form factors reduces to 2 \cite{Mannel:1990vg, Hussain:1990uu, Hussain:1992rb}. Furthermore, in the limit of large recoil,
soft-collinear effective theory (SCET) predicts that only one form factor remains \cite{Feldmann:2011xf, Mannel:2011xg, Wang:2011uv}.
The $\Lambda_b \to \Lambda$ form factors have been estimated using various models or approximations, including
quark models \cite{Cheng:1994kp, Cheng:1995fe, Mohanta:1999id, Mott:2011cx}, perturbative QCD \cite{He:2006ud},
and sum rules \cite{Huang:1998ek, Wang:2008sm, Wang:2009hra, Feldmann:2011xf}, but have not been determined from QCD previously.
In the charm sector, some information on the $\Lambda_c \to \Lambda$ form factors is available from the experimental measurement
of the semileptonic $\Lambda_c \to \Lambda\: e^+ \nu_e$ decay \cite{Crawford:1995wz, Hinson:2004pj}, and this information has been used
to constrain the $\Lambda_b \to \Lambda$ form factors in Refs.~\cite{Cheng:1994kp, Mannel:1997xy}. In summary, several estimates of
$\Lambda_b \to \Lambda$ form factors exist in the literature, but a considerable uncertainty remains, especially in the low-recoil
region where SCET and light-cone sum rules are not applicable. Clearly, first-principles, nonperturbative QCD calculations of the form factors
are needed, and the method for performing such calculations is lattice QCD.

In this paper, we report on the first lattice QCD calculation of $\Lambda_b \to \Lambda$ form factors
(we presented preliminary results of this work in Ref.~\cite{Detmold:2012ug}). We use HQET for the $b$ quark, and compute
the two form factors that appear. Their definitions are given in Sec.~\ref{sec:FFdefs}.
Treating the $b$ quarks with HQET on the lattice \cite{Eichten:1989kb} also leads to several other technical simplifications that make
the calculation feasible, as will become clear in Sec.~\ref{sec:correlationfunctions}.
For the up, down, and strange quarks, we use a domain-wall fermion action \cite{Kaplan:1992bt, Furman:1994ky, Shamir:1993zy},
which is computationally expensive but provides chiral symmetry even at non-zero lattice spacing.
Our calculations make use of gauge field ensembles generated by the RBC/UKQCD collaborations \cite{Aoki:2010dy}.
These ensembles include 2+1 flavors of dynamical sea quarks, and the lattice parameters used in our study are given in Sec.~\ref{sec:latticeparams}.
We use two different lattice spacings and several different values of the light quark masses, which allows us to perform
simultaneous extrapolations of the form factors to the continuum limit and to the physical values of the quark masses.
The data analysis involves several stages, which are explained in Secs.~\ref{sec:resultsR+R-}, \ref{sec:lambda2ptfits}, \ref{sec:srcsnkextrap},
and \ref{sec:chiralcontinuumextrap}, and estimates of the systematic uncertainties in the form factors are given in Sec.~\ref{sec:systerrs}.
As a first application of our form factor results,
in Sec.~\ref{sec:Lambdabdecay} we then calculate the differential branching fractions for the decays $\Lambda_b \to \Lambda\: \ell^+ \ell^-$ with
$\ell=e,\mu,\tau$ in the standard model. The differential branching fraction for $\Lambda_b \to \Lambda\: \mu^+ \mu^-$ can be compared
to the existing Tevatron data and is of immediate interest for LHCb. Further phenomenological
applications of the form factors that we have determined are left for future work.

\FloatBarrier
\section{\label{sec:FFdefs}Definition of form factors}
\FloatBarrier

In QCD, using Lorentz symmetry and the discrete C, P, T symmetries, one can show that the matrix elements
$\langle \Lambda(p', s') | \:\bar{s} \Gamma b\: | \Lambda_b(p, s) \rangle$ with
$\Gamma=\gamma_\mu,\: \gamma_\mu\gamma_5,\: q^\nu\sigma_{\mu\nu},\: q^\nu\sigma_{\mu\nu}\gamma_5$ (where $q=p-p'$)
are parametrized by ten independent form factors (see for example Ref.~\cite{Feldmann:2011xf}). Heavy-quark symmetry, which becomes exact
in the limit $m_b \rightarrow \infty$, reduces the number of independent form factors to 2 \cite{Mannel:1990vg, Hussain:1990uu, Hussain:1992rb}.
In the following, when working with HQET, we denote the heavy quark by $Q$. The $\Lambda_Q \to \Lambda$ matrix element
with an arbitrary Dirac matrix $\Gamma$ is then given by \cite{Mannel:1990vg, Hussain:1990uu, Hussain:1992rb}
\begin{equation}
\langle \Lambda(p', s') | \:\bar{s} \Gamma Q \: | \Lambda_Q(v, 0, s) \rangle =
\overline{u}(p',s')\left[ F_1(p'\cdot v) + \slashed{v}\:F_2(p'\cdot v) \right] \Gamma\: \mathcal{U}(v, s), \label{eq:FFdef}
\end{equation}
where $v$ is the four-velocity of the $\Lambda_Q$, and the two form factors $F_1$ and $F_2$ can be expressed as functions
solely of $p'\cdot v$, the energy of the $\Lambda$ baryon in the $\Lambda_Q$ rest frame. Here we use the following normalization of states and spinors:
\begin{eqnarray}
 \langle \Lambda (p,s) | \Lambda(p',s') \rangle &=&  2 E_\Lambda (2\pi)^3  \delta_{s s'} \delta^3(\mathbf{p}-\mathbf{p'}), \\
 \langle \Lambda_Q (v,k,s) | \Lambda_Q(v,k',s') \rangle &=&  2 v^0 (2\pi)^3  \delta_{s s'} \delta^3(\mathbf{k}-\mathbf{k'}),
\end{eqnarray}
\begin{eqnarray}
\sum_{s'=1}^2 u(p', s') \overline{u}(p', s') &=& m_\Lambda+\slashed{p}', \\
\sum_{s=1}^2 \mathcal{U}(v, s) \overline{\mathcal{U}}(v, s) &=& 1+\slashed{v}. \label{eq:HQETspinors}
\end{eqnarray}
For most of the analysis in this paper, it is convenient to work with the linear combinations
\begin{eqnarray}
\nonumber F_+ &=& F_1 + F_2, \\
 F_- &=& F_1 - F_2,
\end{eqnarray}
instead of $F_1$ and $F_2$. Note that in the limit $m_b\to\infty$, five of the ten helicity-based $\Lambda_b \to \Lambda$
form factors introduced in Ref.~\cite{Feldmann:2011xf} become equal to $F_+$, while the other five become equal to $F_-$
[see Eq.~(2.10) of Ref.~\cite{Feldmann:2011xf}, where $F_1$ is denoted by $A$, and $F_2$ is denoted by $B$].

\FloatBarrier
\section{Lattice calculation}
\FloatBarrier

\FloatBarrier
\subsection{\label{sec:correlationfunctions}Two-point and three-point functions}
\FloatBarrier

For the lattice calculation, we work in the $\Lambda_Q$ rest frame, so that $v=(1,0,0,0)$. The heavy quark, $Q$, is
implemented with the Eichten-Hill lattice HQET action \cite{Eichten:1989kb}, where we use one level of HYP smearing
\cite{Hasenfratz:2001hp} for the temporal gauge links in order to improve the signal-to-noise ratio of the correlation
functions \cite{DellaMorte:2003mn}. For the up, down, and strange quarks, we use a domain-wall action
\cite{Kaplan:1992bt, Furman:1994ky, Shamir:1993zy}, and the gluons are implemented using the Iwasaki action
\cite{Iwasaki:1983ck, Iwasaki:1984cj}. Our calculations are based on gauge field configurations generated by the
RBC/UKQCD collaboration \cite{Aoki:2010dy} using these actions. Further details of the lattices will be given in
Sec.~\ref{sec:latticeparams}.

In order to extract the matrix element (\ref{eq:FFdef}), we need to compute suitable three-point- and two-point-functions as discussed
in the following. We use the following baryon interpolating fields,
\begin{eqnarray}
 \Lambda_{Q\alpha} &=& \epsilon^{abc}\:(C\gamma_5)_{\beta\gamma}\:\tilde{d}^a_\beta\:\tilde{u}^b_\gamma\: Q^c_\alpha, \\
 \Lambda_{\alpha} &=& \epsilon^{abc}\:(C\gamma_5)_{\beta\gamma}\:\tilde{u}^a_\beta\:\tilde{d}^b_\gamma\: \tilde{s}^c_\alpha, \label{eq:Lambdainterpol}
\end{eqnarray}
where $a,b,c$ are color indices, $\alpha,\beta,\gamma$ are spinor indices, and $C$ is the charge conjugation matrix. The tilde
on the up, down, and strange quark fields $u$, $d$, $s$ denotes gauge-covariant three-dimensional Gaussian smearing, intended
to reduce excited-state contamination in the correlation functions.

In the three-point functions, we use the following $\mathcal{O}(a)$-improved discretization of the continuum HQET current \cite{Ishikawa:2011dd},
\begin{equation}
 J_{\Gamma}^{(\rm HQET)}(m_b) = U(m_b, a^{-1})\: \mathcal{Z}
 \left[ J_\Gamma^{\rm (LHQET)}  + c^{(m_s a)}_\Gamma\:\frac{m_s\:a}{1-(w_0^{\rm MF})^2}\: J_\Gamma^{\rm (LHQET)}
 + c^{(p_s a)}_\Gamma\:a\: J_{\Gamma D}^{\rm (LHQET)}   \right], \label{eq:LHQETcurrent}
\end{equation}
where $J_\Gamma^{\rm (LHQET)}$ and $J_{\Gamma D}^{\rm (LHQET)}$ are given by
\begin{eqnarray}
  J_\Gamma^{\rm (LHQET)} & = & \overline{Q}\: \Gamma \: s, \\
  J_{\Gamma D}^{\rm (LHQET)} & = & \overline{Q}\: \Gamma\: \bs{\gamma}\cdot \bs{\nabla} \: s.
\end{eqnarray}
The current $J_{\Gamma}^{(\rm HQET)}$ is renormalized in the $\overline{\rm MS}$ scheme at $\mu=m_b$.
Note that here we match from lattice HQET to continuum HQET, but not yet to QCD. This is important because the form factors
$F_1$ and $F_2$ are defined in continuum HQET, not full QCD. The matching to QCD will lead to radiative corrections to
the simple relationship (\ref{eq:FFdef}) which depend on $\Gamma$. We will return to this issue in Sec.~\ref{sec:Lambdabdecay}
when computing the differential branching fraction for $\Lambda_b \to \Lambda \: \ell^+\ell^-$.

In Eq.~(\ref{eq:LHQETcurrent}), symmetries of the lattice actions and the equations of motion have been used to reduce
the number of operators that appear \cite{Ishikawa:2011dd}. The term with coefficient $c^{(m_s a)}_\Gamma$ provides
$\mathcal{O}(m_s a)$ improvement, while the term with coefficient $c^{(p_s a)}_\Gamma$ provides $\mathcal{O}(p_s a)$
improvement (here $p_s$ denotes the momentum of the strange quark). The quantity $w_0^{\rm MF}$ is related to tadpole
improvement, and is defined as $w_0^{\rm MF} = 1 - a m_5 + 4(1-u_0)$ \cite{Ishikawa:2011dd}, where $a m_5$ is the
domain-wall height and $u_0$ is the 4th root of the average plaquette. The matching coefficients $\mathcal{Z}$,
$c^{(m_s a)}_\Gamma$, and $c^{(p_s a)}_\Gamma$ have been computed to one-loop order in lattice perturbation theory for
the actions used here in Ref.~\cite{Ishikawa:2011dd} and are evaluated at the scale $\mu=a^{-1}$ (the inverse lattice spacing).
The coefficient $\mathcal{Z}$ is independent of $\Gamma$, but $c^{(m_s a)}_\Gamma$ and $c^{(p_s a)}_\Gamma$ change sign
depending on whether $\Gamma$ commutes or anticommutes with $\gamma^0$. We are interested in the matrix element
(\ref{eq:FFdef}) renormalized at $\mu=m_b$, and following Ref.~\cite{Ishikawa:2011dd} we therefore perform a
renormalization-group (RG) evolution from $\mu=a^{-1}$ to $\mu=m_b$, using the two-loop anomalous dimension of the heavy-light
current in HQET, which was derived in Refs.~\cite{Ji:1991pr, Broadhurst:1991fz}. This leads to the multiplicative factor
$U(m_b, a^{-1})$ in Eq.~(\ref{eq:LHQETcurrent}). The RG running is performed with $N_f=3$ flavors from $\mu=a^{-1}$ down
to $\mu=m_c$, and then with $N_f=4$ flavors from $\mu=m_c$ up to $\mu=m_b$. This two-step running is used because the
nonperturbative lattice calculations are done with $N_f=2+1$ dynamical flavors, and with $a^{-1}>m_c$. However, note that
doing a simple $N_f=4$ running from $\mu=a^{-1}$ to $\mu=m_b$ gives a result that differs only by 0.5\%.
Numerical values for $U(m_b, a^{-1})$,  $\mathcal{Z}$, $c^{(m_s a)}_\Gamma$, $c^{(p_s a)}_\Gamma$, and $u_0$ will be
given in Table \ref{tab:renparams} in the next section.

\begin{figure}
\hfill \includegraphics[height=4.5cm]{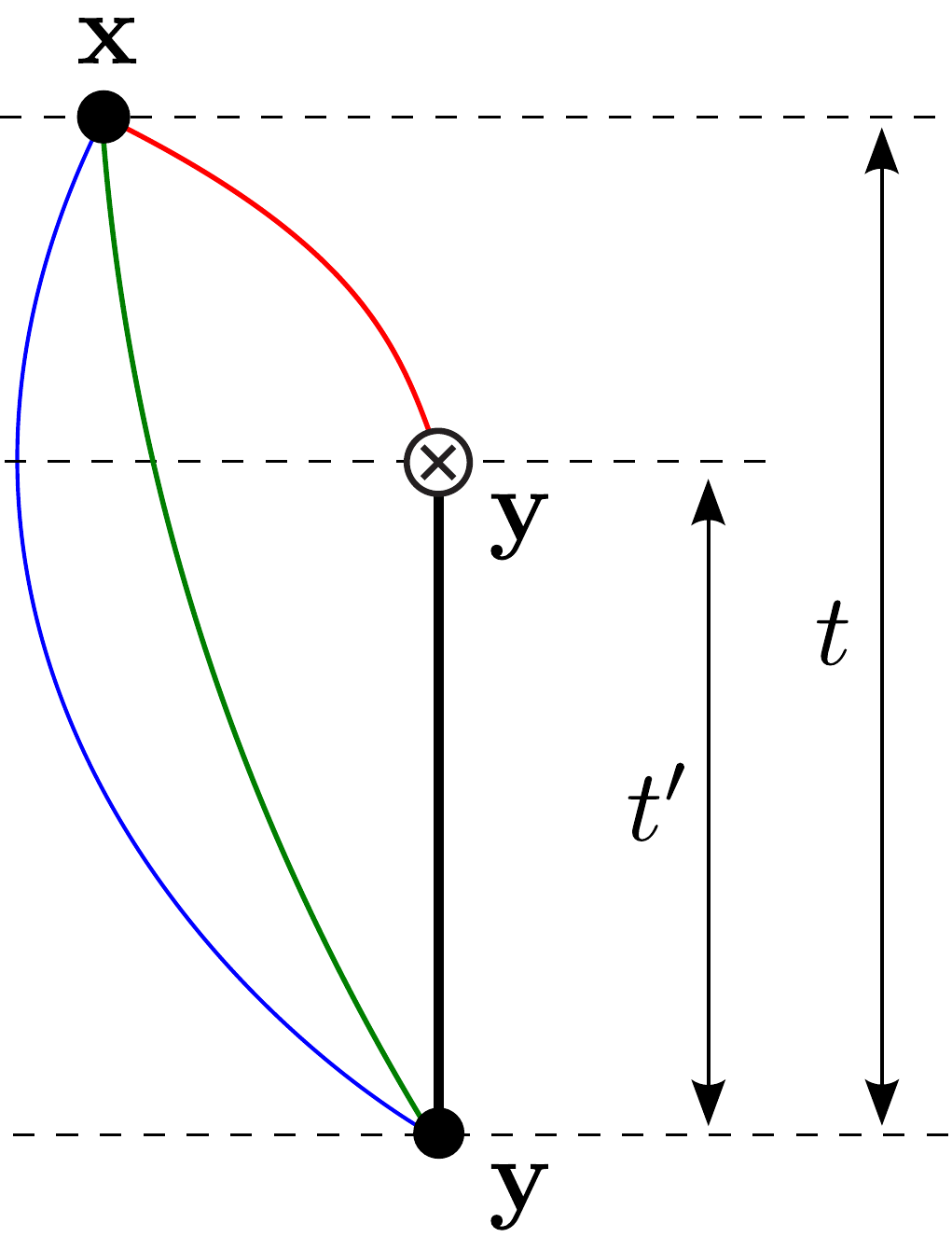}\hfill\includegraphics[height=4.5cm]{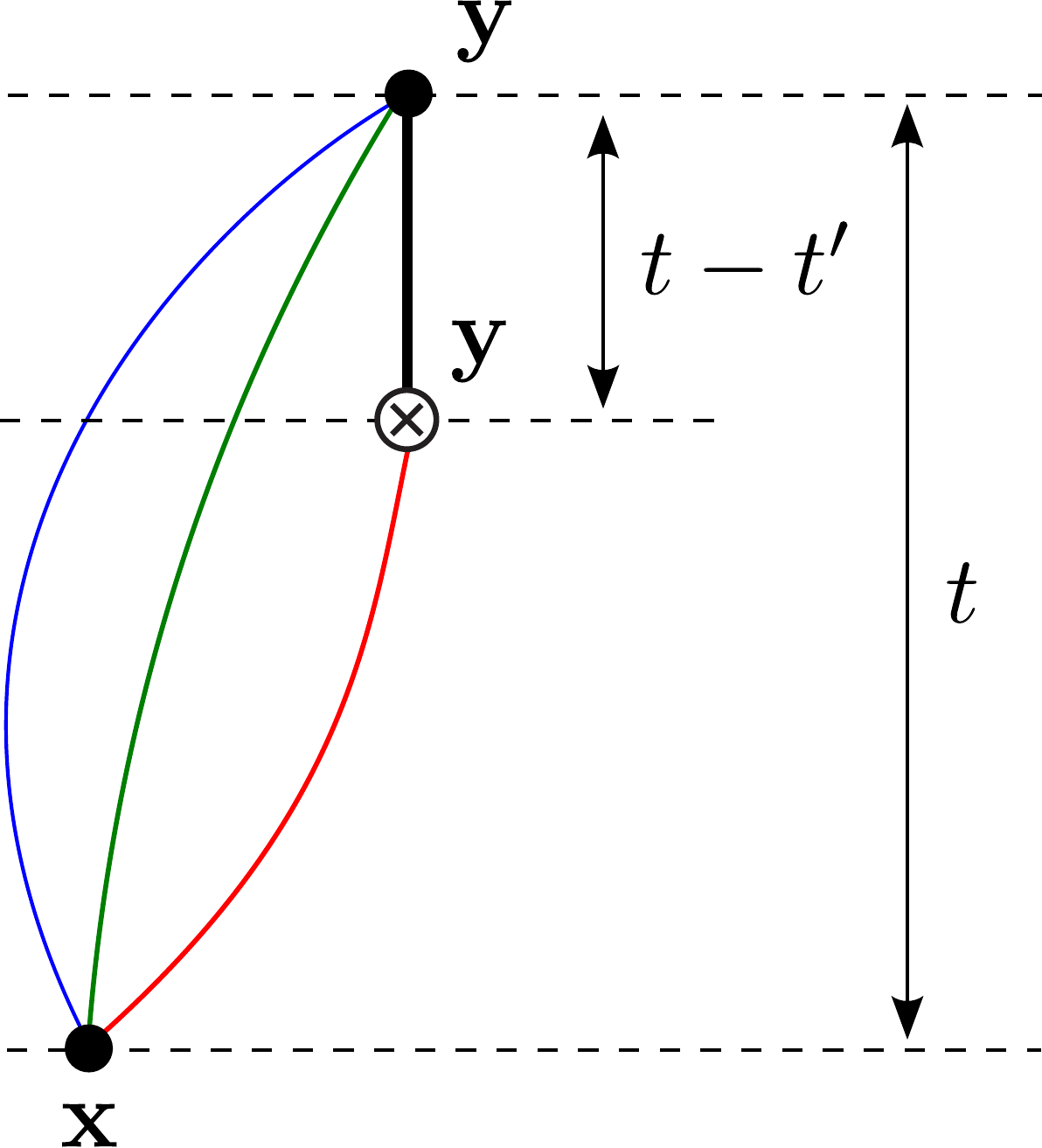} \hfill \null
\caption{\label{fig:threept}Propagator contractions for the forward three-point functions (left) and backward three-point
functions (right). The thick vertical line at the spatial point $\mathbf{y}$ indicates the static heavy-quark propagator.
The source for all light and strange-quark propagators is located at the fixed point $(x_0, \mathbf{x})$. We sum over
all points $\mathbf{y}$, with the appropriate momentum phases as in Eqs.~(\protect\ref{eq:threept}) and (\protect\ref{eq:threeptbw}).}
\end{figure}

Having defined the interpolating fields and the current, we will now discuss the correlation functions.
We compute ``forward'' and ''backward'' two-point functions for the $\Lambda$ and $\Lambda_Q$ as follows:
\begin{eqnarray}
 C^{(2,\Lambda)}_{\delta\alpha}(\mathbf{p'}, t) &=&  \sum_{\mathbf{y}} e^{-i\mathbf{p'}\cdot (\mathbf{y}-\mathbf{x})}
 \left\langle \Lambda_{\delta}(x_0+t,\mathbf{y})\: \overline{\Lambda}_{\alpha}(x_0,\mathbf{x})  \right\rangle, \label{eq:Lambda2pt}\\
C^{(2,\Lambda,\mathrm{bw})}_{\delta\alpha}(\mathbf{p'}, t) &=& \sum_{\mathbf{y}} e^{-i\mathbf{p'}\cdot (\mathbf{x}-\mathbf{y})}
\left\langle \Lambda_{\delta}(x_0,\mathbf{x})\: \overline{\Lambda}_{\alpha}(x_0-t,\mathbf{y}) \right\rangle,\\
C^{(2,\Lambda_Q)}_{\delta\alpha}(t) &=&
\left \langle \Lambda_{Q \delta}(x_0+t,\mathbf{x})\:\overline{\Lambda}_{Q \alpha}(x_0,\mathbf{x}) \right\rangle, \label{eq:Lambdab2pt} \\
C^{(2,\Lambda_Q,\mathrm{bw})}_{\delta\alpha}(t) &=&
\left \langle \Lambda_{Q \delta}(x_0,\mathbf{x})\:\overline{\Lambda}_{Q \alpha}(x_0-t,\mathbf{x}) \right\rangle, \label{eq:Lambdab2ptbw}
\end{eqnarray}
where the superscript ``bw'' denotes the backward correlator. In Eqs.~(\ref{eq:Lambdab2pt}) and (\ref{eq:Lambdab2ptbw}),
the $\Lambda_Q$ interpolating fields at source and sink are required to be at the same spatial point $\mathbf{x}$ because
of the static heavy-quark propagator. Finally, the forward and backward three-point functions for a given gamma matrix
$\Gamma$ in the current are defined as
\begin{eqnarray}
 C^{(3)}_{\delta\alpha}(\Gamma,\:\mathbf{p'}, t, t') &=& \sum_{\mathbf{y}} e^{-i\mathbf{p'}\cdot(\mathbf{x}-\mathbf{y})}
 \left\langle \Lambda_{\delta}(x_0,\mathbf{x})\:\: J_\Gamma^{(\rm HQET)\dag}(x_0-t+t',\mathbf{y})
 \:\:\: \overline{\Lambda}_{Q\alpha} (x_0-t,\mathbf{y}) \right\rangle, \label{eq:threept} \\
C^{(3,\mathrm{bw})}_{\alpha\delta}(\Gamma,\:\mathbf{p'}, t, t-t') &=& \sum_{\mathbf{y}}
e^{-i\mathbf{p'}\cdot(\mathbf{y}-\mathbf{x})} \left\langle \Lambda_{Q\alpha}(x_0+t,\mathbf{y})\:\: J_\Gamma^{(\rm HQET)}(x_0+t',\mathbf{y})
\:\:\: \overline{\Lambda}_{\delta} (x_0,\mathbf{x}) \right\rangle. \label{eq:threeptbw}
\end{eqnarray}
All of the above correlation functions (\ref{eq:Lambda2pt})-(\ref{eq:threeptbw}) can be computed using light and strange
quark propagators with a Gaussian-smeared source located at $(x_0,\mathbf{x})$. For the three-point functions, the quark
propagator contractions are illustrated schematically in Fig.~\ref{fig:threept}. Because no additional domain-wall propagators
are required, we can efficiently compute the three-point functions for arbitrary values of $t$ and $t'$,
only limited by statistical precision.

In order to discuss the spectral decomposition of the correlation functions, we introduce the following definitions for the overlap
factors:
\begin{eqnarray}
 \langle 0 | \Lambda_{Q\alpha} (0) | \Lambda_Q(s) \rangle &=& Z_{\Lambda_Q} \: \mathcal{U}_\alpha(s), \\\
 \langle 0 | \Lambda_{\alpha} (0) | \Lambda(p',s) \rangle &=& [(Z_{\Lambda}^{(1)}+Z_{\Lambda}^{(2)}\gamma^0) \: u(p', s)]_\alpha,
\end{eqnarray}
where $Z_{\Lambda}^{(1)}$ and $Z_{\Lambda}^{(2)}$ depend on $\mathbf{p'}$. Here we need two different overlap factors
$Z_{\Lambda}^{(1)}$ and $Z_{\Lambda}^{(2)}$ for the $\Lambda$, because the spatial-only smearing of the quarks in the
interpolating field (\ref{eq:Lambdainterpol}) breaks hypercubic symmetry \cite{Bowler:1997ej}. The spectral decompositions
of the two-point and three-point functions then read
\begin{eqnarray}
 C^{(2,\Lambda)}_{\delta\alpha}(\mathbf{p'}, t) = C^{(2,\Lambda,\mathrm{bw})}_{\delta\alpha}(\mathbf{p'}, t)
 &=&  \frac{1}{2E_\Lambda} \: e^{-E_\Lambda t}  \left[(Z_{\Lambda}^{(1)}+Z_{\Lambda}^{(2)}\gamma^0)(m_\Lambda+\slashed{p}')
 (Z_{\Lambda}^{(1)}+Z_{\Lambda}^{(2)}\gamma^0) \right]_{\delta\alpha} + \hdots, \\
C^{(2,\Lambda_Q)}_{\delta\alpha}(t) = C^{(2,\Lambda_Q,\mathrm{bw})}_{\delta\alpha}(t)
&=& \frac12 e^{-E_{\Lambda_Q} t} \:\: Z_{\Lambda_Q}^2 \left[1+\gamma^0 \right]_{\delta\alpha} + \hdots ,
\end{eqnarray}
\begin{eqnarray}
\nonumber C^{(3)}_{\delta\alpha}(\Gamma,\:\mathbf{p'}, t, t') &=& Z_{\Lambda_Q} \: \frac{1}{2E_\Lambda}\frac12 
 \: e^{-E_\Lambda(t-t')} \:e^{-E_{\Lambda_Q}t'}  \left[(Z_{\Lambda}^{(1)}+Z_{\Lambda}^{(2)}\gamma^0)(m_\Lambda+\slashed{p}')
 \big( F_1 + \gamma^0 F_2 \big)\: \Gamma\: (1+\gamma^0)  \right]_{\delta\alpha} + \hdots, \\
\\
\nonumber C^{(3,\mathrm{bw})}_{\alpha\delta}(\Gamma,\:\mathbf{p'}, t, t-t') &=& Z_{\Lambda_Q}  \: \frac{1}{2E_\Lambda}
\frac12  \: e^{-E_{\Lambda_Q}(t-t')} \:e^{-E_\Lambda t'} \left[(1+\gamma^0) \: \overline{\Gamma}\:
\big( F_1 + \gamma^0 F_2 \big) (m_\Lambda+\slashed{p}')(Z_{\Lambda}^{(1)}+Z_{\Lambda}^{(2)}\gamma^0) \right]_{\alpha\delta} + \hdots, \\
\end{eqnarray}
where we have only shown the ground-state contributions, and the ellipsis denote excited-state contributions that decay
exponentially faster with the Euclidean time separations. For the three-point functions, we have used Eq.~(\ref{eq:FFdef})
to express the current matrix element in terms of the form factors $F_1$ and $F_2$. 

Using the three-point and two-point functions, we then define the following ratio,
\begin{equation}
 \mathcal{R}(\Gamma, \mathbf{p'}, t, t') = \frac{4 \:\mathrm{Tr}\left[  C^{(3)}(\Gamma,\:\mathbf{p'}, t, t')
 \:\: C^{(3,\mathrm{bw})}(\Gamma,\:\mathbf{p'}, t, t-t') \right] }{ \mathrm{Tr}[ C^{(2,\Lambda,{\rm av})}(\mathbf{p'}, t)]
 \:\mathrm{Tr}[ C^{(2,\Lambda_Q,{\rm av})}(t) ] }, \label{eq:doubleratio}
\end{equation}
where the traces are over spinor indices, and the two-point functions in the denominator are the averages of the
forward- and backward two-point functions (to increase statistics). For the ground-state contributions, the product of
forward and backward three-point functions in the numerator of Eq.~(\ref{eq:doubleratio}) eliminates the $t'$-dependence,
and dividing by the two-point functions evaluated at the same $t$ also cancels the $t$-dependence and $Z$-factors.
For gamma matrices $\Gamma$ that commute (anticommute) with $\gamma^0$, the ground-state contribution in the ratio
$\mathcal{R}(\Gamma, \mathbf{p'}, t, t')$  will be proportional to $[F_+]^2$ ($[F_-]^2$). Thus, we form the combinations
\begin{eqnarray}
 \mathcal{R}_+(\mathbf{p'}, t, t') &=& \frac14 \left[ \mathcal{R}(1, \mathbf{p'}, t, t') + \mathcal{R}(\gamma^2\gamma^3, \mathbf{p'}, t, t')
 + \mathcal{R}(\gamma^3\gamma^1, \mathbf{p'}, t, t') + \mathcal{R}(\gamma^1\gamma^2, \mathbf{p'}, t, t') \right],  \label{eq:curlyRplus} \\
 \mathcal{R}_-(\mathbf{p'}, t, t') &=& \frac14 \left[ \mathcal{R}(\gamma^1, \mathbf{p'}, t, t') + \mathcal{R}(\gamma^2, \mathbf{p'}, t, t')
 + \mathcal{R}(\gamma^3, \mathbf{p'}, t, t') + \mathcal{R}(\gamma_5, \mathbf{p'}, t, t') \right],  \label{eq:curlyRminus}
\end{eqnarray}
which are equal to
\begin{eqnarray}
 \mathcal{R}_+(\mathbf{p'}, t, t') &=& \frac{ E_\Lambda+m_\Lambda}{E_\Lambda} [F_+]^2 + \hdots,\\
 \mathcal{R}_-(\mathbf{p'}, t, t') &=& \frac{ E_\Lambda-m_\Lambda}{E_\Lambda} [F_-]^2 + \hdots, \label{eq:curlyRminusvalue}
\end{eqnarray}
where, as before, the ellipsis denote excited-state contributions. Note that multiplying the gamma matrices in
Eqs.~(\ref{eq:curlyRplus}) and (\ref{eq:curlyRminus}) with $\gamma^0$ would not give any new information, because
$\gamma^0 Q = Q$. Next, we average (\ref{eq:curlyRplus}) and (\ref{eq:curlyRminus}) over momenta $\mathbf{p'}$ with fixed magnitude
$|\mathbf{p'}|$, and replace the label $\mathbf{p'}$ by $|\mathbf{p'}|^2$ to denote the direction-averaged quantities,
\begin{equation}
 \mathcal{R}_\pm (|\mathbf{p'}|^2, t, t'). \label{eq:curlyRmomav}
\end{equation}
Finally, we evaluate $\mathcal{R}_\pm(|\mathbf{p'}|^2, t, t')$ at $t'=t/2$ (or average it over $(t-a)/2$ and $(t+a)/2$ for odd values of $t/a$)
where the excited-state contamination is smallest, rescale using $E_\Lambda(|\mathbf{p'}|^2)$ and $m_\Lambda$ obtained from
fits to the two-point functions, and take the square root to obtain
\begin{eqnarray}
 R_+(|\mathbf{p'}|^2, t) &=& \sqrt{\frac{E_\Lambda}{E_\Lambda+m_\Lambda} \mathcal{R}_+(|\mathbf{p'}|^2,\: t,\: t/2)}, \label{eq:Rplus}\\
 R_-(|\mathbf{p'}|^2, t) &=& \sqrt{\frac{E_\Lambda}{E_\Lambda-m_\Lambda} \mathcal{R}_-(|\mathbf{p'}|^2,\: t,\: t/2)}. \label{eq:Rminus}
\end{eqnarray}
For $t \rightarrow \infty$, the quantities $R_\pm(|\mathbf{p'}|^2, t)$ become equal to the form factors $F_\pm(E_\Lambda)$
where $E_\Lambda=E_\Lambda(|\mathbf{p'}|^2)$ .

\FloatBarrier
\subsection{\label{sec:latticeparams}Lattice parameters}
\FloatBarrier

The details of the domain-wall/Iwasaki gauge field ensembles generated by the RBC/UKQCD collaboration can be found in Ref.~\cite{Aoki:2010dy}.
In Table \ref{tab:params}, we summarize the main properties of the subset of ensembles used here, as well as some parameters of the domain-wall propagators
that we computed on them. There are ensembles with two different lattice spacings $a\approx0.11$ fm and $a\approx0.085$ fm, with lattice
dimensions of $24^3\times64$ and $32^3\times64$, respectively, so that the spatial box size is $L\approx 2.7$ fm
in both cases. We will refer to these two lattice spacings as ``coarse'' and ``fine''. At the coarse
lattice spacing, we use only one ensemble with the lightest available up/down sea-quark masses. At the fine lattice
spacing, we use two different ensembles.

\begin{table}
\begin{tabular}{cccccccccccccccccccccc}
\hline\hline
Set & \hspace{1ex} & $N_s^3\times N_t\times N_5$ & \hspace{1ex} & $a m_5$ & \hspace{1ex} & $am_{s}^{(\mathrm{sea})}$
& \hspace{1ex} & $am_{u,d}^{(\mathrm{sea})}$   & \hspace{1ex} & $a$ (fm) & \hspace{1ex} & $am_{s}^{(\mathrm{val})}$
& \hspace{1ex} & $am_{u,d}^{(\mathrm{val})}$  & \hspace{1ex} & $m_\pi^{(\mathrm{vv})}$ (MeV)
& \hspace{1ex} & $m_{\eta_s}^{(\mathrm{vv})}$ (MeV)  & \hspace{1ex} & $N_{\rm meas}$ \\
\hline
\texttt{C14} && $24^3\times64\times16$ && $1.8$ && $0.04$ && $0.005$ && $0.1119(17)$ && $0.04$ && $0.001$ && 245(4) && 761(12) && 2705 \\
\texttt{C24} && $24^3\times64\times16$ && $1.8$ && $0.04$ && $0.005$ && $0.1119(17)$ && $0.04$ && $0.002$ && 270(4) && 761(12) && 2683 \\
\texttt{C54} && $24^3\times64\times16$ && $1.8$ && $0.04$ && $0.005$ && $0.1119(17)$ && $0.04$ && $0.005$ && 336(5) && 761(12) && 2780 \\
\texttt{C53} && $24^3\times64\times16$ && $1.8$ && $0.04$ && $0.005$ && $0.1119(17)$ && $0.03$ && $0.005$ && 336(5) && 665(10) && 1192 \\
\texttt{F23} && $32^3\times64\times16$ && $1.8$ && $0.03$ && $0.004$ && $0.0849(12)$ && $0.03$ && $0.002$ && 227(3) && 747(10) && 1918 \\
\texttt{F43} && $32^3\times64\times16$ && $1.8$ && $0.03$ && $0.004$ && $0.0849(12)$ && $0.03$ && $0.004$ && 295(4) && 747(10) && 1919 \\
\texttt{F63} && $32^3\times64\times16$ && $1.8$ && $0.03$ && $0.006$ && $0.0848(17)$ && $0.03$ && $0.006$ && 352(7) && 749(14) && 2785 \\
\hline\hline
\end{tabular}
\caption{\label{tab:params} Parameters of the gauge field ensembles and quark propagators. Here, $N_5$ is the extent of the 5th
dimension of the lattice, and $a m_5$ is the domain-wall height \protect\cite{Aoki:2010dy}. The sea quark masses $am_q^{(\mathrm{sea})}$
were used in the generation of the ensembles, and we use the valence-quark masses $am_q^{(\mathrm{val})}$ when
computing domain-wall propagators. The values for the lattice spacings, $a$, are taken from Ref.~\protect\cite{Meinel:2010pv}.
We denote the valence pion masses by $m_\pi^{(\mathrm{vv})}$, and $m_{\eta_s}^{(\mathrm{vv})}$ is defined as the mass of the
pseudoscalar meson with valence strange-antistrange quarks, but without any disconnected contributions (we use $m_{\eta_s}^{(\mathrm{vv})}$
 to tune the strange-quark mass, using the approach of Ref.~\protect\cite{Davies:2009tsa}). Finally, $N_{\rm meas}$ is the number of
 light/strange domain-wall propagator pairs computed on each ensemble.}
\end{table}

In order to construct the correlation functions discussed in Sec.~\ref{sec:correlationfunctions}, we require domain-wall propagators
with Gaussian-smeared sources at $(x_0, \mathbf{x})$, and with masses corresponding to the strange quark as well as the
(degenerate) up/down quarks. As shown in Table \ref{tab:params}, we have seven different combinations of parameters,
which we denote as \texttt{C14}, \texttt{C24}, \texttt{C54}, \texttt{C53}, \texttt{F23}, \texttt{F43}, \texttt{F63}
(where \texttt{C}, \texttt{F} stand for ``coarse''and ``fine'', and the two digits indicate the light and strange valence quark masses).
In four of these combinations, the valence-quark masses are chosen to be lighter than the sea-quark masses (``partially quenched''),
while the other three combinations have valence-quark masses equal to the sea-quark masses (unitary case).
On each gauge configuration, we use $\mathcal{O}(10)$ source locations $(x_0, \mathbf{x})$ to increase statistics. The resulting total
numbers of ``measurements'', $N_{\rm meas}$, are listed in Table \ref{tab:params}. On each configuration, we average the correlators
over the source locations prior to further analysis.

In the static heavy-quark action, we use gauge links with one level of HYP smearing with the parameters
$(\alpha_1,\alpha_2,\alpha_3)=(1.0,\,1.0,\,0.5)$ as introduced in Ref.~\cite{Della Morte:2005yc}. The numerical values
of the matching coefficients needed for the current (\ref{eq:LHQETcurrent}) are taken from Ref.~\cite{Ishikawa:2011dd}
and are given in Table \ref{tab:renparams} for the choice of HYP smearing parameters used here.

\begin{table}
\begin{tabular}{cccccccccccc}
\hline\hline
$a$ (fm) & \hspace{1ex} & $U(m_b, a^{-1})$ & \hspace{1ex} & $u_0$  & \hspace{1ex} & $\mathcal{Z}$
& \hspace{1ex} & $c^{(m_s a)}_\Gamma$ & \hspace{1ex} & $c^{(p_s a)}_\Gamma$ \\
\hline
0.112  && 1.09964 &&  0.875789 && 0.9383 && $-0.1660 \: G_\Gamma$ && $-0.1374 \: G_\Gamma$ \\
0.085  && 1.06213 &&  0.885778 && 0.9526 && $-0.1482 \: G_\Gamma$ && $-0.1294 \: G_\Gamma$ \\
\hline\hline
\end{tabular}
\caption{\label{tab:renparams} Renormalization parameters for the matching of LHQET to HQET in the $\overline{\rm MS}$
scheme, from Ref.~\protect\cite{Ishikawa:2011dd}. Here, $G_\Gamma$ is defined by $\gamma^0 \Gamma \gamma^0 = G_\Gamma \Gamma$,
so that $G_\Gamma=+1$ if $\Gamma$ commutes with $\gamma^0$, and $G_\Gamma=-1$ if $\Gamma$ anticommutes with $\gamma^0$.}
\end{table}

\FloatBarrier
\subsection{\label{sec:resultsR+R-}Results for $\mathcal{R}_+$ and $\mathcal{R}_-$}
\FloatBarrier

\begin{figure}[!b]
\includegraphics[height=4.5cm]{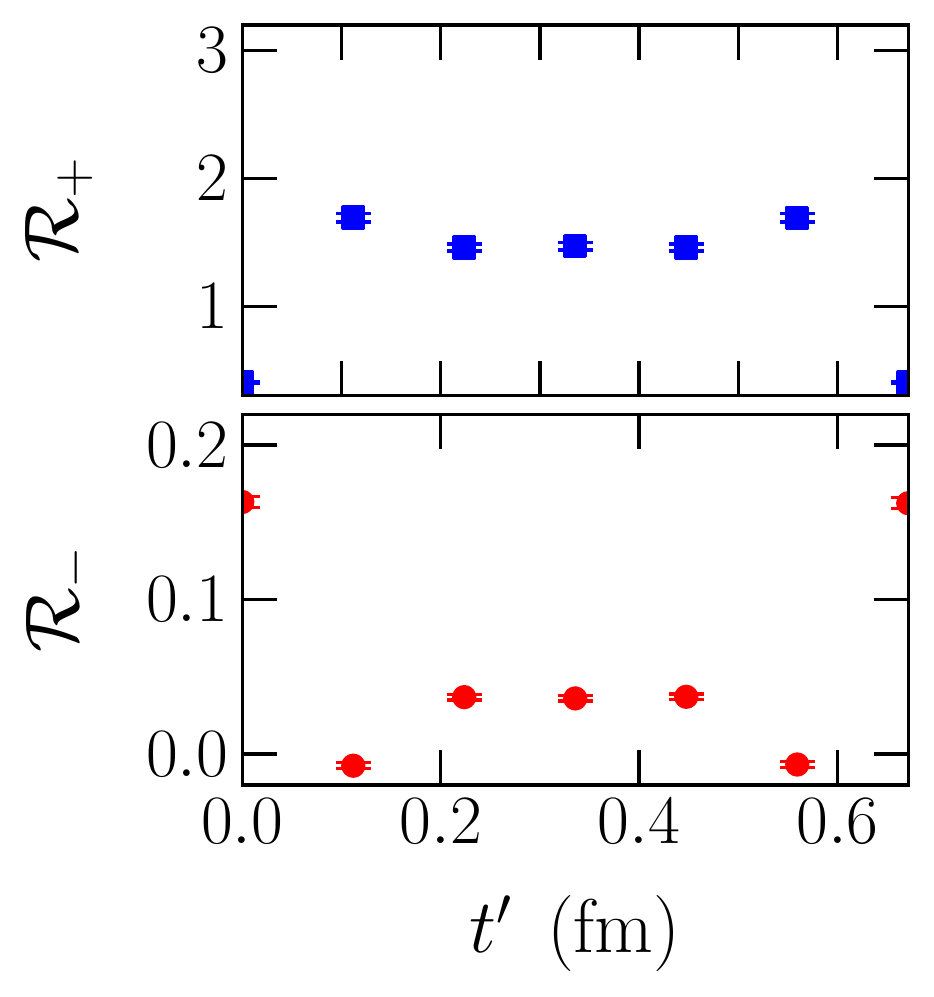}\includegraphics[height=4.5cm]{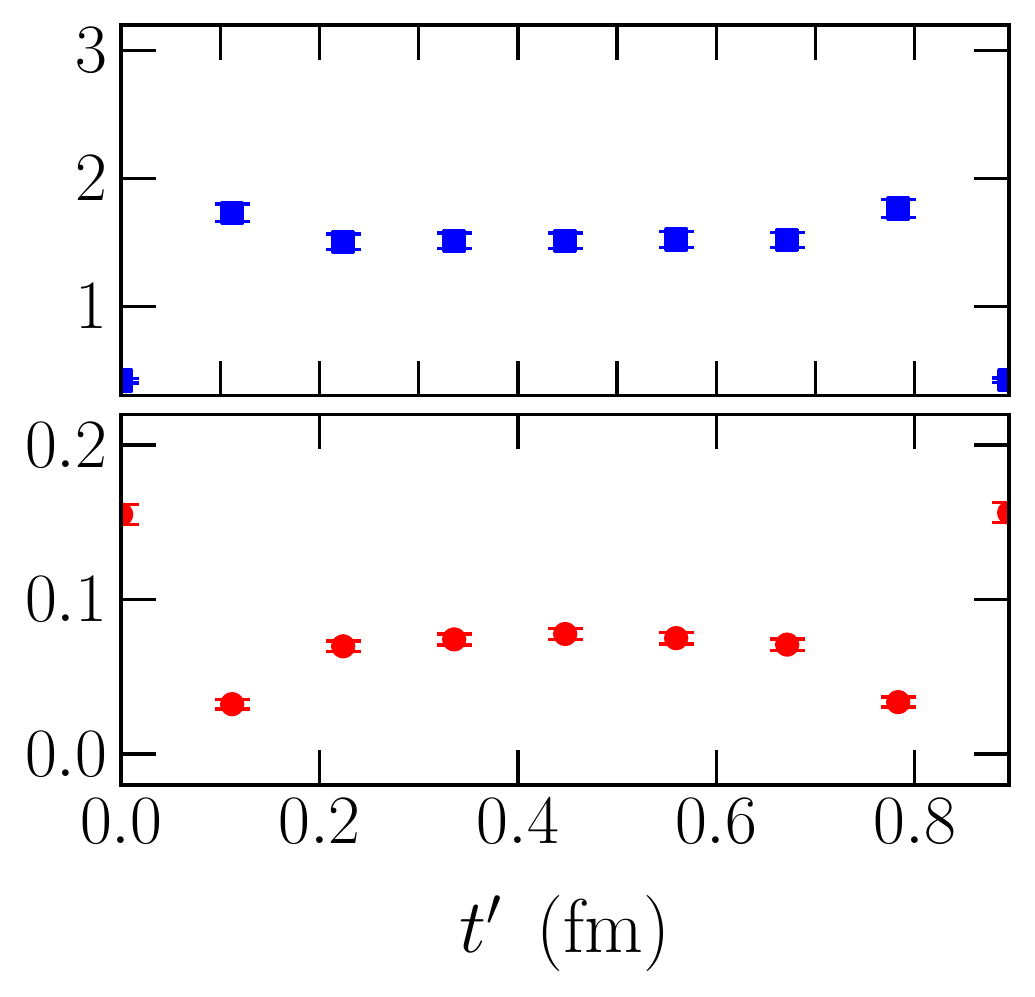}\includegraphics[height=4.5cm]{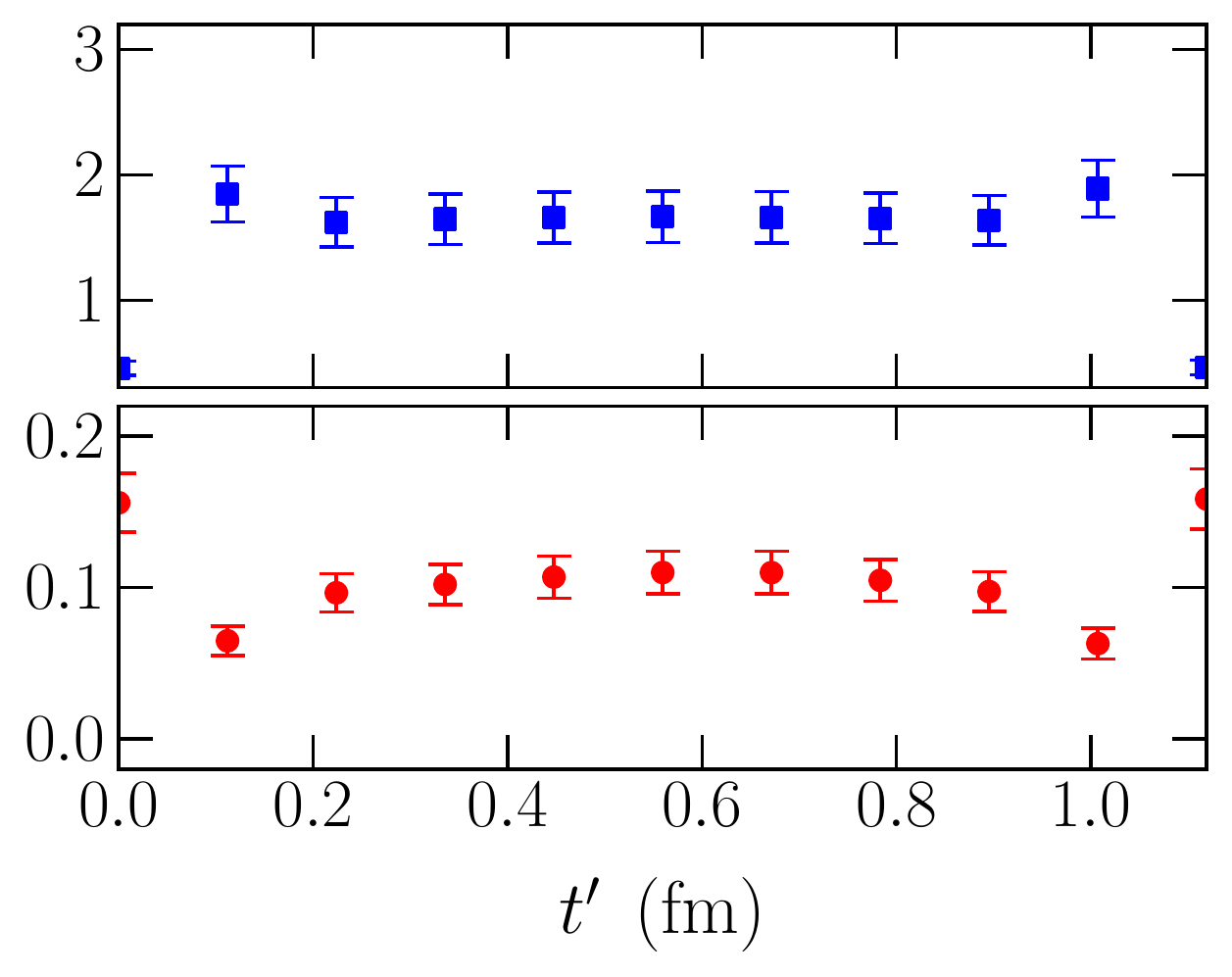}
\caption{\label{fig:ratios_L24_005_psqr1}Numerical results for $\mathcal{R}_\pm (|\mathbf{p'}|^2, t, t')$ at $|\mathbf{p'}|^2=1\cdot(2\pi/L)^2$
from the \texttt{C54} data set ($a=0.112$ fm, $am_{s}^{(\mathrm{val})}=0.04$, $am_{u,d}^{(\mathrm{val})}=0.005$).
The source-sink separations shown here are (from left to right) $t/a=6,8,10$.}
\end{figure}

\begin{figure}[!b]
\includegraphics[height=4.5cm]{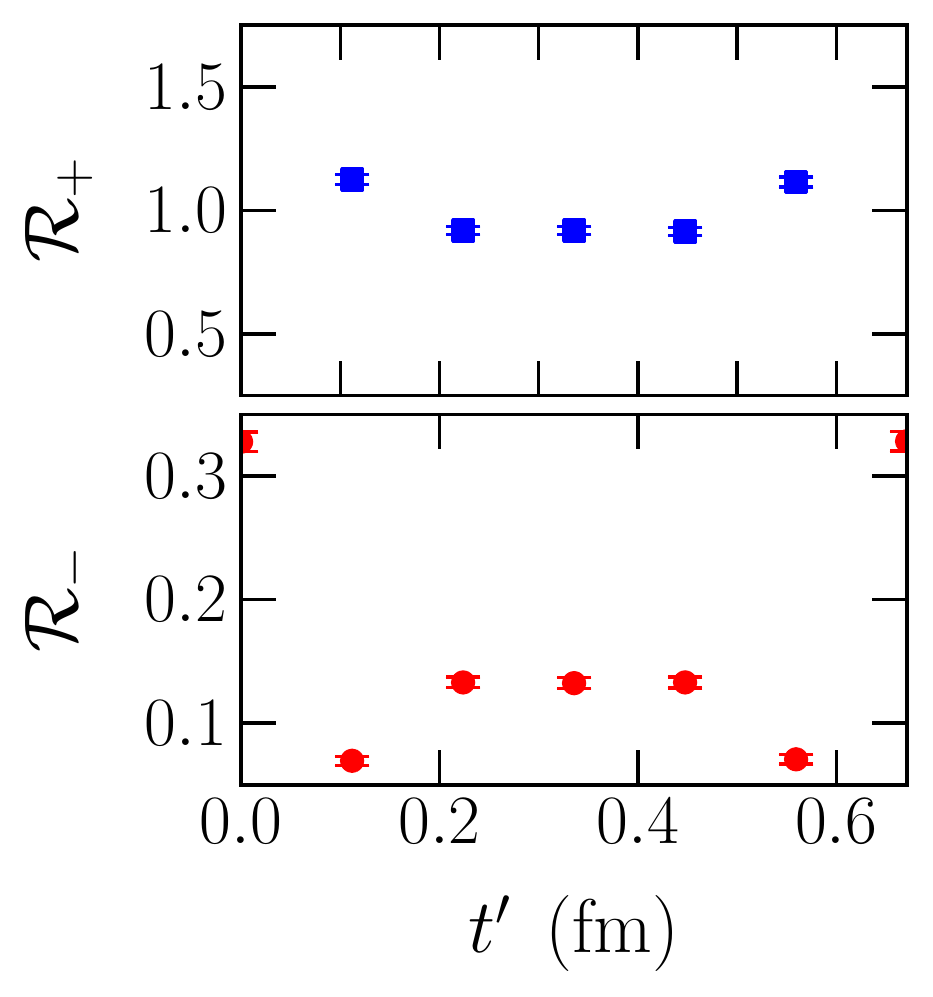}\includegraphics[height=4.5cm]{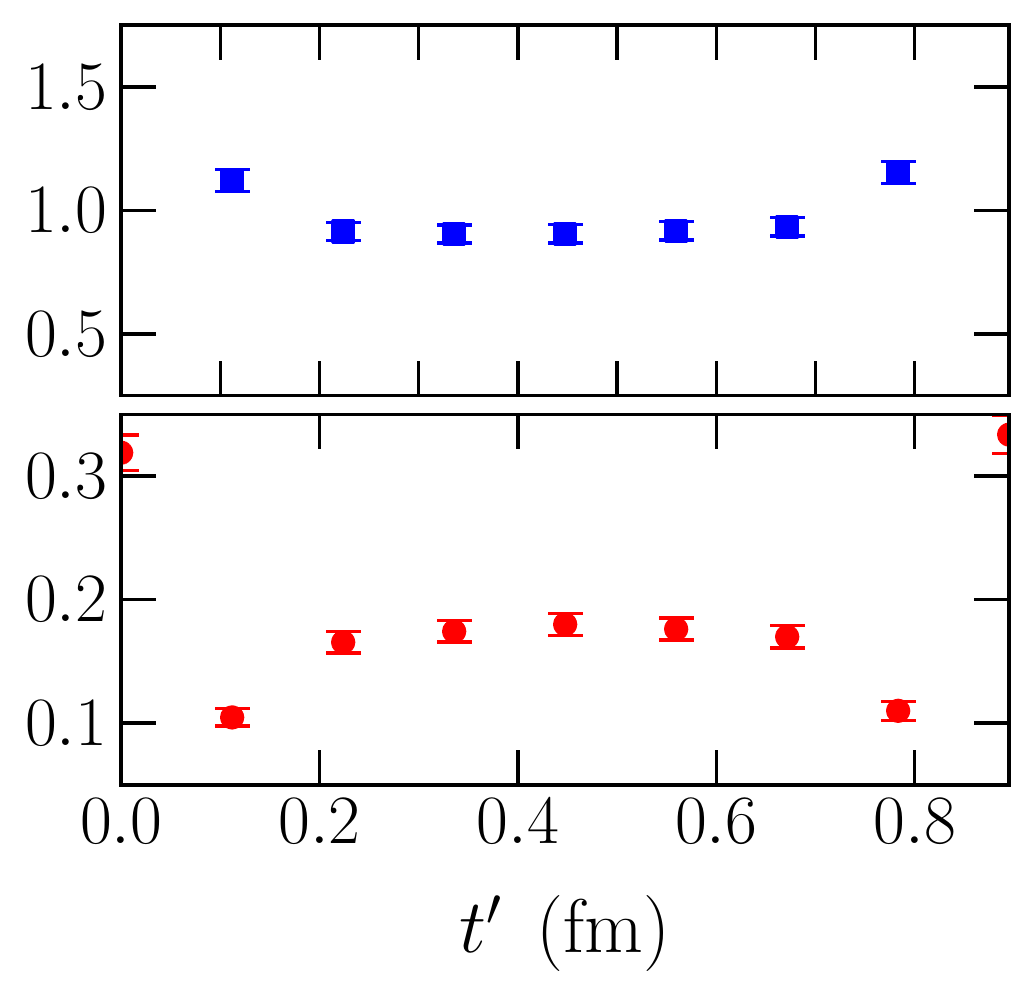}\includegraphics[height=4.5cm]{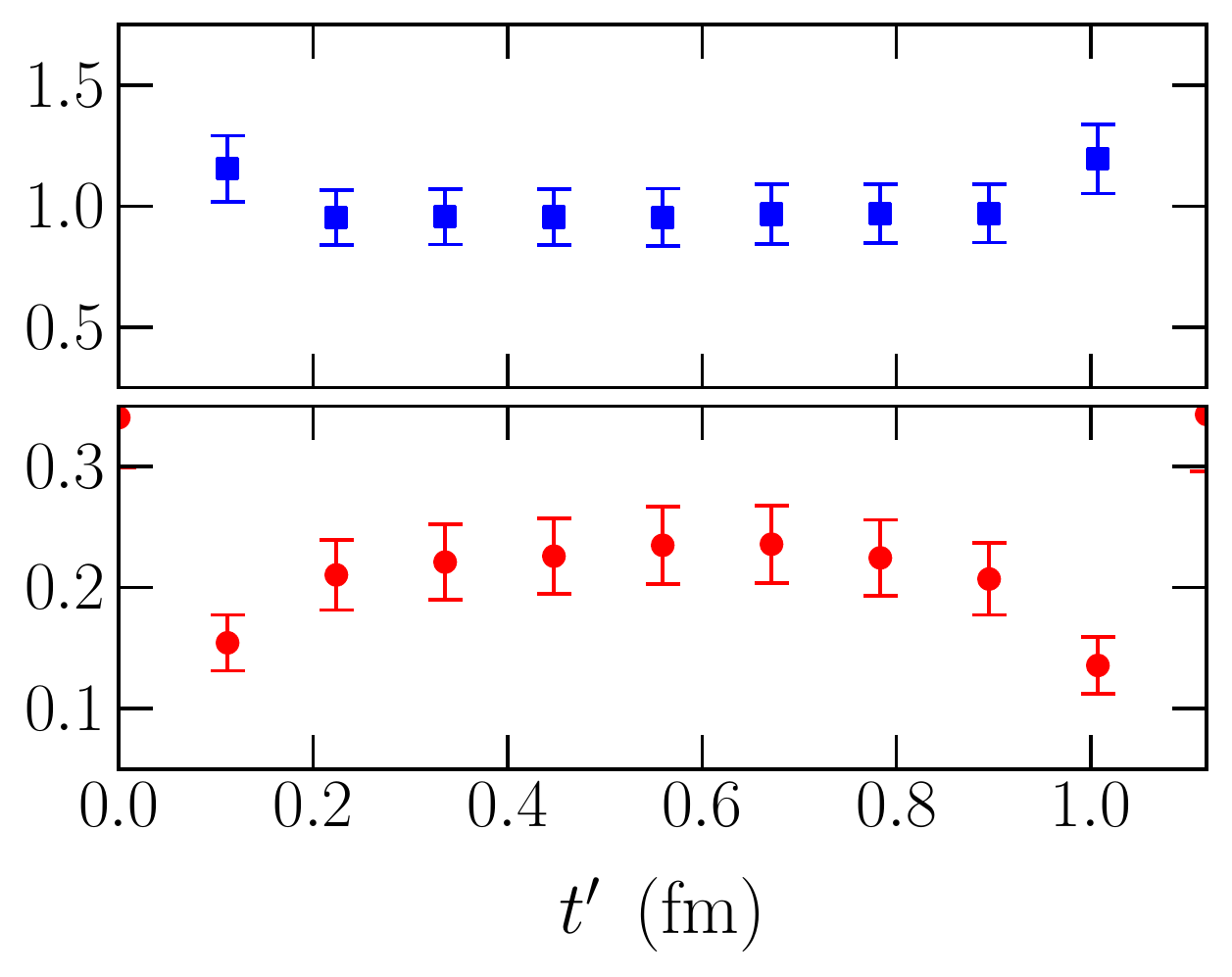}
\caption{\label{fig:ratios_L24_005_psqr4}Like Fig.~\protect\ref{fig:ratios_L24_005_psqr1}, but at $|\mathbf{p'}|^2=4\cdot(2\pi/L)^2$.}
\end{figure}

\begin{figure}[!b]
\includegraphics[height=4.5cm]{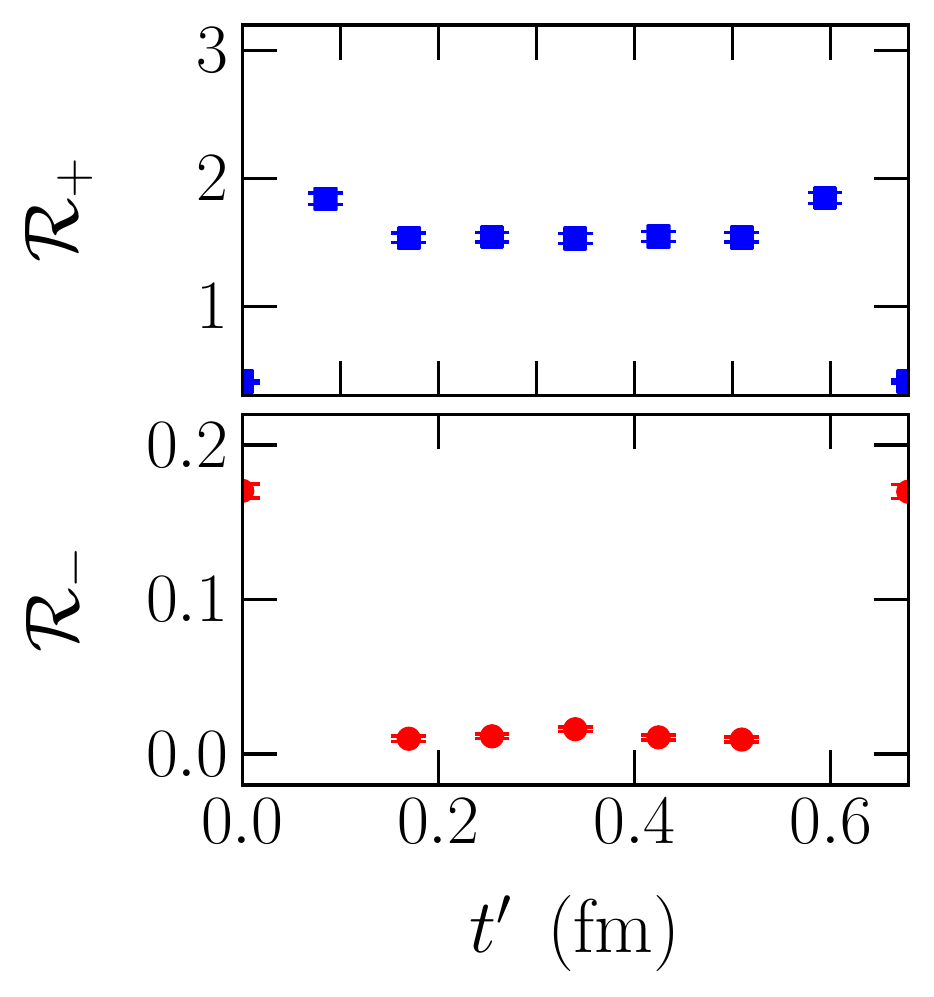}\includegraphics[height=4.5cm]{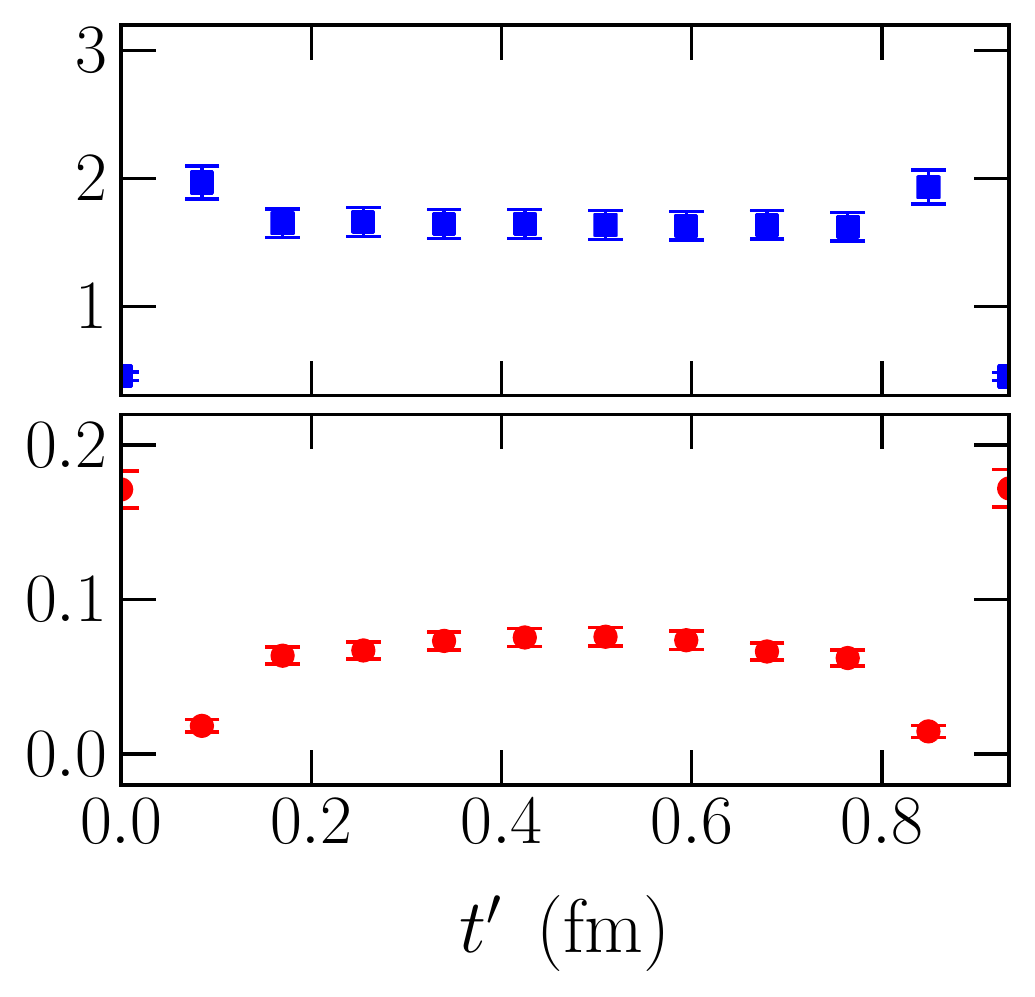}\includegraphics[height=4.5cm]{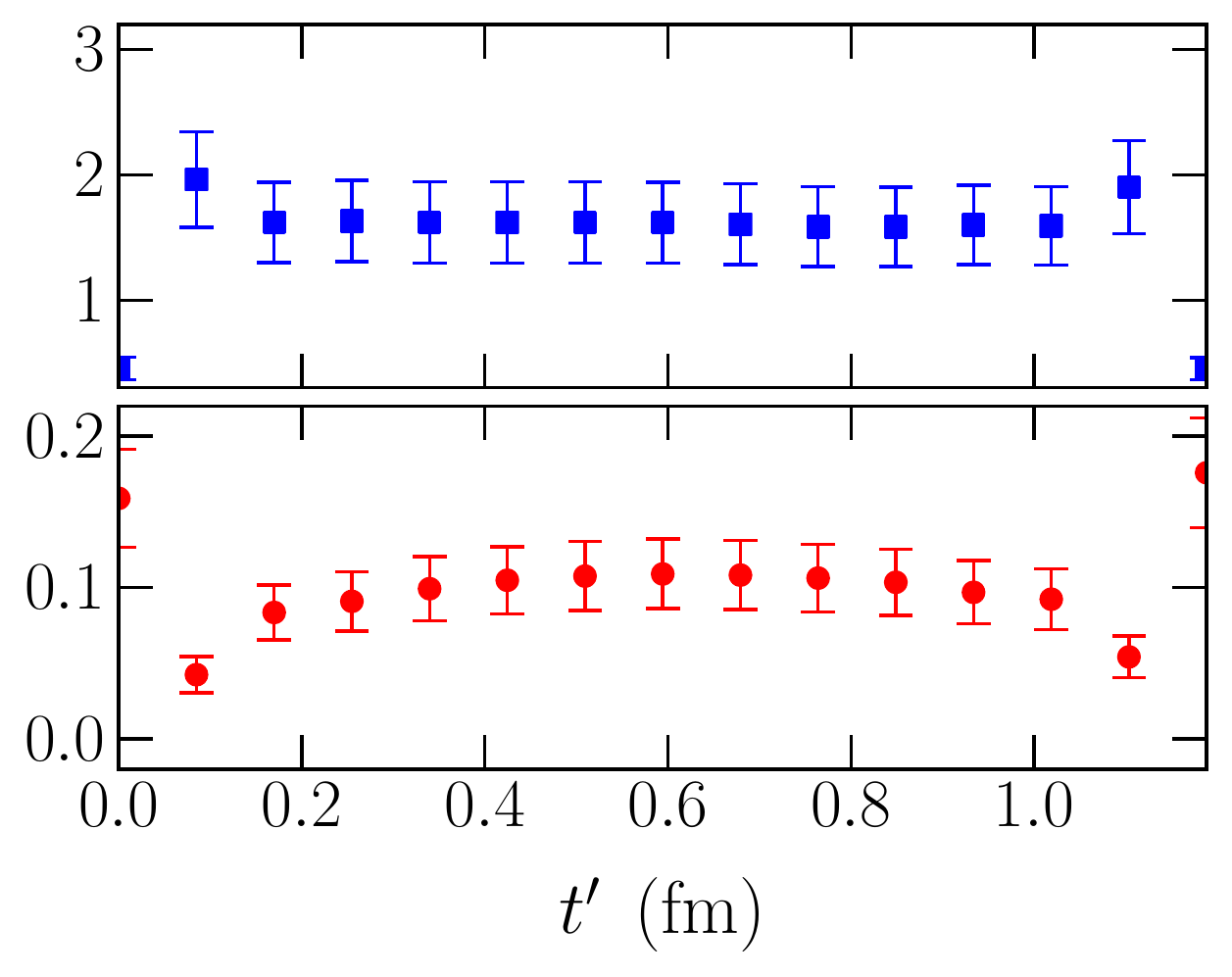}
\caption{\label{fig:ratios_L32_004_psqr1}Numerical results for $\mathcal{R}_\pm (|\mathbf{p'}|^2, t, t')$ at $|\mathbf{p'}|^2=1\cdot(2\pi/L)^2$
from the \texttt{F43} data set ($a=0.085$ fm, $am_{s}^{(\mathrm{val})}=0.03$, $am_{u,d}^{(\mathrm{val})}=0.004$).
The source-sink separations shown here are (from left to right) $t/a=8,11,14$.}
\end{figure}

\begin{figure}
\includegraphics[height=4.5cm]{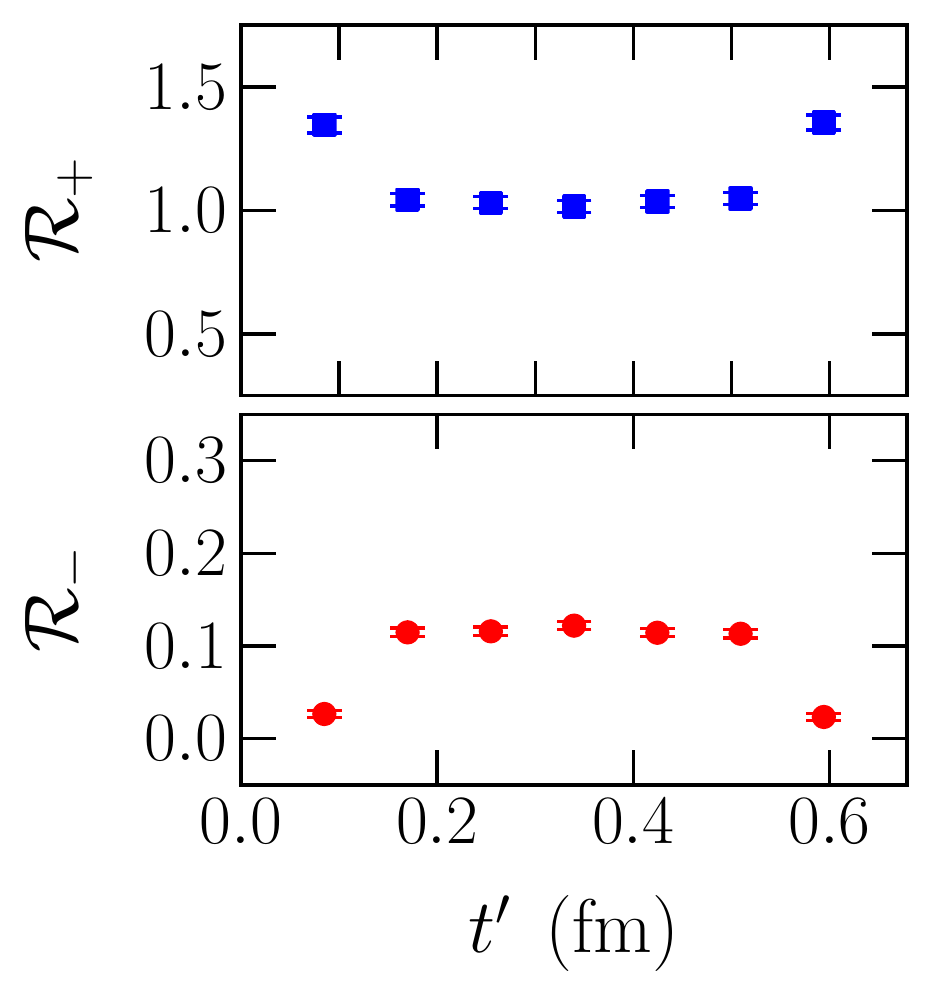}\includegraphics[height=4.5cm]{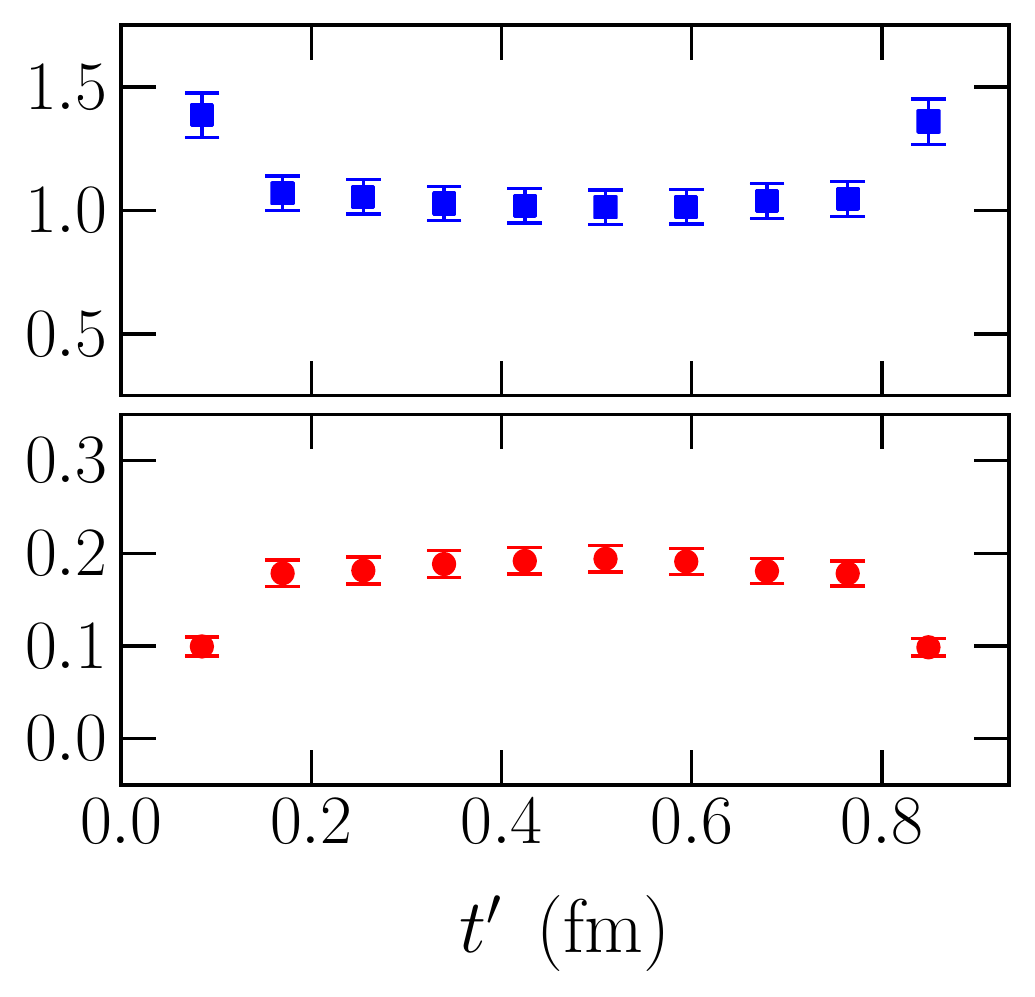}\includegraphics[height=4.5cm]{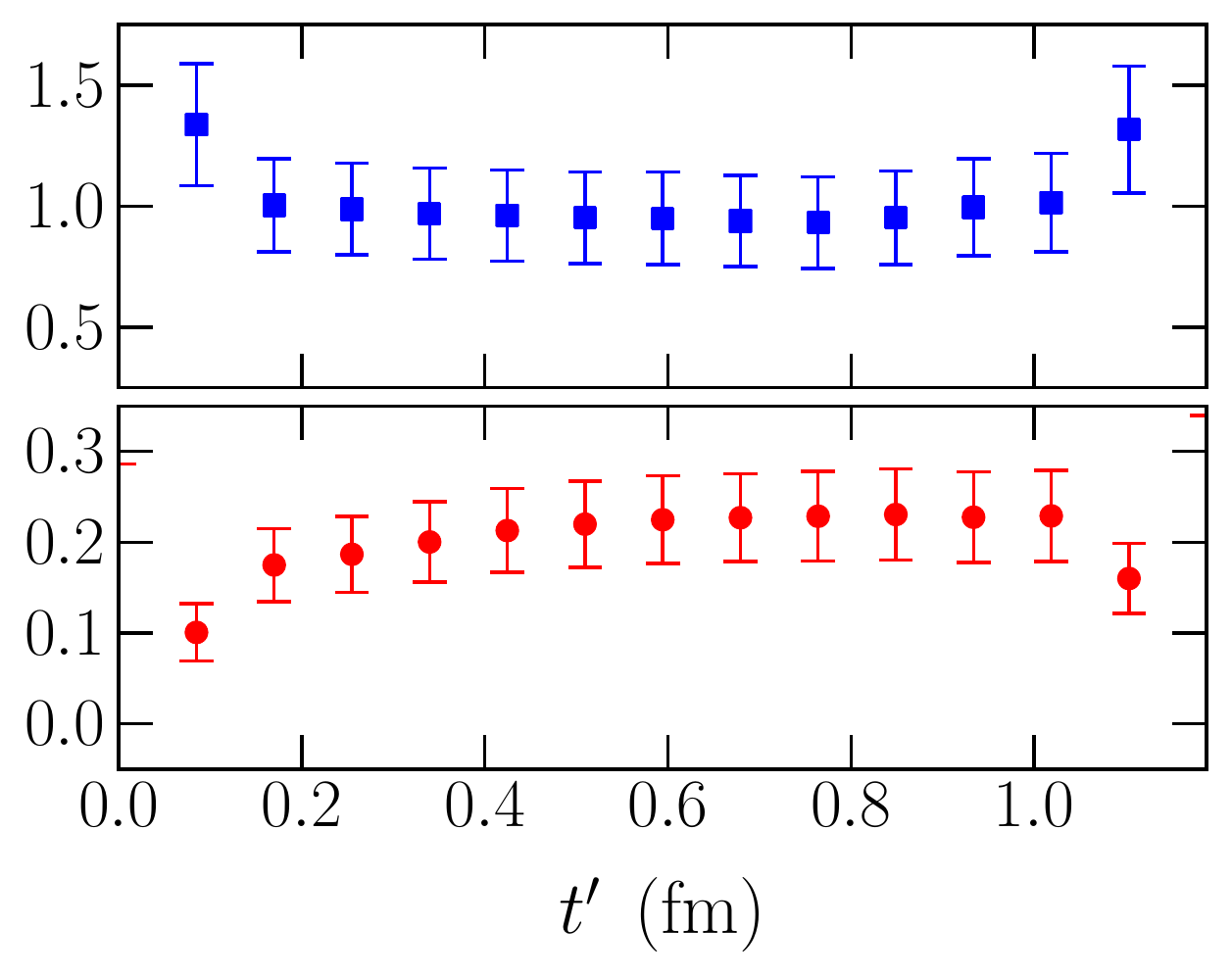}
\caption{\label{fig:ratios_L32_004_psqr4}Like Fig.~\protect\ref{fig:ratios_L32_004_psqr1}, but at $|\mathbf{p'}|^2=4\cdot(2\pi/L)^2$.}
\end{figure}

At the coarse lattice spacing, we computed the three-point functions (\ref{eq:threept}), (\ref{eq:threeptbw})
for the source-sink separations $t/a=4,\:5,\:...,\:15$, and at the fine lattice spacing for $t/a=5,\:6,\:...,\:20$.
We computed these three-point functions for lattice momenta $\mathbf{p'}$ with $0 \leq |\mathbf{p'}|^2 \leq 9\cdot (2\pi)^2/L^2$.
We then constructed the quantities (\ref{eq:curlyRplus}) and (\ref{eq:curlyRminus}) using statistical bootstrap with 1000 samples.
When performing the momentum direction average for the largest momentum $|\mathbf{p'}|^2 = 9\cdot (2\pi)^2/L^2$,
we used only $\mathbf{p'}=(2,2,1)\cdot 2\pi/L$ and lattice symmetries applied to that (for $|\mathbf{p'}|^2<9\cdot (2\pi)^2/L^2$,
all possible $\mathbf{p'}$ with the same magnitude are related by cubic rotations and reflections, and we average over all of them).
Examples of numerical results for the quantities $\mathcal{R}_\pm (|\mathbf{p'}|^2, t, t')$ defined in
Sec.~\ref{sec:correlationfunctions} are shown in Figs. \ref{fig:ratios_L24_005_psqr1}-\ref{fig:ratios_L32_004_psqr4}.
Except in the immediate neighborhood of $t'=0$ and $t'=t$, the results for $\mathcal{R}_\pm (|\mathbf{p'}|^2, t, t')$
show only a weak dependence on the current-insertion time $t'$. However, a significant dependence on the source-sink
separation $t$ is seen, in particular for $\mathcal{R}_-$. Consequently, we need to extrapolate the results to
infinite source-sink separation in order to remove the excited-state contamination. We perform these extrapolations
for $R_\pm(|\mathbf{p'}|^2, t)$ [defined in Eqs.~(\ref{eq:Rplus}), (\ref{eq:Rminus})], as discussed in
Sec.~\ref{sec:srcsnkextrap}.

\FloatBarrier
\subsection{\label{sec:lambda2ptfits} Fitting the $\Lambda$ two-point functions}
\FloatBarrier

\begin{figure}
\includegraphics[width=0.48\linewidth]{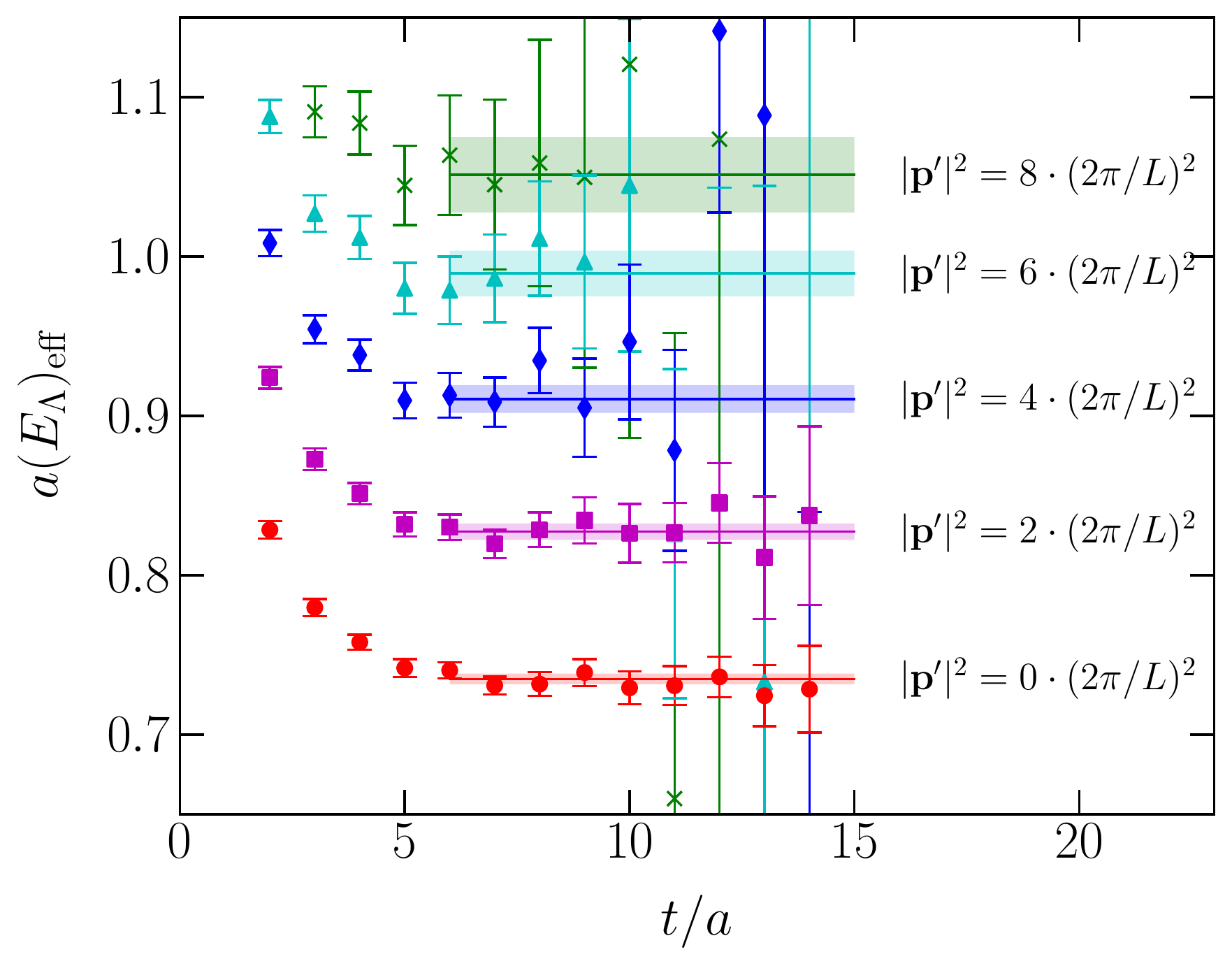}\hfill\includegraphics[width=0.48\linewidth]{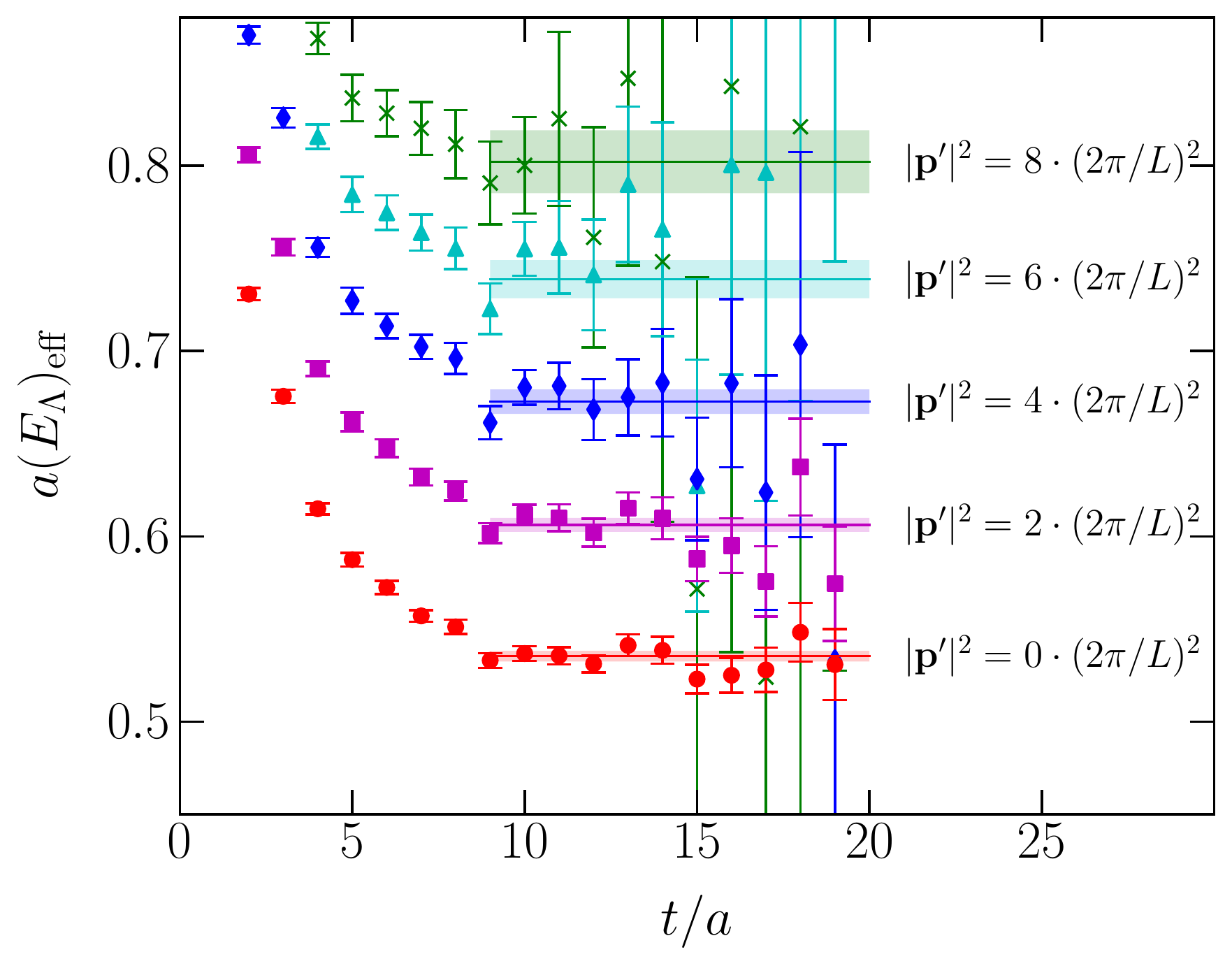}
\caption{\label{fig:lambda2ptem}Effective-energy plots for the $\Lambda$ two-point functions at selected momenta.
The shaded bands indicate the energies extracted from exponential fits of the two-point functions, as well as the $t$-range
used for these fits. Left panel: $a=0.112$ fm, $am_{s}^{(\mathrm{val})}=0.04$, $am_{u,d}^{(\mathrm{val})}=0.005$;
right panel: $a=0.085$ fm, $am_{s}^{(\mathrm{val})}=0.03$, $am_{u,d}^{(\mathrm{val})}=0.004$.}
\end{figure}

In order to compute the quantities $R_\pm(|\mathbf{p'}|^2, t)$, which were defined in Eqs.~(\ref{eq:Rplus}), (\ref{eq:Rminus}),
we use the energy $E_\Lambda(|\mathbf{p'}|^2)$ and mass $m_\Lambda=E_\Lambda(0)$ of the $\Lambda$ baryon computed on the
lattice for the same data set. To this end, we average the two-point function $\mathrm{Tr}[ C^{(2,\Lambda,{\rm av})}(\mathbf{p'}, t)]$
over momenta $\mathbf{p'}$ with the same $|\mathbf{p'}|^2$, and perform correlated fits of the form $A\: e^{-E_\Lambda t}$
for sufficiently large $t$ so that the excited-state contamination in the fitted $E_\Lambda$ is negligible
compared to the statistical uncertainty. To give an idea of the signal quality, we show the effective energy
$(a E_\Lambda)_{\rm eff}=\ln[C(t)/C(t+a)]$ for selected momenta and two data sets in Fig.~\ref{fig:lambda2ptem}.
The complete fit results for $a E_\Lambda$ are listed in Table \ref{tab:ELambda}. When computing $R_\pm(|\mathbf{p'}|^2, t)$
via Eqs.~(\ref{eq:Rplus}), (\ref{eq:Rminus}), we used bootstrap to fully take into account the correlations between
$E_\Lambda$, $m_\Lambda$, and $\mathcal{R}_\pm(|\mathbf{p'}|^2,\: t,\: t/2)$.

\begin{table}
\begin{tabular}{cccccccccccccccc}
\hline\hline
$|\mathbf{p'}|^2/(2\pi/L)^2$ & \hspace{1ex} & \texttt{C14} & \hspace{1ex} & \texttt{C24} & \hspace{1ex} & \texttt{C54}
& \hspace{1ex} & \texttt{C53}   & \hspace{1ex} & \texttt{F23} & \hspace{1ex} & \texttt{F43} & \hspace{1ex} & \texttt{F63} \\
\hline
0 &&  $0.7046(41)$ && $0.7135(45)$ && $0.7350(34)$ && $0.7150(42)$ && $0.5191(41)$ && $0.5353(29)$ && $0.5540(17)$ \\
1 &&  $0.7542(46)$ && $0.7613(53)$ && $0.7821(40)$ && $0.7653(49)$ && $0.5575(46)$ && $0.5717(32)$ && $0.5894(19)$ \\
2 &&  $0.8027(58)$ && $0.8061(69)$ && $0.8274(52)$ && $0.8151(62)$ && $0.5931(53)$ && $0.6062(38)$ && $0.6240(24)$ \\
3 &&  $0.8460(76)$ && $0.8495(97)$ && $0.8709(69)$ && $0.8612(84)$ && $0.6288(67)$ && $0.6422(50)$ && $0.6568(31)$ \\
4 &&  $0.889(10)$  && $0.896(11)$  && $0.9106(87)$ && $0.892(11)$  && $0.6652(89)$ && $0.6727(66)$ && $0.6892(43)$ \\
5 &&  $0.921(11)$  && $0.929(13)$  && $0.948(11)$  && $0.928(13)$  && $0.695(10)$  && $0.7033(79)$ && $0.7201(52)$ \\
6 &&  $0.954(16)$  && $0.965(19)$  && $0.989(14)$  && $0.966(17)$  && $0.730(13)$  && $0.739(10)$  && $0.7479(64)$ \\
8 &&  $1.006(27)$  && $1.021(31)$  && $1.051(24)$  && $1.014(30)$  && $0.800(23)$  && $0.802(17)$  && $0.804(11)$  \\
9 &&  $1.033(35)$  && $1.066(45)$  && $1.087(32)$  && $1.058(40)$  && $0.839(30)$  && $0.837(21)$  && $0.822(14)$  \\
\hline\hline
\end{tabular}
\caption{\label{tab:ELambda} Fit results for $a E_\Lambda(|\mathbf{p'}|^2)$ from the different data sets
(see Table \protect\ref{tab:params}).}
\end{table}

\FloatBarrier
\subsection{\label{sec:srcsnkextrap}Extrapolation of $R_+$ and $R_-$ to infinite source-sink separation}
\FloatBarrier

Using the methods outlined in the previous sections, we have obtained numerical results for $R_\pm(|\mathbf{p'}|^2, t)$
for multiple momenta $|\mathbf{p'}|^2$, for a wide range of source-sink separations $t$, and for seven different
combinations of quark masses and lattice spacings. The next step is to compute the ground-state form factors $F_\pm$
by extrapolating $R_\pm$ to infinite source-sink separation. In the following, we use the notation $R^{i,n}_\pm(t)$,
where  $i=\mathtt{C14}, \mathtt{C24}, ..., \mathtt{F63}$ labels the data set (see Table \ref{tab:params}),
and $n=0,1,2,3,4,5,6,8,9$ labels the momentum of the $\Lambda$, writing $|\mathbf{p'}|^2=n\cdot (2\pi)^2/L^2$.

\begin{figure}
\includegraphics[width=0.47\linewidth]{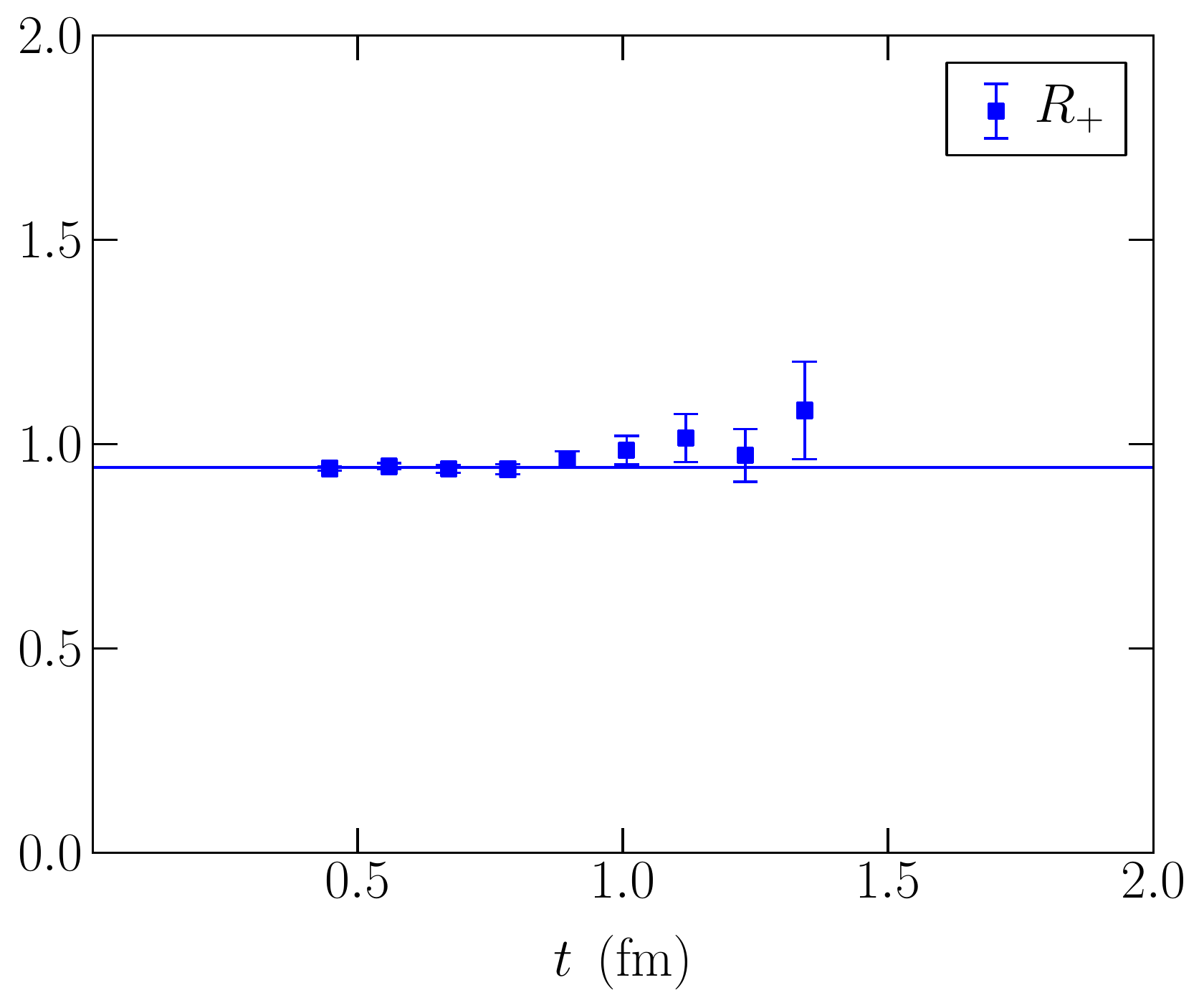}\hfill\includegraphics[width=0.47\linewidth]{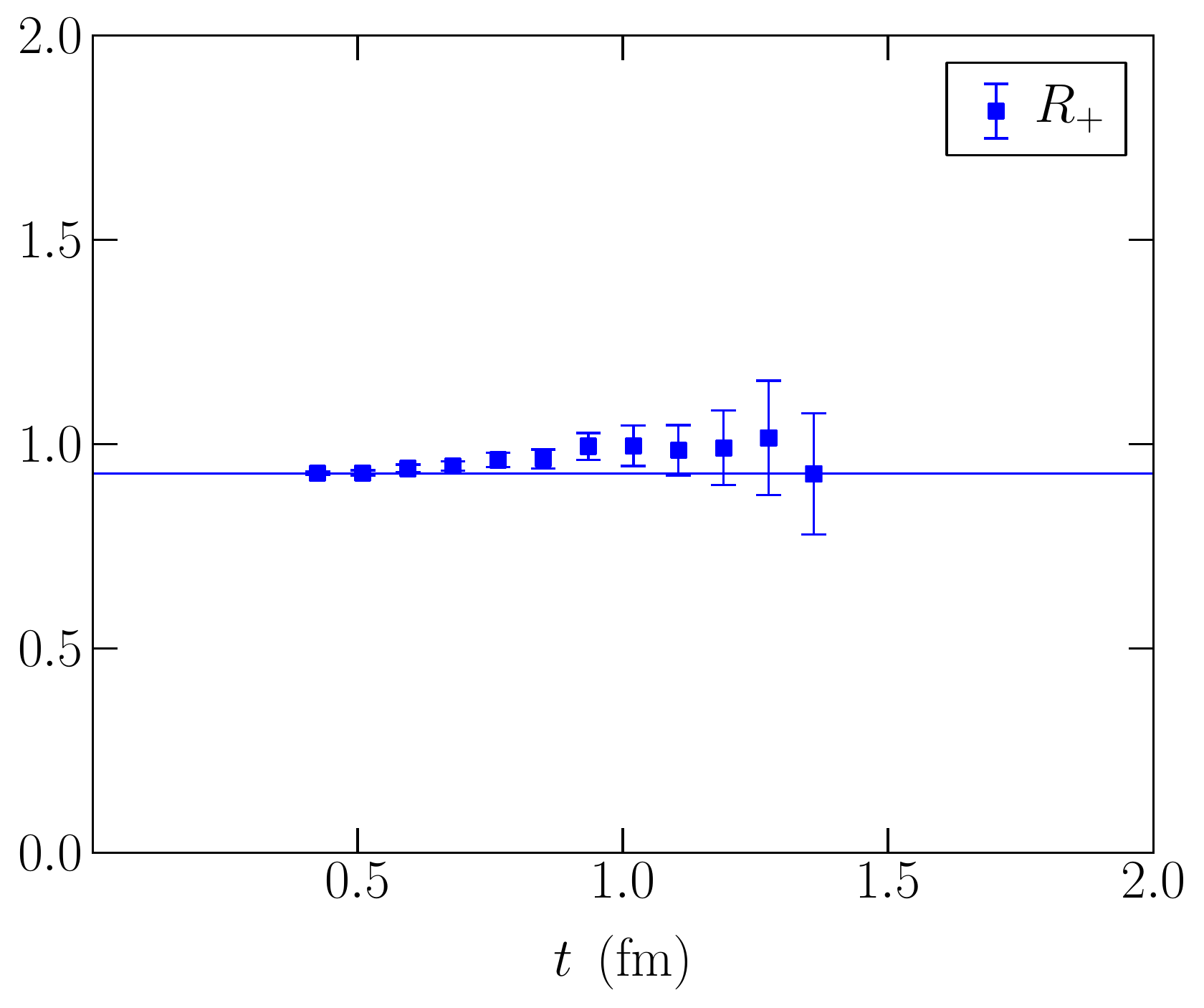}
\caption{\label{fig:tsep_dependence_psqr0}Constant fits to the $t$-dependence of $R_+(t)$ at $|\mathbf{p'}|^2=0$.
Left panel: $a=0.112$ fm, $am_{s}^{(\mathrm{val})}=0.04$, $am_{u,d}^{(\mathrm{val})}=0.005$;
right panel: $a=0.085$ fm, $am_{s}^{(\mathrm{val})}=0.03$, $am_{u,d}^{(\mathrm{val})}=0.004$.
The fits have $\chi^2/{\rm dof}$ values of 0.92 and 0.78, respectively.}
\end{figure}

At zero momentum, we can only compute $R_+$ because $\mathcal{R}_-$ vanishes for $E_\Lambda=m_\Lambda$
[see Eq.~(\ref{eq:curlyRminusvalue})]. Results for $R^{i,0}_+(t)$ from the two data sets $i=\mathtt{C54},\:\mathtt{F43}$
are plotted in Fig.~\ref{fig:tsep_dependence_psqr0} as a function of the source-sink separation $t$. Remarkably, no
significant $t$-dependence is found in this quantity, allowing constant fits of the form
\begin{equation}
R^{i,0}_+(t) = F^{i,0}_+,
\end{equation}
which are also shown in Fig.~\ref{fig:tsep_dependence_psqr0}. The fits fully take into account correlations and have
$\chi^2/{\rm dof}<1$, showing that there is indeed no evidence for deviation from a constant. Note that $R_+$ can be obtained
from the vector current $\bar{s} \gamma^0 b$ (in our calculation we have replaced $\gamma^0$ by $1$ because $\gamma^0 Q=Q$
for static heavy quarks). In relativistic QCD, neglecting mass effects, charge conservation would then prevent any
$t$-dependence at zero momentum, because excited states have the same charge as ground states. It appears that some
remnant of this symmetry also remains in our case.

At non-zero $\Lambda$-momentum, both $R_+$ and $R_-$ can be extracted, and significant $t$-dependence
is seen. To extract the ground-state contributions $F_+$ and $F_-$, we perform fits including an exponential
correction term that describes the leading effects of excited states,
\begin{equation}
R^{i,n}_\pm(t) = F^{i,n}_\pm + A^{i,n}_\pm \:\exp[-\delta^{i,n}_\pm\:t], \label{eq:tdepsep}
\end{equation}
where $F^{i,n}_\pm$, $A^{i,n}_\pm$, and $\delta^{i,n}_\pm$ are the fit parameters, which explicitly depend on the
data set $i$ and the momentum $n$. Because the energy gaps $\delta^{i,n}_\pm$ are positive by definition, we write
\begin{equation}
 \delta^{i,n}_\pm /(1\: {\rm GeV}) = \exp(l^{i,n}_\pm), \label{eq:expgap}
\end{equation}
and use $l^{i,n}_\pm$ instead of $\delta^{i,n}_\pm$ as the fit parameters. 
The fits using Eq.~(\ref{eq:expgap}) are performed separately for each momentum $n$, but simultaneously for
the different data sets $i$. Note that the size of the momentum unit, $(2\pi)/L$ (in GeV),
is the same at the coarse and fine lattice spacings within uncertainties, because the box sizes (in physical units)
are equal within uncertainties. Performing the fits simultaneously
for the different data sets at the same momentum allows us to use the prior knowledge that
the hadron spectrum does not change dramatically when the lattice spacing or quark masses are varied by small amounts.
To this end, we augment the $\chi^2$ function used to perform the fits to Eq.~(\ref{eq:tdepsep}) as follows:
\begin{eqnarray}
\nonumber \chi^2 &\rightarrow \chi^2 & + \frac{(l^{\mathtt{C14},n}_\pm-l^{\mathtt{C24},n}_\pm)^2}{[\sigma_m^{\mathtt{C14},\mathtt{C24}}]^2}
+ \frac{(l^{\mathtt{C24},n}_\pm-l^{\mathtt{C54},n}_\pm)^2}{[\sigma_m^{\mathtt{C24},\mathtt{C54}}]^2}
+ \frac{(l^{\mathtt{C54},n}_\pm-l^{\mathtt{C53},n}_\pm)^2}{[\sigma_m^{\mathtt{C54},\mathtt{C53}}]^2} \\
&& + \frac{(l^{\mathtt{F23},n}_\pm-l^{\mathtt{F43},n}_\pm)^2}{[\sigma_m^{\mathtt{F23},\mathtt{F43}}]^2}
+ \frac{(l^{\mathtt{F43},n}_\pm-l^{\mathtt{F63},n}_\pm)^2}{[\sigma_m^{\mathtt{F43},\mathtt{F63}}]^2}
+ \frac{(l^{\mathtt{C54},n}_\pm-l^{\mathtt{F63},n}_\pm)^2}{[\sigma_m^{\mathtt{C54},\mathtt{F63}}]^2+\sigma_a^2}, \label{eq:chisqr}
\end{eqnarray}
where
\begin{equation}
 [\sigma_m^{i,j}]^2 = w_m^2 [ (m_\pi^i)^2-(m_\pi^j)^2 ]^2 + w_m^2 [ (m_{\eta_s}^i)^2-(m_{\eta_s}^j)^2 ]^2,
\end{equation}
with $w_m = 4\:{\rm GeV}^{-2}$, and $\sigma_a=0.1$. With these parameters, Eq.~(\ref{eq:chisqr}) implements the constraint
that the energy gaps, at given $\Lambda$-momentum $n$, should not change by more than 10\% when going from the fine to the
coarse lattice spacing, and not more than 400\% times the change in $m_\pi^2$ or $m_{\eta_s}^2$ (in ${\rm GeV}^2$).
Note that absolute variations of $l^{i,n}_\pm$ translate to relative variations of $\delta^{i,n}_\pm$, because
$\mathrm{d}[\exp(l^{i,n}_\pm)]/\exp(l^{i,n}_\pm) = \mathrm{d} l^{i,n}_\pm$.

\begin{figure}
\includegraphics[width=0.47\linewidth]{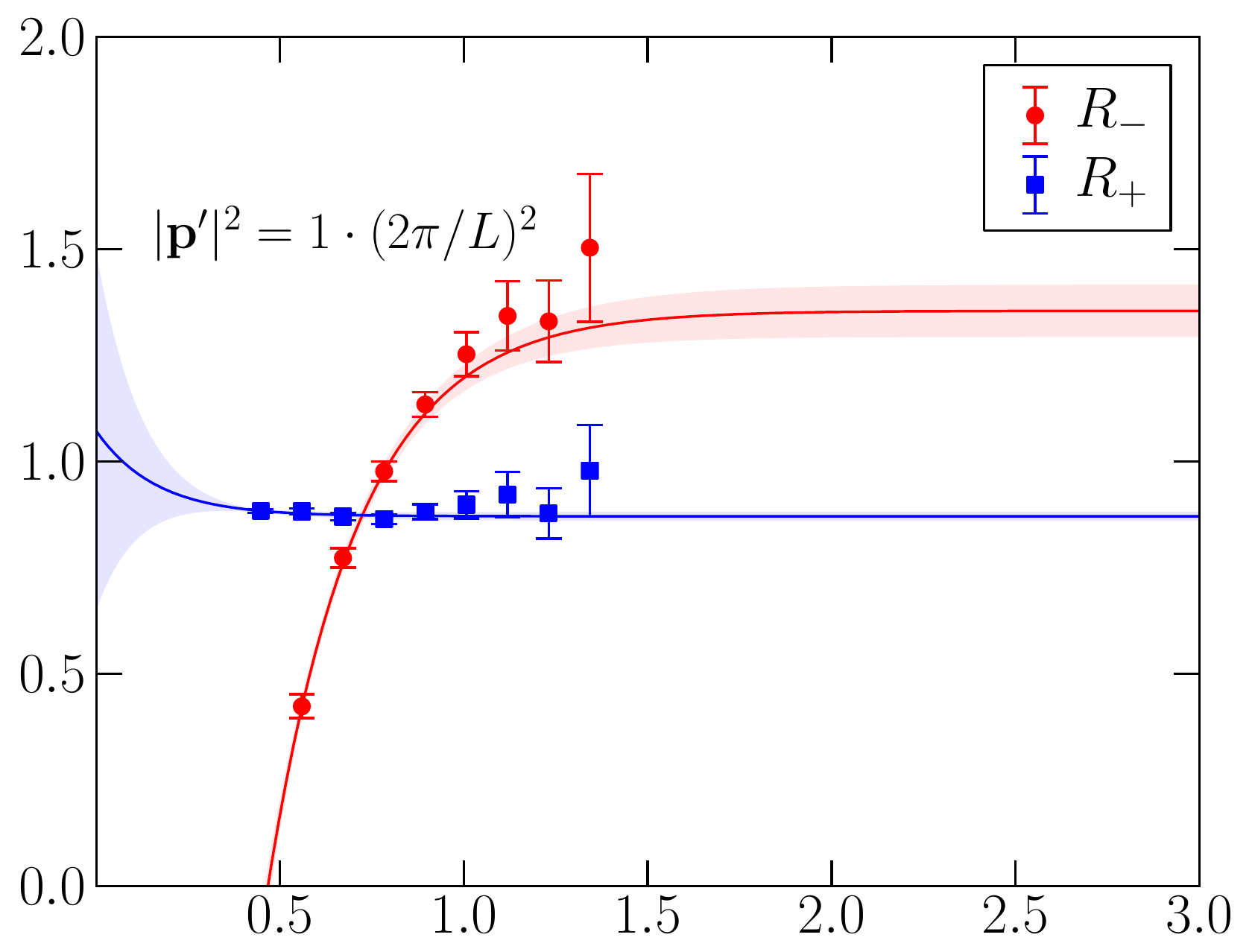}\hfill\includegraphics[width=0.47\linewidth]{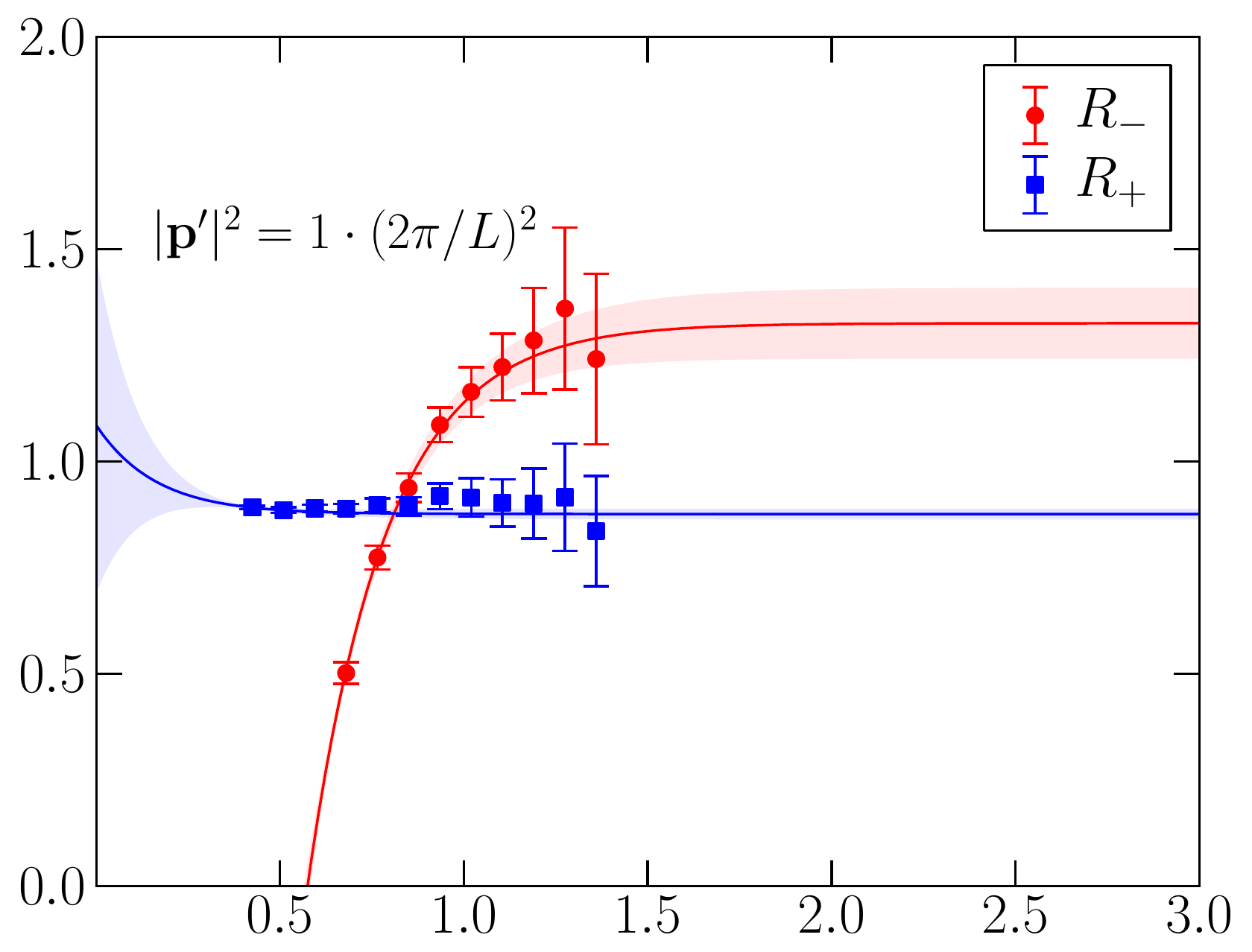}

\includegraphics[width=0.47\linewidth]{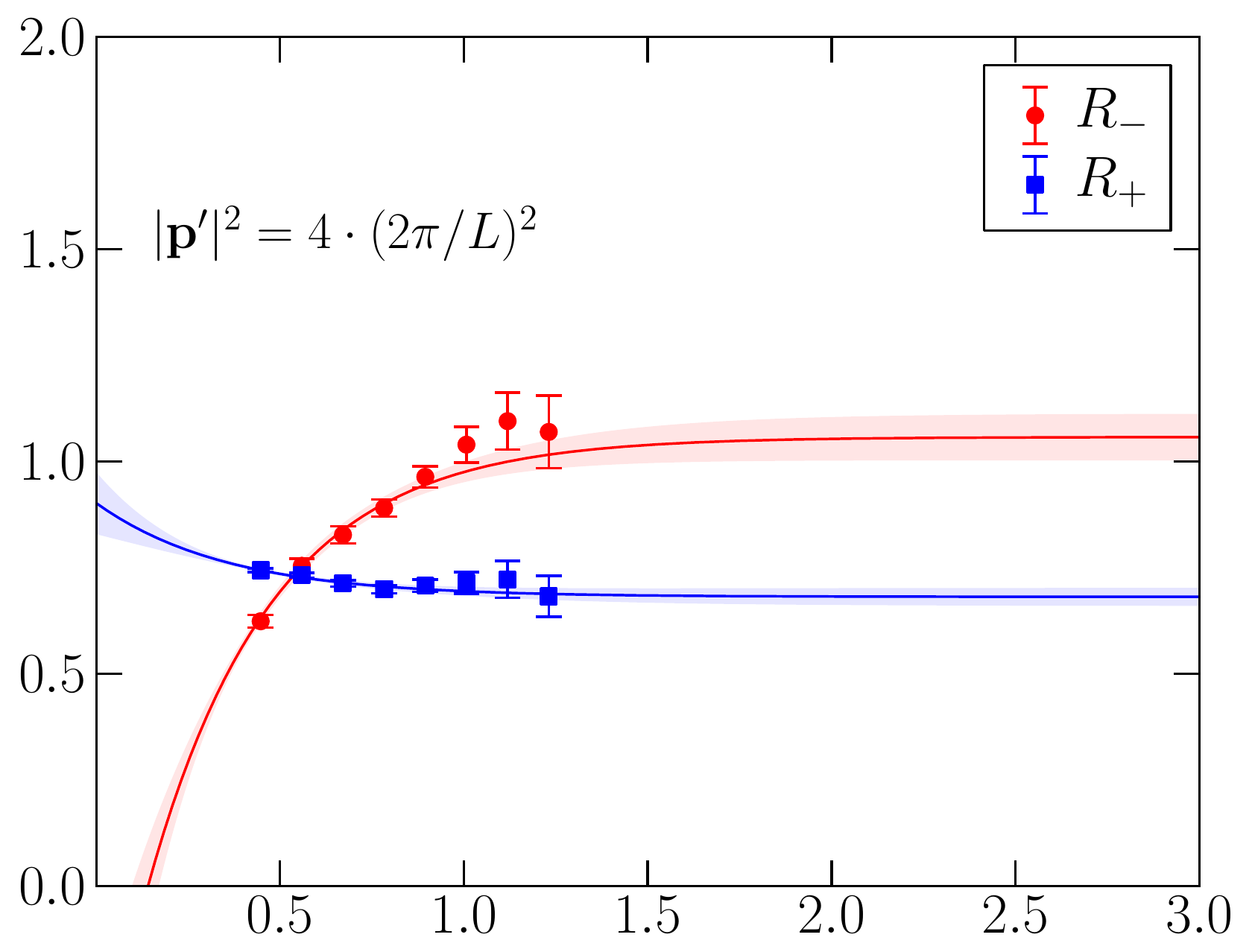}\hfill\includegraphics[width=0.47\linewidth]{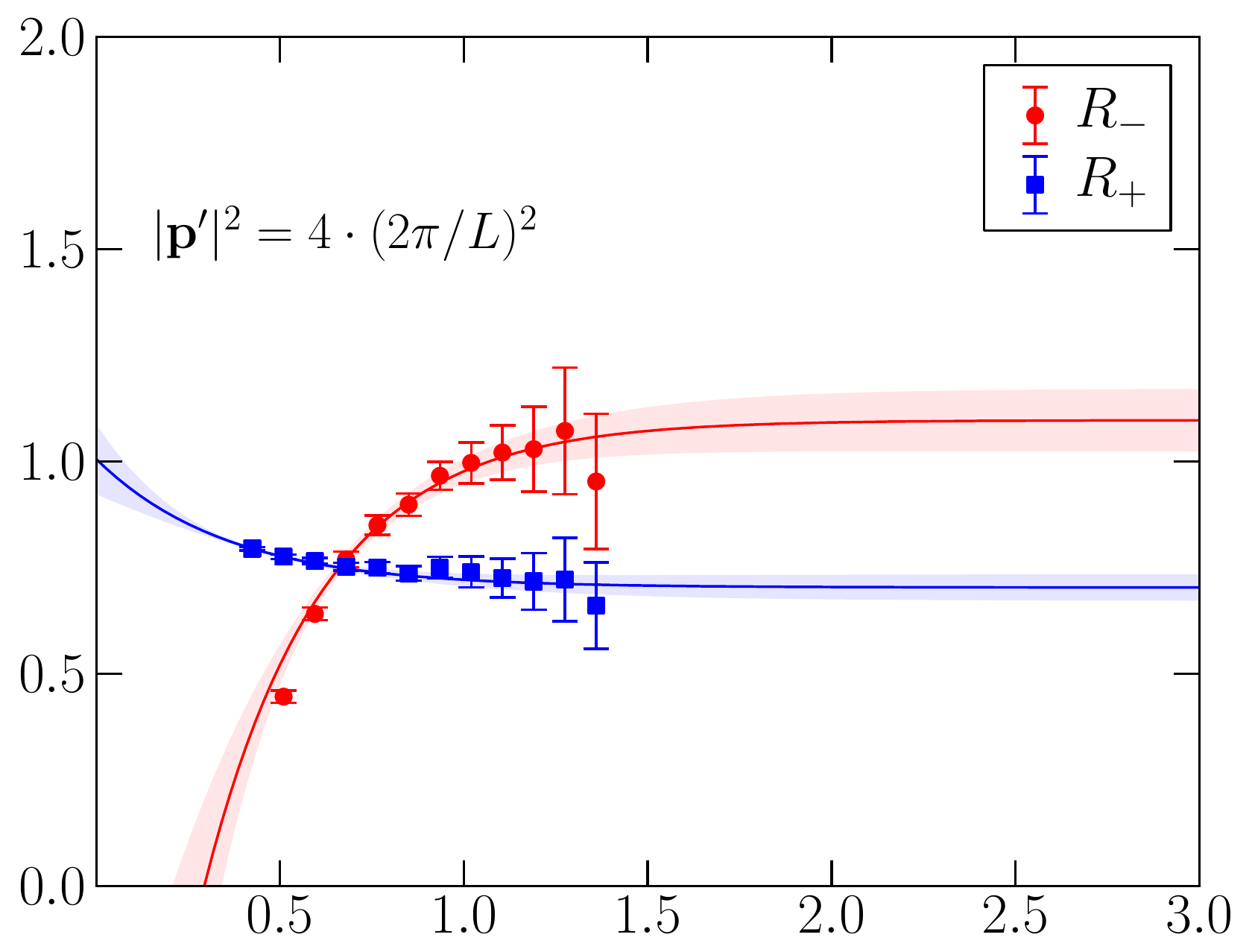}

\includegraphics[width=0.47\linewidth]{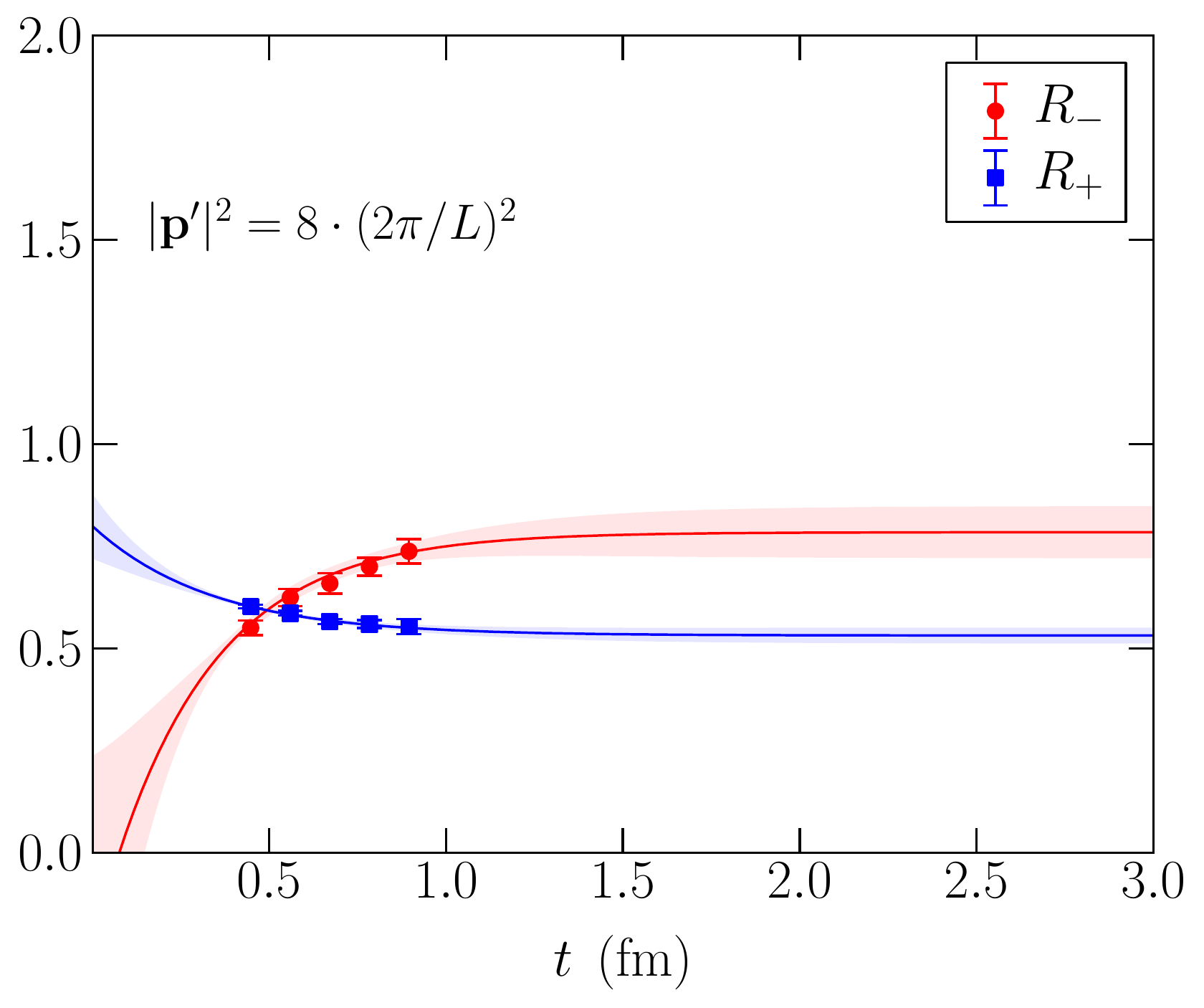}\hfill\includegraphics[width=0.47\linewidth]{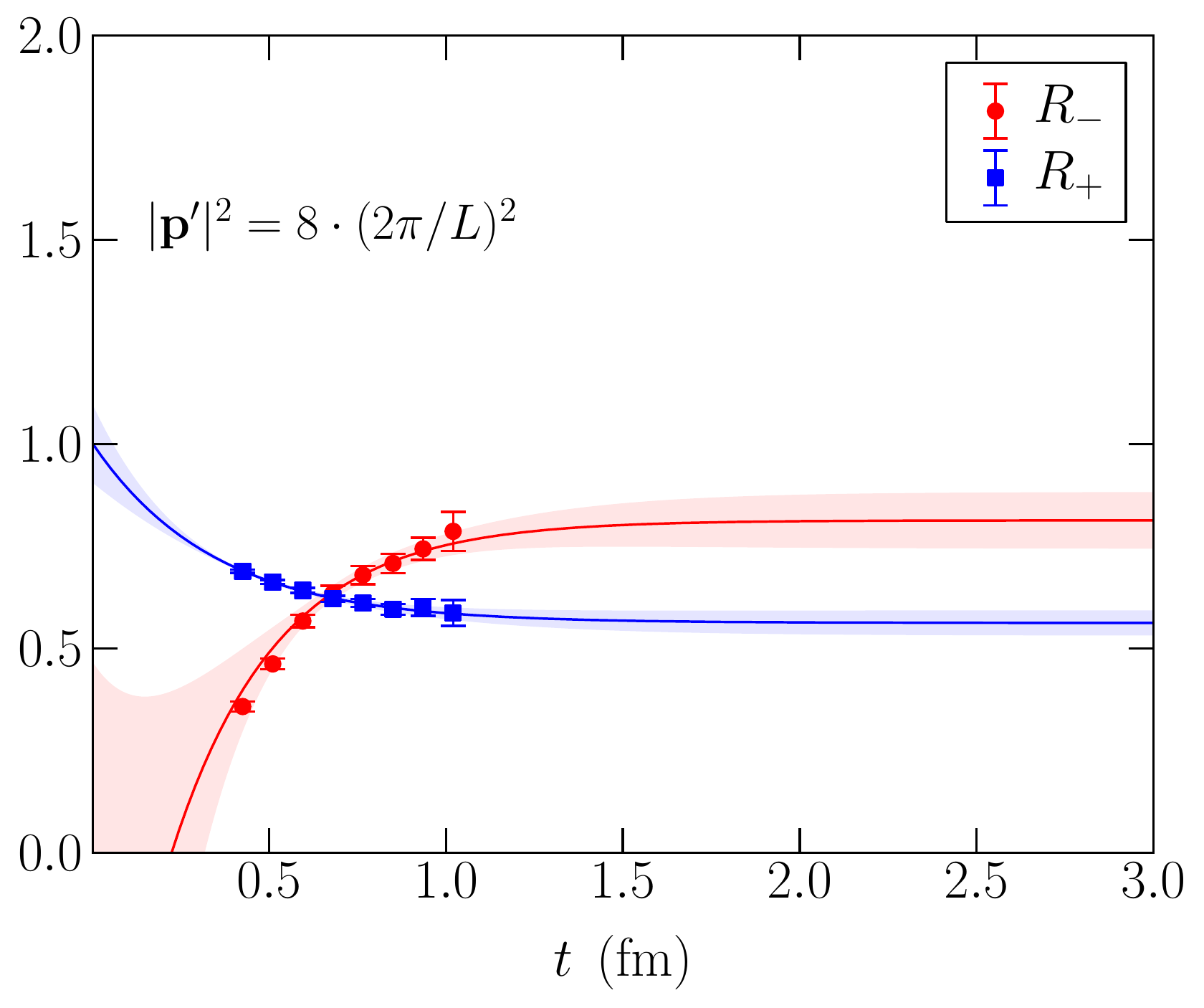}

\caption{\label{fig:tsep_dependence}Fits of the $t$-dependence of $R_\pm(t)$ using Eq.~(\protect\ref{eq:tdepsep}).
Left panels: $a=0.112$ fm, $am_{s}^{(\mathrm{val})}=0.04$, $am_{u,d}^{(\mathrm{val})}=0.005$;
right panels: $a=0.085$ fm, $am_{s}^{(\mathrm{val})}=0.03$, $am_{u,d}^{(\mathrm{val})}=0.004$.
At the fine lattice spacing, only the points with $t>0.6$ fm are included in the fit of $R_-$.
The maximum values of $t$ for the data were limited by statistical noise in the two-point and
three-point functions; for $t$ larger than the right-most points in the plots,
the statistical fluctuations were too large to compute the square roots in Eqs.~(\ref{eq:Rplus}) and (\ref{eq:Rminus}).}
\end{figure}

Example fits of $R^{i,n}_\pm(t)$ using Eq.~(\ref{eq:tdepsep}) are shown in Fig.~\ref{fig:tsep_dependence}. The fits are
fully correlated, using covariance matrices computed from the bootstrap ensembles for $R^{i,n}_\pm(t)$.
Note that the excited-state contribution in $R_+$, which is negligible at $\mathbf{p'}=0$ (cf.~Fig.~\ref{fig:tsep_dependence_psqr0}),
gradually increases with the momentum. In contrast, $R_-$ shows the strongest excited-state overlap at the smallest momentum,
and this overlap decreases as the momentum increases. The excited-state overlap is slightly stronger at the fine lattice
spacing when compared to the coarse lattice spacing. This is expected because the quark smearing width in the baryon
operators was different for the two lattice spacings (we used the same width in lattice units). We only computed the correlators
for $t/a \geq4$ at the coarse lattice spacing and $t/a\geq5$ at the fine lattice spacing. At the fine lattice spacing,
it was necessary to exclude the points with $t/a<8$ from the fits to $R_-$. Once these points were excluded, all fits had $\chi^2/{\rm dof} \approx 1.0$.
Given the limited time range and the limited statistical precision of the available data, it was not possible (and not necessary)
to perform fits with more than one exponential. As a check, we have also performed fits without the constraints (\ref{eq:chisqr}), which
give consistent results but are less stable.

\begin{figure}
\includegraphics[width=0.47\linewidth]{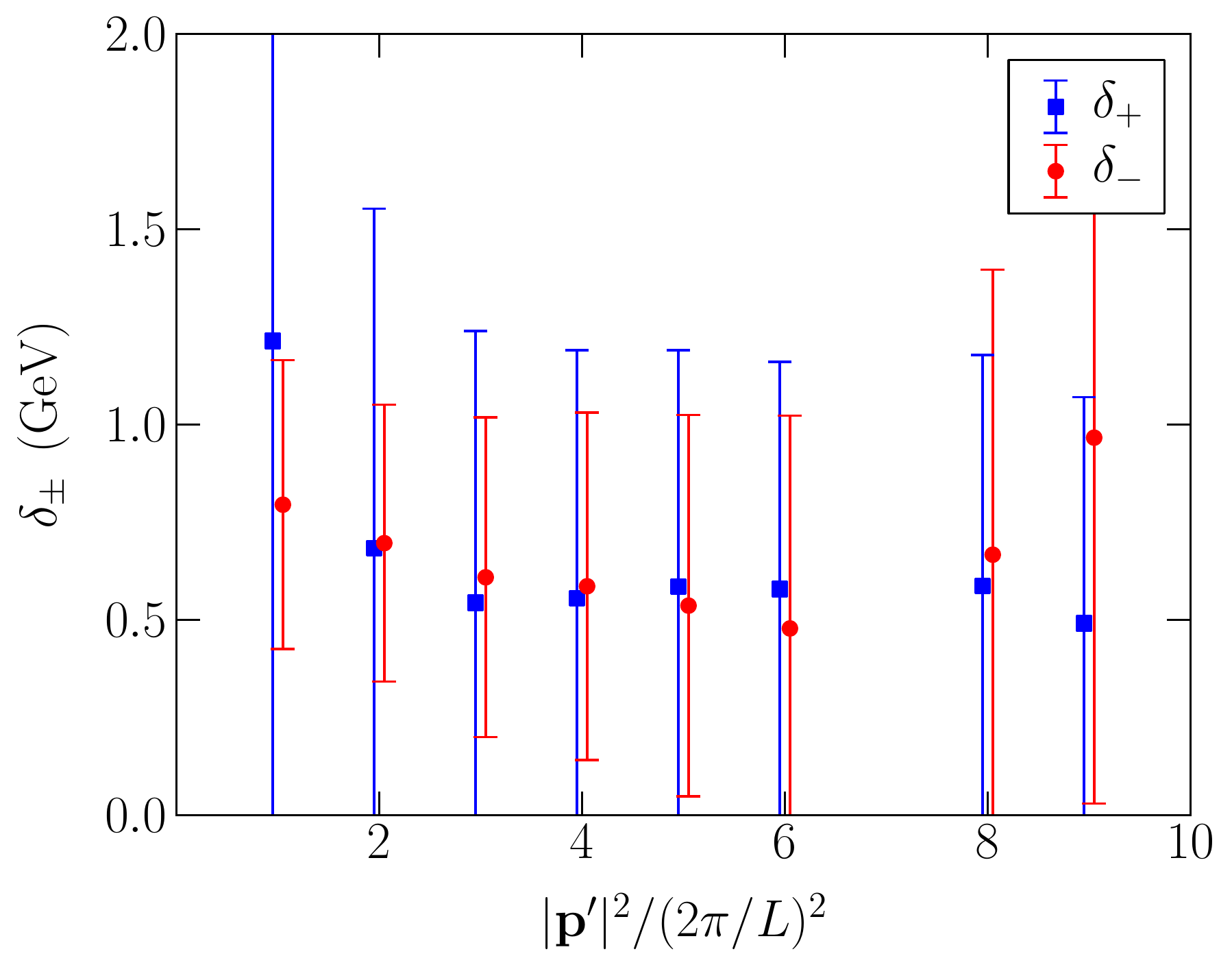}\hfill\includegraphics[width=0.47\linewidth]{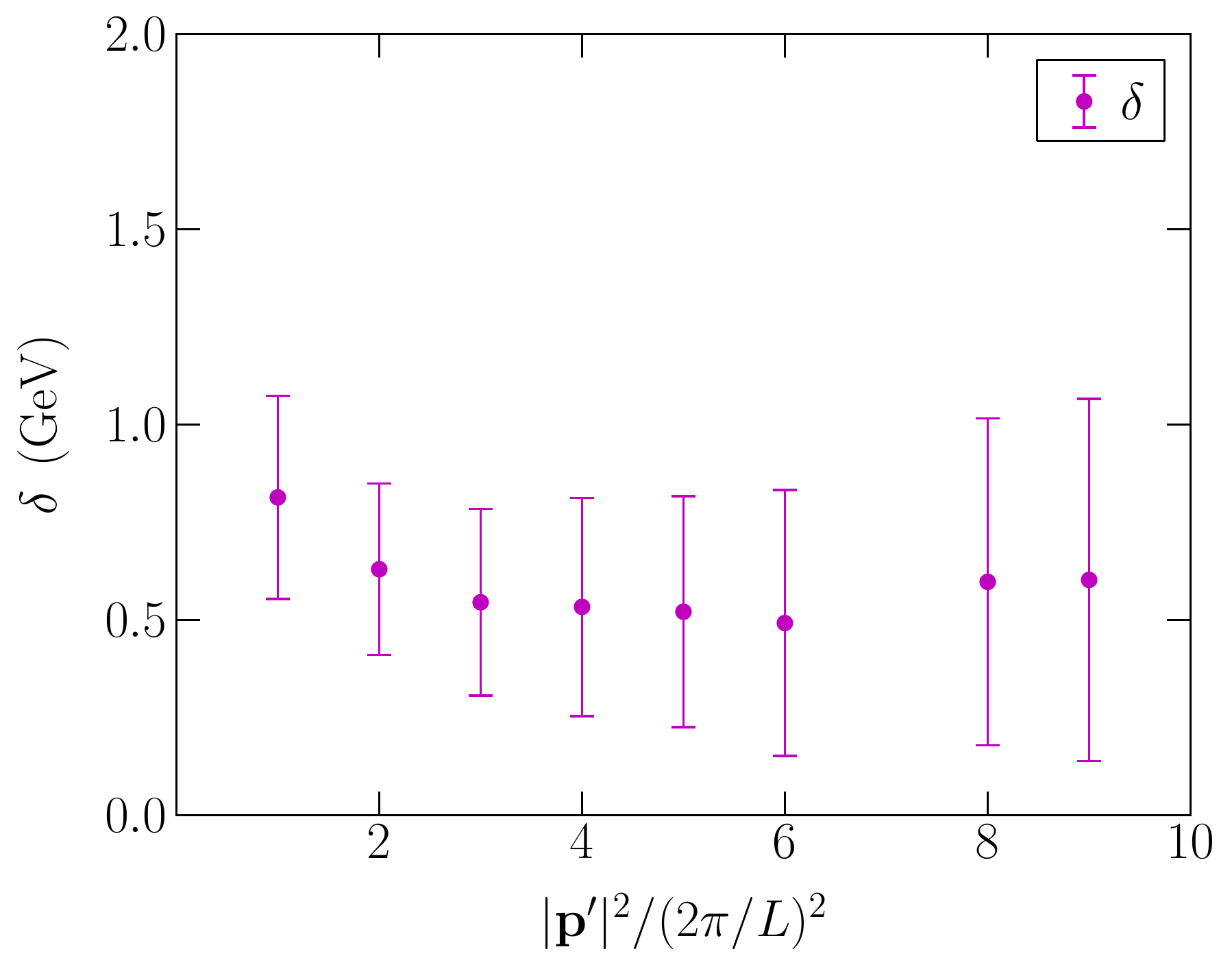}
\caption{\label{fig:energygap}Left: results for the energy gap parameters from separate fits of $R_+$ and $R_-$ using
Eq.~(\protect\ref{eq:tdepsep}). The points are offset horizontally for clarity. Right: results for the energy gap parameters
from coupled fits of $R_+$ and $R_-$ using Eq.~(\protect\ref{eq:tdepsepcoupled}).
The data shown here are for $a=0.112$ fm, $am_{s}^{(\mathrm{val})}=0.04$, $am_{u,d}^{(\mathrm{val})}=0.005$.}
\end{figure}

The fitted values of the energy gap parameters, $\delta^{i,n}_\pm = \exp(l^{i,n}_\pm) \cdot (1\:\: {\rm GeV})$, are shown as a function of
the $\Lambda$-momentum for one ensemble in Fig.~\ref{fig:energygap} (left panel). Within uncertainties, we find that
\begin{equation}
 \delta^{i,n}_+ = \delta^{i,n}_-, \label{Eq:delta+eqdelta-}
\end{equation}
for all data sets $i$ and momenta $n$. The energy spectrum is a property of the QCD Hamiltonian and is independent of
the correlation function considered, so the result (\ref{Eq:delta+eqdelta-}) is not surprising. However, one possible
situation in which $\delta^{i,n}_+$ and $\delta^{i,n}_-$ would be different is when an excited state has negligible
overlap in $R_+$ but significant overlap in $R_-$ (or vice versa). Furthermore, by using only a single exponential, we may be
effectively averaging over multiple excited states which we cannot resolve individually, but which may appear with different
sets of weights in $R_+$ and $R_-$. Having said that, the values of $\delta^{i,n}_+$ and $\delta^{i,n}_-$ from our fits
are in complete agreement and it is evident that the separate parameters $\delta^{i,n}_+$ and $\delta^{i,n}_-$ may be
replaced by a single parameter $\delta^{i,n}$. Thus, we performed new, coupled fits of $R_+$ and $R_-$ of the form
\begin{equation}
R^{i,n}_\pm(t) = F^{i,n}_\pm + A^{i,n}_\pm \:\exp[-\delta^{i,n}\:t], \label{eq:tdepsepcoupled}
\end{equation}
with energy gap parameters $\delta^{i,n}= \exp(l^{i,n})\cdot (1\: {\rm GeV})$ that are shared between $R_+$ and $R_-$.
The results for $\delta^{i,n}$ from the coupled fits are shown in the right panel of Fig.~\ref{fig:energygap}. We note
that the coupled fits had values of $\chi^2/{\rm dof}$ that were as good or better than the values from the separate fits,
confirming that the assumption (\ref{Eq:delta+eqdelta-}) is justified.

\begin{figure}
\includegraphics[width=0.47\linewidth]{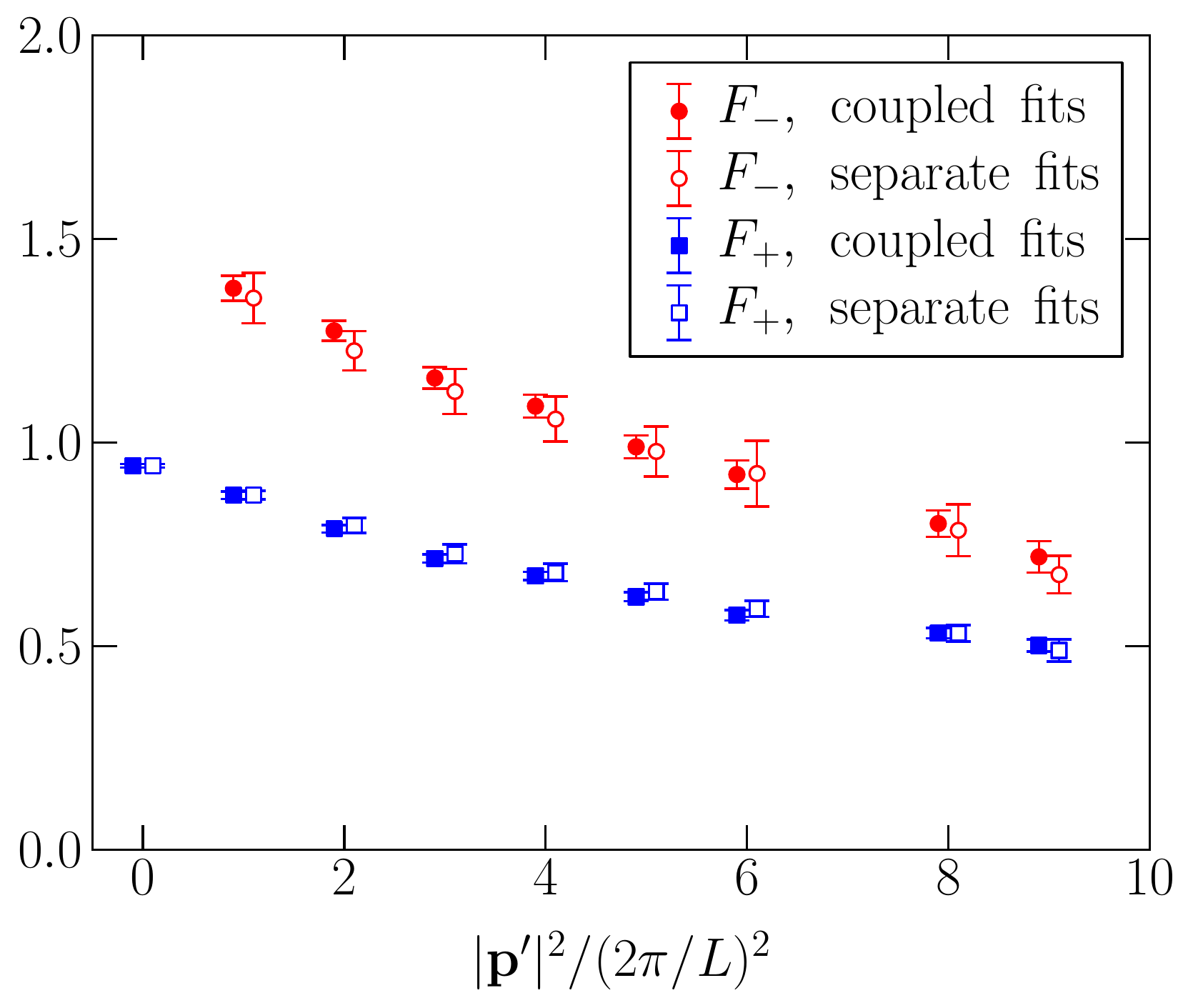}\hfill\includegraphics[width=0.47\linewidth]{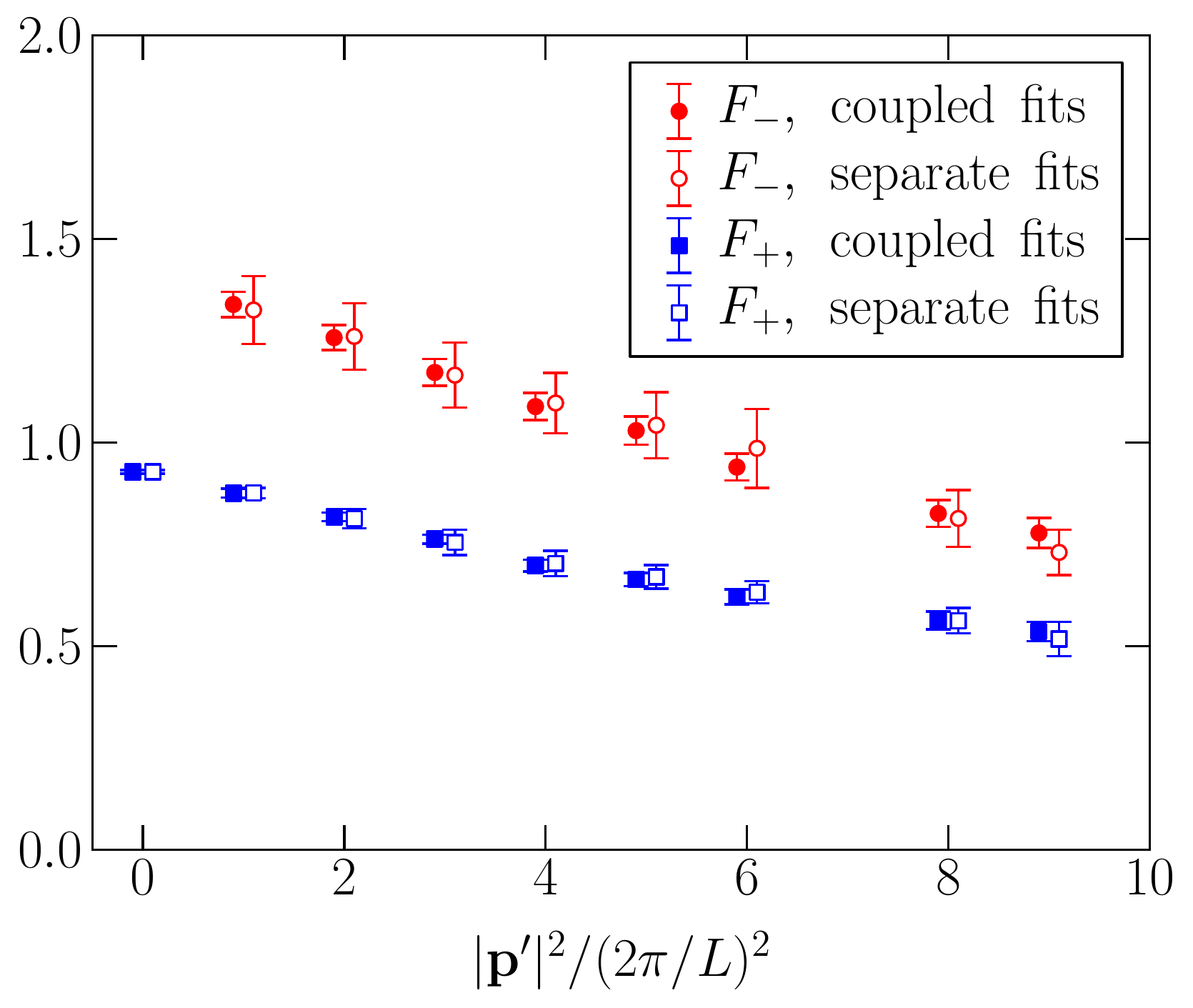}
\caption{\label{fig:resultsinfinitesepcoupledvsseparate}Fit results for the ground-state contributions $F_+$ and $F_-$
from separate fits of $R_+$ and $R_-$ using Eq.~(\protect\ref{eq:tdepsep}) vs results from coupled fits using
Eq.~(\protect\ref{eq:tdepsepcoupled}). Left panel: $a=0.112$ fm, $am_{s}^{(\mathrm{val})}=0.04$,
$am_{u,d}^{(\mathrm{val})}=0.005$; right panel: $a=0.085$ fm, $am_{s}^{(\mathrm{val})}=0.03$, $am_{u,d}^{(\mathrm{val})}=0.004$.
The points are offset horizontally for clarity.}
\end{figure}

The results for the extracted ground-state form factors $F^{i,n}_\pm$ from separate and coupled fits are compared in
Fig.~\ref{fig:resultsinfinitesepcoupledvsseparate}. As can be seen there, the results are consistent with each other,
but the coupled fits with shared energy gap parameters give significantly smaller uncertainties, in particular for $F_-$.
Therefore, we use the results from the coupled fits in our further analysis.

\begin{figure}
\includegraphics[width=0.48\linewidth]{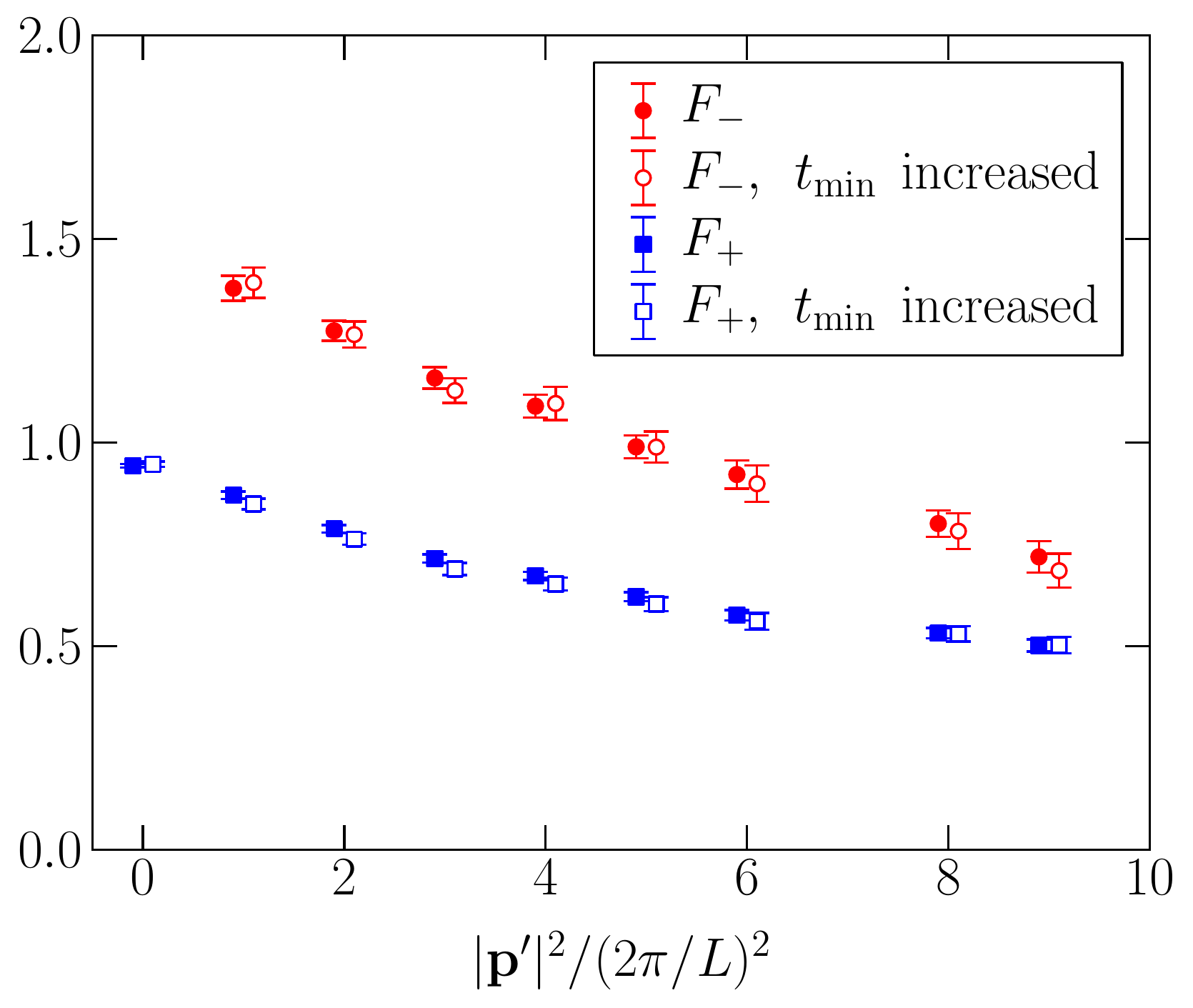}\hfill\includegraphics[width=0.48\linewidth]{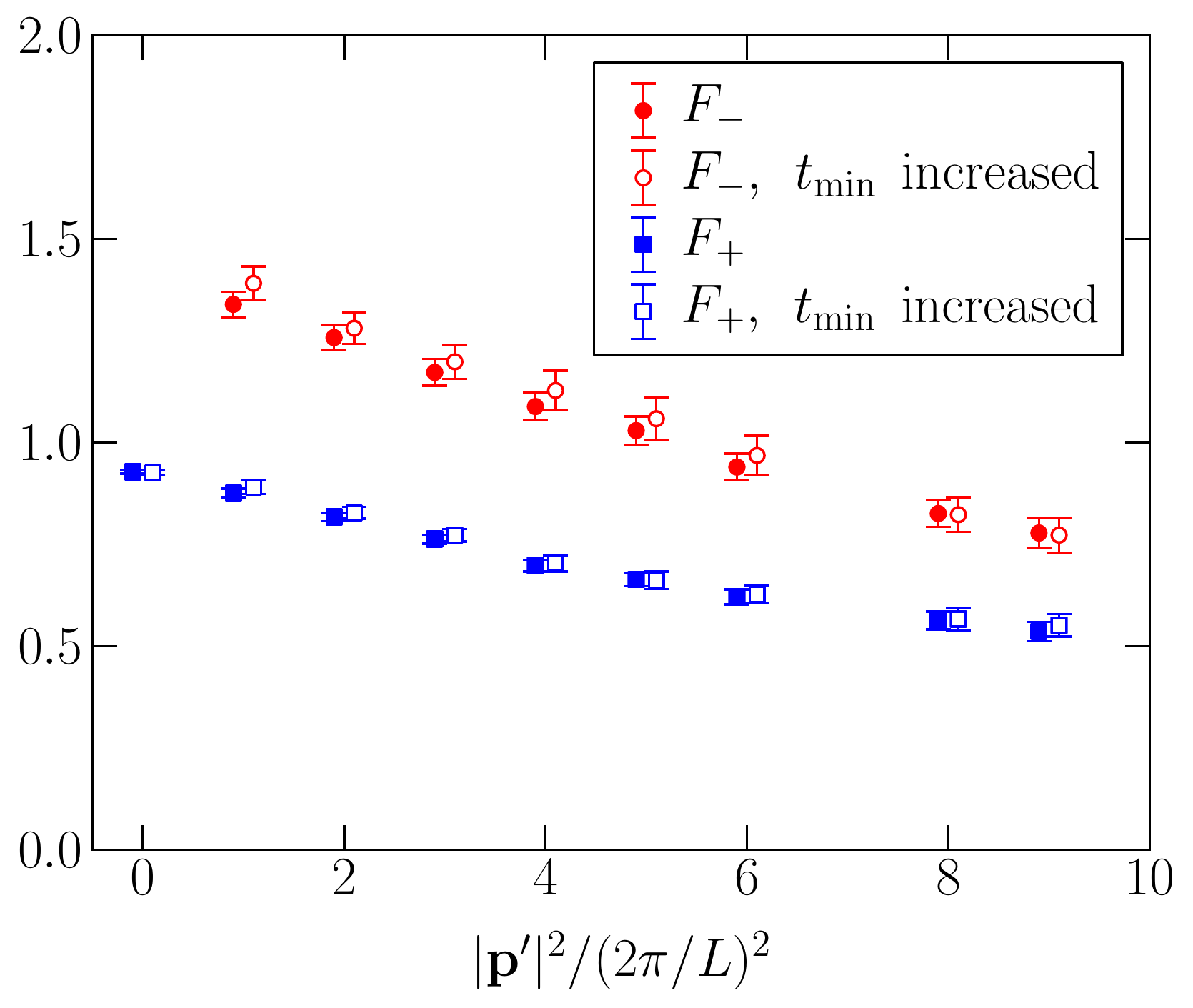}
\caption{\label{fig:resultsinfinitesepcoupledtshift}Fit results for the ground-state contributions $F_+$ and $F_-$ from
coupled fits of $R_+$ and $R_-$ using Eq.~(\protect\ref{eq:tdepsepcoupled}): effect of increasing all $t_{\rm min}$
values by one unit. Left panel: $a=0.112$ fm, $am_{s}^{(\mathrm{val})}=0.04$, $am_{u,d}^{(\mathrm{val})}=0.005$;
right panel: $a=0.085$ fm, $am_{s}^{(\mathrm{val})}=0.03$, $am_{u,d}^{(\mathrm{val})}=0.004$.
The points are offset horizontally for clarity.}
\end{figure}

To estimate the systematic uncertainty resulting from the choice of the fit range in $t$, we computed the changes in
$F^{i,n}_\pm$ when increasing $t_{\rm min}$ by one unit simultaneously for all data sets, thereby removing the points
with the most contamination from additional higher excited states. As can be seen in Fig.~\ref{fig:resultsinfinitesepcoupledtshift},
the resulting change in the fitted $F^{i,n}_\pm$ is small, and in most cases consistent with zero. Nevertheless,
we add this shift in quadrature to the original statistical uncertainty of $F^{i,n}_\pm$. The final results for
$F^{i,n}_\pm$, including this systematic uncertainty, are given in Tables \ref{tab:Fplus} and \ref{tab:Fminus}.
For convenience, we also provide results for $F^{i,n}_1 = (F^{i,n}_+ + F^{i,n}_-)/2$ and $F^{i,n}_2=(F^{i,n}_+ - F^{i,n}_-)/2$,
computed using bootstrap to take into account the correlations between $F^{i,n}_+$ and $F^{i,n}_-$,
in Tables \ref{tab:F1} and \ref{tab:F2}.

\begin{table}
\begin{tabular}{cccccccccccccccc}
\hline\hline
$|\mathbf{p'}|^2/(2\pi/L)^2$ & \hspace{1ex} & \texttt{C14} & \hspace{1ex} & \texttt{C24} & \hspace{1ex} &
\texttt{C54}  & \hspace{1ex} & \texttt{C53}   & \hspace{1ex} & \texttt{F23} & \hspace{1ex} &
\texttt{F43} & \hspace{1ex} & \texttt{F63} \\
\hline
0   && $0.9416(56)$ && $0.9443(66)$ && $0.9430(59)$ && $0.9390(71)$ && $0.9320(58)$ && $0.9281(49)$ && $0.920(13)$ \\
1   && $0.868(14)$  && $0.874(27)$  && $0.871(23)$  && $0.873(26)$  && $0.878(33)$  && $0.876(18)$  && $0.807(25)$ \\
2   && $0.785(15)$  && $0.786(34)$  && $0.788(27)$  && $0.791(25)$  && $0.823(25)$  && $0.818(14)$  && $0.738(28)$ \\
3   && $0.717(15)$  && $0.705(36)$  && $0.715(28)$  && $0.723(24)$  && $0.770(26)$  && $0.763(14)$  && $0.680(28)$ \\
4   && $0.671(15)$  && $0.669(26)$  && $0.672(22)$  && $0.668(17)$  && $0.697(32)$  && $0.698(15)$  && $0.633(28)$ \\
5   && $0.618(15)$  && $0.612(26)$  && $0.621(21)$  && $0.622(15)$  && $0.667(27)$  && $0.664(16)$  && $0.587(27)$ \\
6   && $0.586(16)$  && $0.580(20)$  && $0.576(20)$  && $0.577(15)$  && $0.625(30)$  && $0.621(19)$  && $0.554(29)$ \\
8   && $0.537(14)$  && $0.545(16)$  && $0.532(13)$  && $0.516(24)$  && $0.569(27)$  && $0.563(22)$  && $0.507(39)$ \\
9   && $0.508(17)$  && $0.516(16)$  && $0.502(15)$  && $0.491(26)$  && $0.546(31)$  && $0.536(28)$  && $0.484(40)$ \\
\hline\hline
\end{tabular}
\caption{\label{tab:Fplus} Results for $F_+$ from the different data sets (see Table \protect\ref{tab:params}). The uncertainties
presented combine statistical and systematic fitting uncertainties in quadrature.}
\end{table}

\begin{table}
\begin{tabular}{cccccccccccccccc}
\hline\hline
$|\mathbf{p'}|^2/(2\pi/L)^2$ & \hspace{1ex} & \texttt{C14} & \hspace{1ex} & \texttt{C24} & \hspace{1ex} &
\texttt{C54}  & \hspace{1ex} & \texttt{C53}   & \hspace{1ex} & \texttt{F23} & \hspace{1ex} &
\texttt{F43} & \hspace{1ex} & \texttt{F63} \\
\hline
1   && $1.355(53)$ && $1.422(52)$ && $1.379(33)$ && $1.346(41)$ && $1.352(57)$ && $1.339(60)$ && $1.296(51)$ \\
2   && $1.233(58)$ && $1.278(36)$ && $1.274(26)$ && $1.230(35)$ && $1.241(45)$ && $1.258(38)$ && $1.185(49)$ \\
3   && $1.143(36)$ && $1.163(43)$ && $1.159(41)$ && $1.135(72)$ && $1.183(52)$ && $1.172(42)$ && $1.094(47)$ \\
4   && $1.032(51)$ && $1.084(40)$ && $1.089(29)$ && $1.071(44)$ && $1.094(81)$ && $1.088(52)$ && $1.004(49)$ \\
5   && $0.973(34)$ && $0.990(35)$ && $0.990(28)$ && $0.967(49)$ && $1.035(73)$ && $1.029(45)$ && $0.928(47)$ \\
6   && $0.904(35)$ && $0.905(43)$ && $0.922(42)$ && $0.907(65)$ && $0.938(82)$ && $0.940(44)$ && $0.872(45)$ \\
8   && $0.752(33)$ && $0.790(42)$ && $0.801(37)$ && $0.820(74)$ && $0.805(58)$ && $0.826(33)$ && $0.742(48)$ \\
9   && $0.700(43)$ && $0.706(51)$ && $0.719(52)$ && $0.712(76)$ && $0.748(58)$ && $0.778(37)$ && $0.707(41)$ \\
\hline\hline
\end{tabular}
\caption{\label{tab:Fminus} Results for $F_-$ from the different data sets (see Table \protect\ref{tab:params}).
 The uncertainties presented combine statistical and systematic fitting uncertainties in quadrature.}
\end{table}

\begin{table}
\begin{tabular}{cccccccccccccccc}
\hline\hline
$|\mathbf{p'}|^2/(2\pi/L)^2$ & \hspace{1ex} & \texttt{C14} & \hspace{1ex} & \texttt{C24} & \hspace{1ex} &
\texttt{C54}  & \hspace{1ex} & \texttt{C53}   & \hspace{1ex} & \texttt{F23} & \hspace{1ex} &
\texttt{F43} & \hspace{1ex} & \texttt{F63} \\
\hline
1   && $1.111(31)$ && $1.148(32)$ && $1.125(18)$ && $1.110(23)$ && $1.115(40)$ && $1.107(38)$ && $1.051(35)$ \\
2   && $1.009(35)$ && $1.032(26)$ && $1.031(23)$ && $1.011(26)$ && $1.032(30)$ && $1.038(24)$ && $0.961(37)$ \\
3   && $0.930(21)$ && $0.934(35)$ && $0.937(32)$ && $0.929(46)$ && $0.976(33)$ && $0.968(25)$ && $0.887(36)$ \\
4   && $0.852(29)$ && $0.876(20)$ && $0.881(17)$ && $0.870(25)$ && $0.895(50)$ && $0.893(28)$ && $0.818(35)$ \\
5   && $0.795(18)$ && $0.801(22)$ && $0.805(18)$ && $0.795(27)$ && $0.851(41)$ && $0.846(22)$ && $0.757(34)$ \\
6   && $0.745(19)$ && $0.743(26)$ && $0.749(26)$ && $0.742(33)$ && $0.782(48)$ && $0.780(24)$ && $0.713(34)$ \\
8   && $0.644(17)$ && $0.667(24)$ && $0.666(19)$ && $0.668(31)$ && $0.687(32)$ && $0.695(17)$ && $0.624(41)$ \\
9   && $0.604(23)$ && $0.611(27)$ && $0.611(24)$ && $0.602(32)$ && $0.647(35)$ && $0.657(19)$ && $0.595(35)$ \\
\hline\hline
\end{tabular}
\caption{\label{tab:F1} Results for $F_1$ from the different data sets (see Table \protect\ref{tab:params}).
 The uncertainties presented combine statistical and systematic fitting uncertainties in quadrature.}
\end{table}

\begin{table}
\begin{tabular}{cccccccccccccccc}
\hline\hline
$|\mathbf{p'}|^2/(2\pi/L)^2$ & \hspace{1ex} & \texttt{C14} & \hspace{1ex} & \texttt{C24} & \hspace{1ex} &
\texttt{C54}  & \hspace{1ex} & \texttt{C53}   & \hspace{1ex} & \texttt{F23} & \hspace{1ex} &
\texttt{F43} & \hspace{1ex} & \texttt{F63} \\
\hline
1   && $-0.244(23)$ && $-0.274(26)$ && $-0.254(23)$ && $-0.236(25)$ && $-0.237(23)$ && $-0.232(23)$ && $-0.244(19)$ \\
2   && $-0.224(25)$ && $-0.246(23)$ && $-0.243(14)$ && $-0.220(16)$ && $-0.209(21)$ && $-0.220(16)$ && $-0.224(16)$ \\
3   && $-0.213(17)$ && $-0.229(20)$ && $-0.222(13)$ && $-0.206(27)$ && $-0.207(24)$ && $-0.205(19)$ && $-0.207(15)$ \\
4   && $-0.181(24)$ && $-0.208(27)$ && $-0.209(19)$ && $-0.202(22)$ && $-0.199(35)$ && $-0.195(26)$ && $-0.186(18)$ \\
5   && $-0.177(19)$ && $-0.189(21)$ && $-0.184(17)$ && $-0.173(24)$ && $-0.184(37)$ && $-0.183(26)$ && $-0.171(18)$ \\
6   && $-0.159(20)$ && $-0.163(21)$ && $-0.173(20)$ && $-0.165(34)$ && $-0.157(38)$ && $-0.159(24)$ && $-0.159(17)$ \\
8   && $-0.108(19)$ && $-0.123(21)$ && $-0.135(21)$ && $-0.152(46)$ && $-0.118(32)$ && $-0.131(22)$ && $-0.118(16)$ \\
9   && $-0.096(23)$ && $-0.095(27)$ && $-0.109(29)$ && $-0.111(47)$ && $-0.101(30)$ && $-0.121(27)$ && $-0.111(20)$ \\
\hline\hline
\end{tabular}
\caption{\label{tab:F2} Results for $F_2$ from the different data sets (see Table \protect\ref{tab:params}).
 The uncertainties presented combine statistical and systematic fitting uncertainties in quadrature.}
\end{table}

\FloatBarrier
\subsection{\label{sec:chiralcontinuumextrap}Chiral and continuum extrapolation of the form factors}
\FloatBarrier

The last step in our analysis of the lattice data is to fit the dependence of $F_\pm^{i,n}$ on the quark masses, the lattice spacing,
and on $E_\Lambda$. The form of the dependence is unknown; low-energy effective field theory combining heavy-baryon chiral perturbation theory
for the $\Lambda$ sector and heavy-hadron chiral perturbation theory for the $\Lambda_Q$ sector may be useful over some range
of $E_\Lambda$, but not in the region with $|\mathbf{p'}| \gtrsim \Lambda_\chi$, where $\Lambda_\chi\sim 1\:\:{\rm GeV}$
is the chiral symmetry-breaking scale. We therefore use a simple ansatz that fits our data well at the level of statistical
precision that we have. In the following it is advantageous to express the form factors as functions of the energy difference
$E_\Lambda-m_\Lambda$ instead of $E_\Lambda$, as this depends less on the quark masses. We find that this dependence can be described
well using a dipole function of the form $F_\pm = N_\pm/(X_\pm + E_\Lambda-m_\Lambda)^2$. We generalize this ansatz to allow for
dependence on the light and strange quark masses, as well as the lattice spacing, in the following way:
\begin{eqnarray}
 F_\pm^{i,n} &=& \frac{N_\pm}{(X_\pm^i+E_\Lambda^{i,n}-m_\Lambda^i)^2}\cdot [1 + d_\pm (a^i E_\Lambda^{i,n})^2], \label{eq:dipole}
\end{eqnarray}
where the functions $X_\pm^i$ are defined as
\begin{equation}
X_\pm^i = X_\pm + c_{l,\pm}\cdot \left[ (m_\pi^i)^2-(m_\pi^{{\rm phys}})^2\right]
+ c_{s,\pm}\cdot \left[ (m_{\eta_s}^i)^2-(m_{\eta_s}^{{\rm phys}})^2 \right]. \label{eq:polemqdep}
\end{equation}
As before, we use the notation where $i=\mathtt{C14}, \mathtt{C24}, ..., \mathtt{F63}$ labels the data set (see Table \ref{tab:params}),
and $n$ labels the momentum of the $\Lambda$. The free fit parameters in Eq.~(\ref{eq:dipole}) are $N_\pm$, $X_\pm$, $d_\pm$, $c_{l,\pm}$,
and $c_{s,\pm}$. The dependence of the form factors on the light and strange quark masses is described by allowing $X_\pm^i$
to depend linearly on $(m_\pi^i)^2$ and $(m_{\eta_s}^i)^2$, where $m_\pi^i$ and $m_{\eta_s}^i$ are the valence $\pi$ and
$\eta_s$ masses for each data set $i$, as given in Table \ref{tab:params}. We wrote Eq.~(\ref{eq:polemqdep}) in terms of
the differences between the lattice and physical masses for convenience, with $m_\pi^{{\rm phys}}=138$ MeV and
$m_{\eta_s}^{{\rm phys}}=686$ MeV \cite{Davies:2009tsa}. The leading dependence of the form factors on the lattice spacing
is expected to be quadratic in $a$, owing to the chiral symmetry of the domain-wall action and the use of the order-$a$-improved
current (\ref{eq:LHQETcurrent}). Discretization errors are expected to grow as the momentum of the $\Lambda$ increases.
We therefore incorporate the $a$-dependence using the factor $[1 + d_\pm (a^i E_\Lambda^{i,n})^2]$ in Eq.~(\ref{eq:dipole}).

In our fits, we take into account the correlations between the results for $F_\pm^{i,n}$ at different momenta $n$ and
different data sets $i$ (in the case where the data sets correspond to the same underlying ensemble of gauge fields).
The fits are performed independently for $F_+^{i,n}$ and $F_-^{i,n}$. To account for the uncertainties and correlations
of the $\Lambda$ baryon energies $E_\Lambda^{i,n}$ (including the masses $m_\Lambda^{i}=E_\Lambda^{i,0}$) in Eq.~(\ref{eq:dipole}),
we promote $E_\Lambda^{i,n}$ to additional parameters of the fit, and add the term
$\sum_{i,n,i',n'} [{\rm Cov(E_\Lambda)}^{-1}]_{i,n,i',n'}(E_\Lambda^{i,n}-\overline{E}_\Lambda^{i,n})  (E_\Lambda^{i',n'}-\overline{E}_\Lambda^{i',n'})$
to the $\chi^2$ function, where $\overline{E}_\Lambda^{i,n}$ are the previous results from the fits to the two-point functions, and the energy
correlation matrix ${\rm Cov}(E_\Lambda)$ was computed from the bootstrap ensemble of the two-point fit results. Using a similar term,
we investigated the inclusion of the further correlations between the $\Lambda$ energies and the form factor values $F_\pm^{i,n}$,
but with the current level of statistics, such fits did not converge to a stable minimum of $\chi^2$.

\begin{figure}
\includegraphics[width=0.48\linewidth]{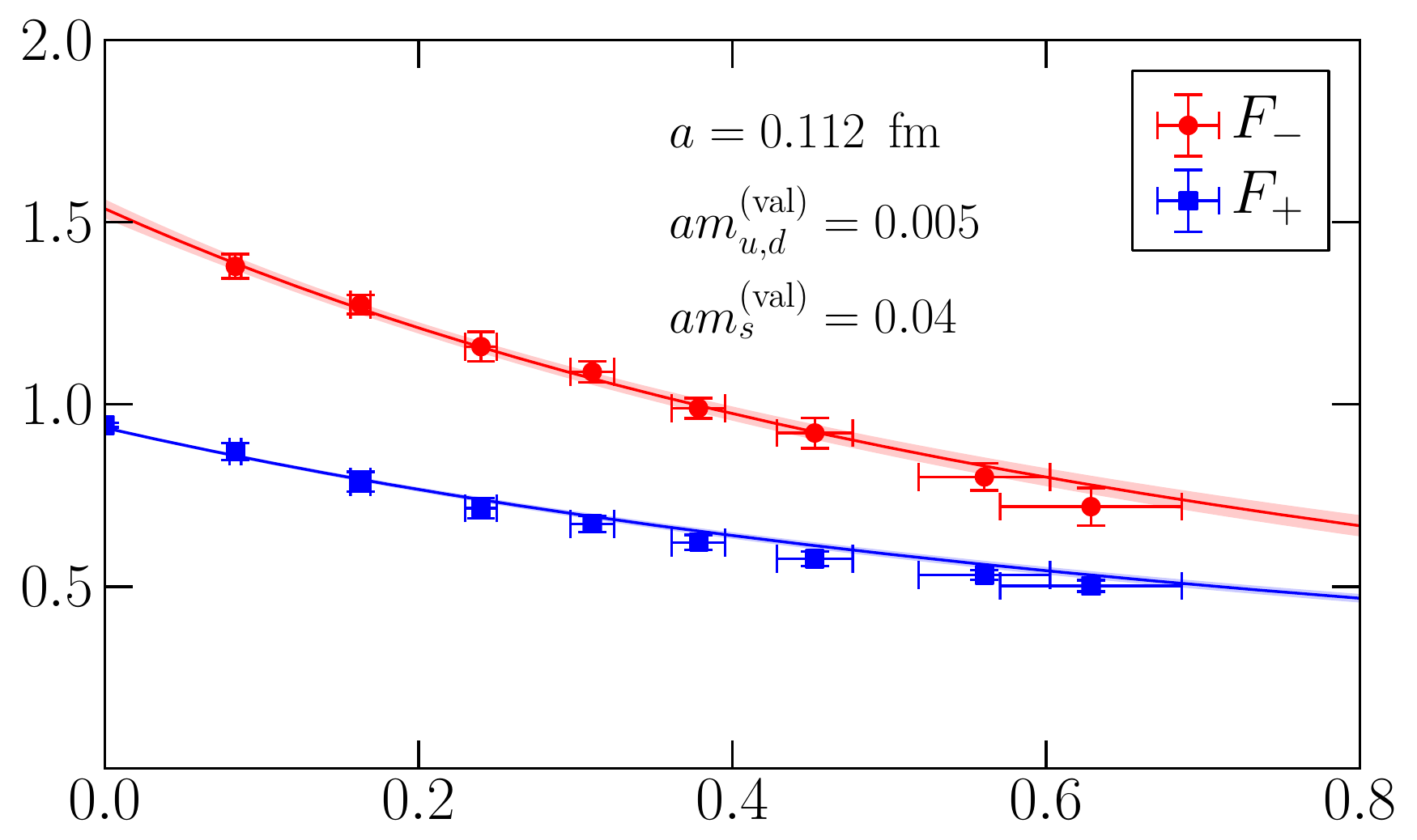}\hfill\includegraphics[width=0.48\linewidth]{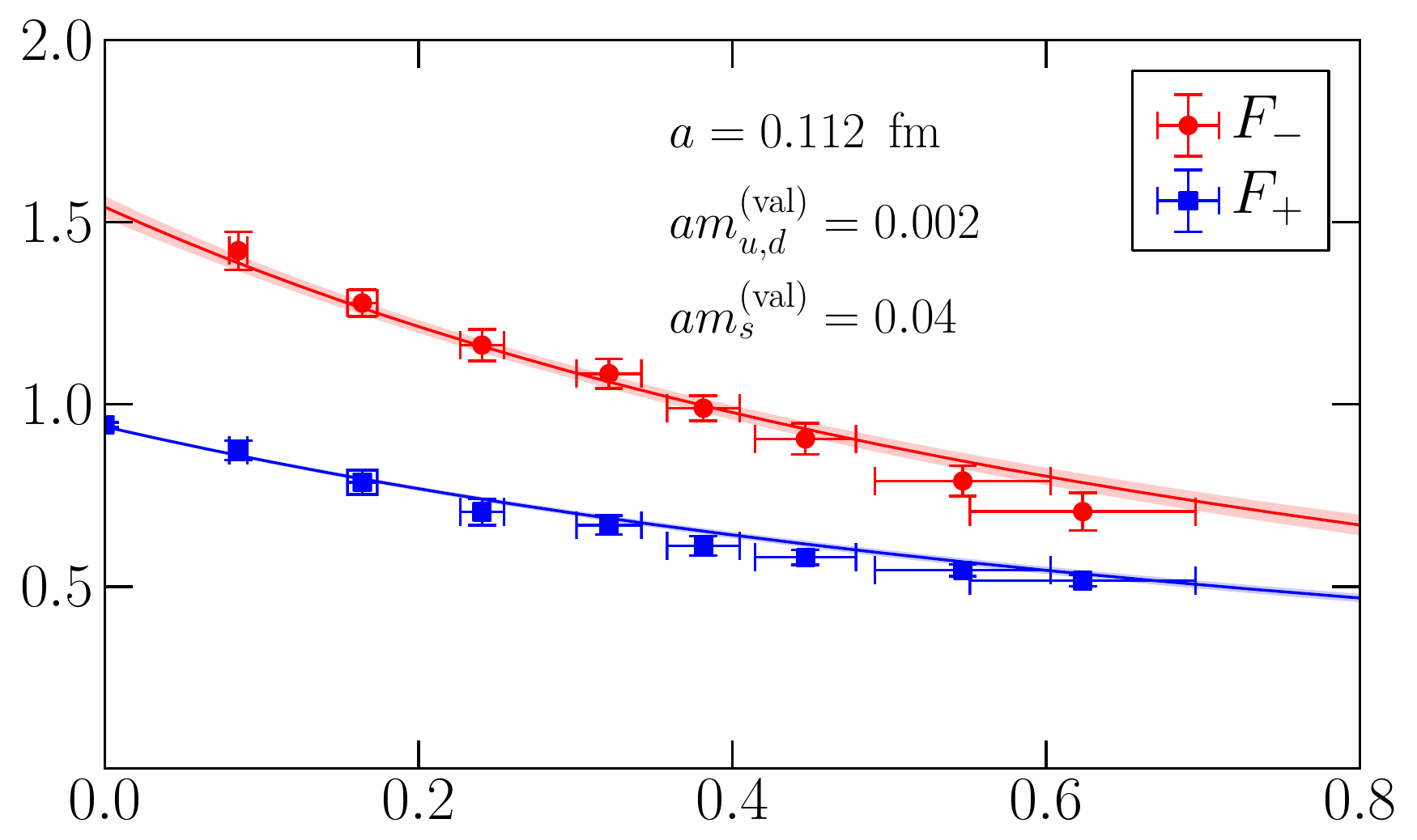} \\
\includegraphics[width=0.48\linewidth]{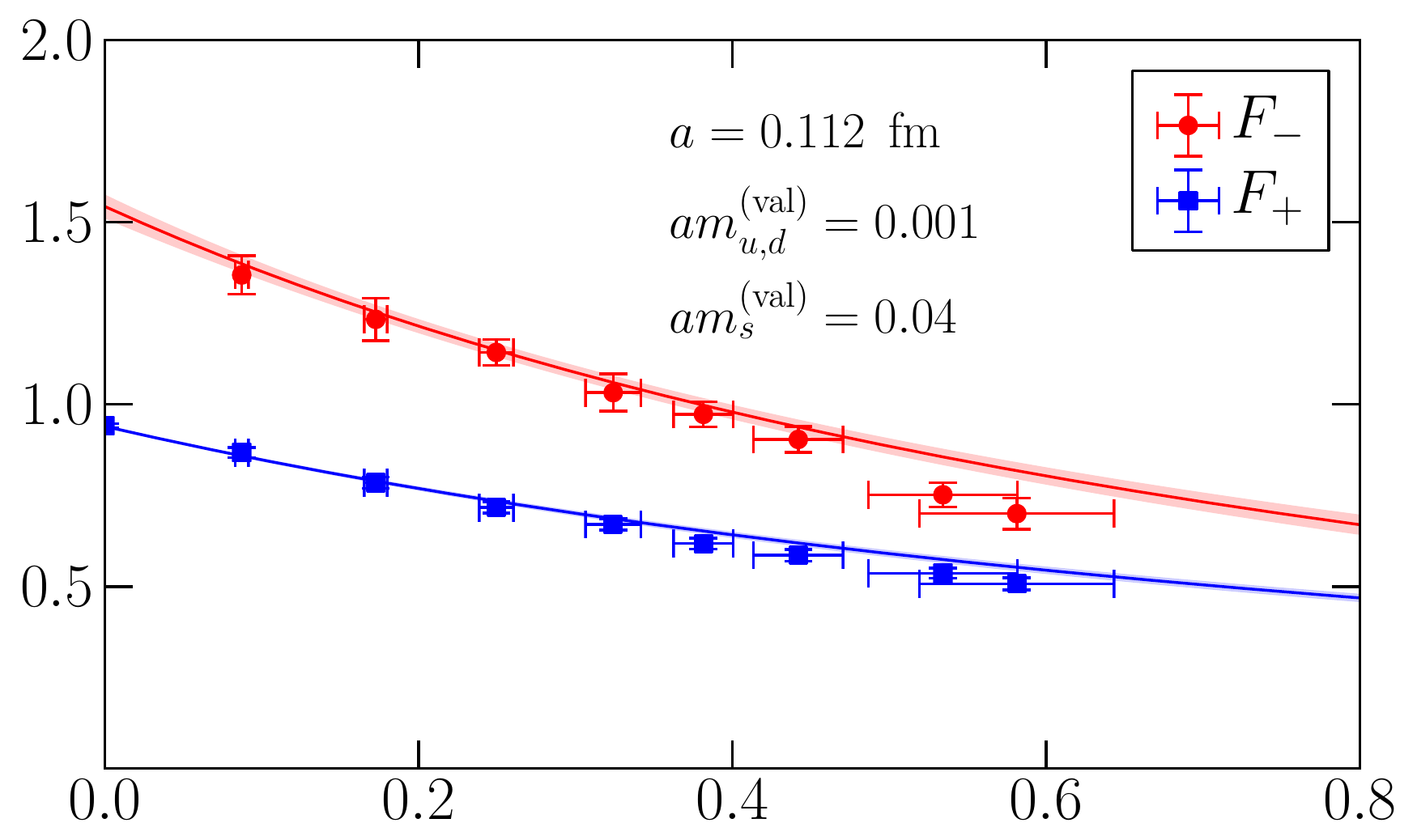}\hfill\includegraphics[width=0.48\linewidth]{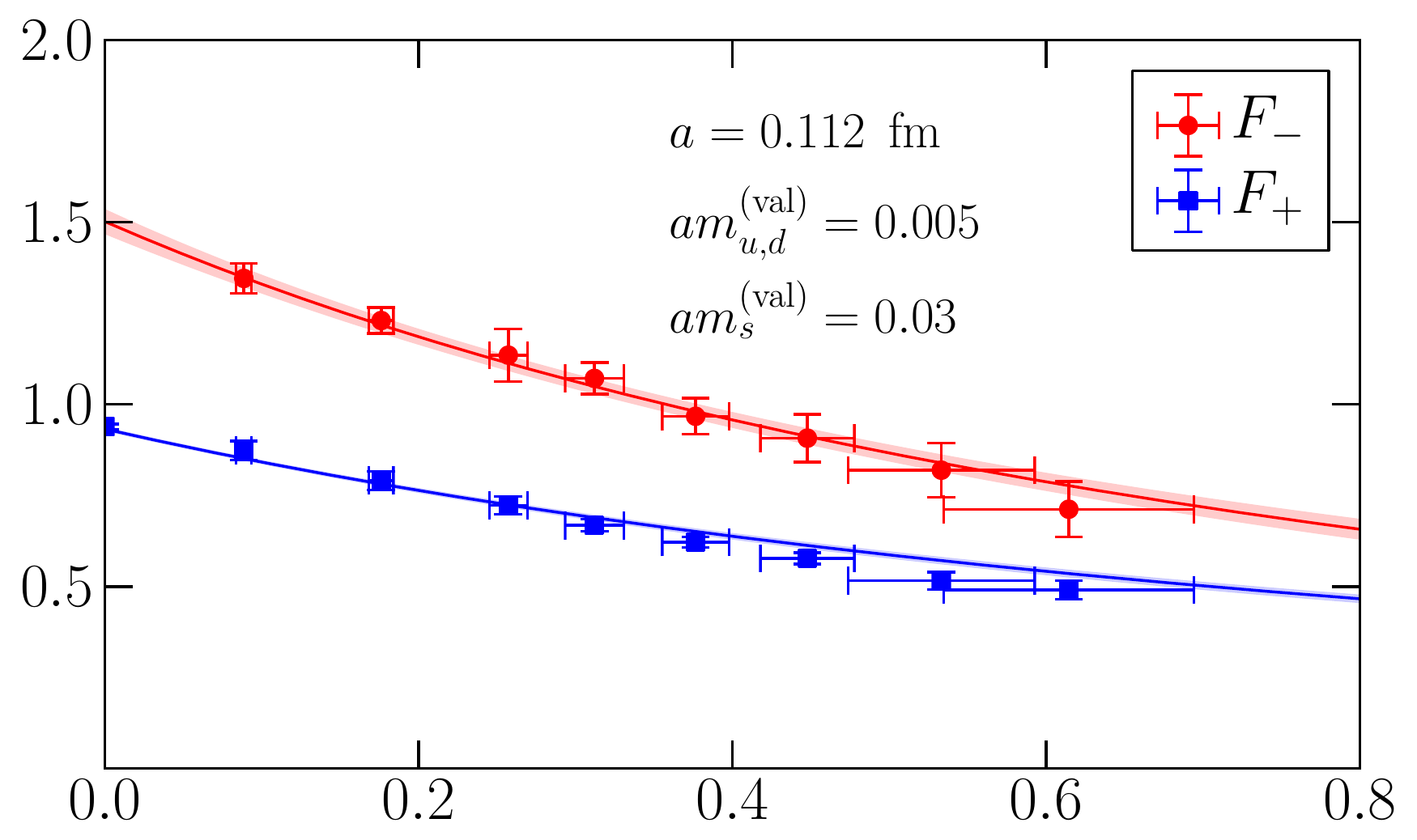} \\
\includegraphics[width=0.48\linewidth]{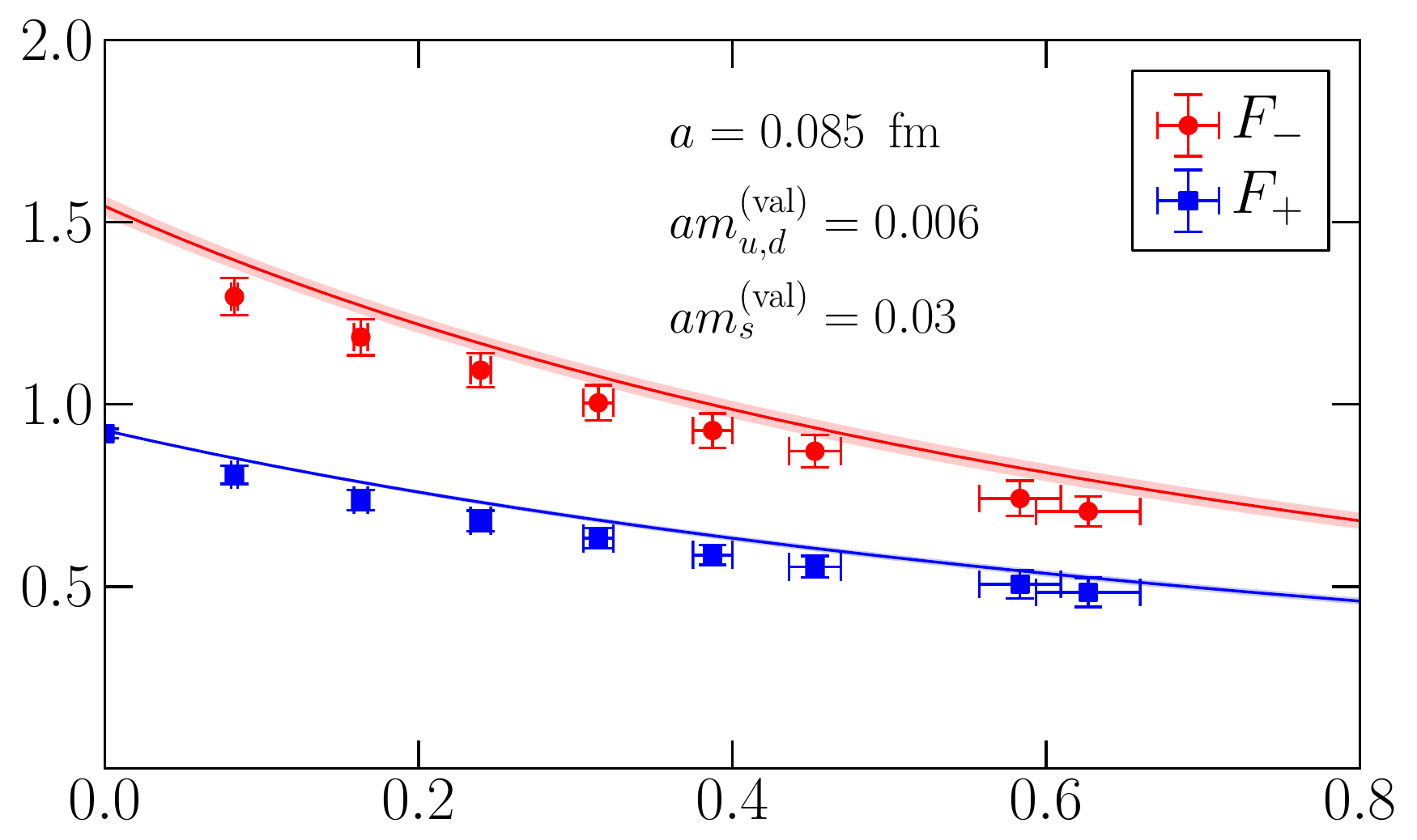}\hfill\includegraphics[width=0.48\linewidth]{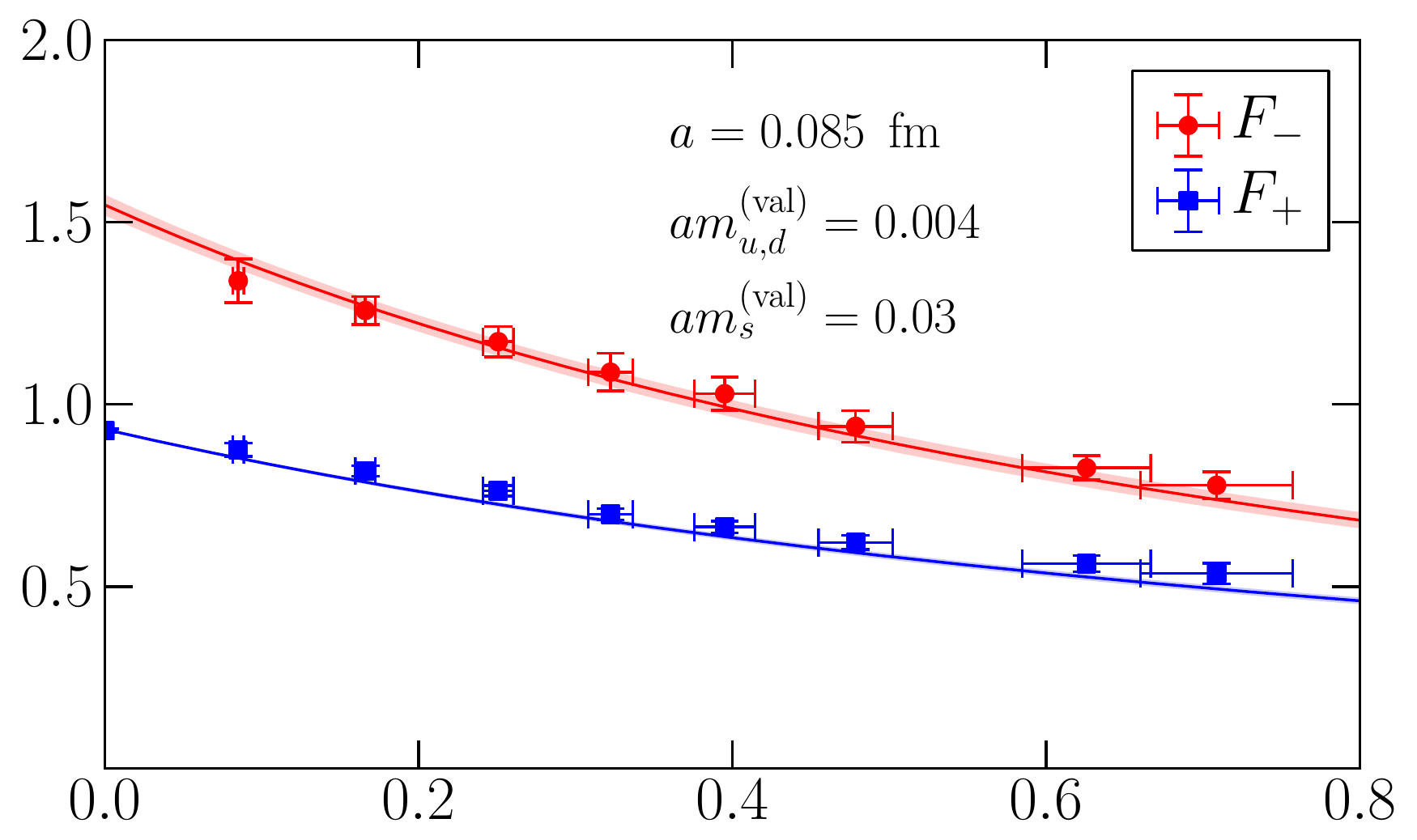} \\
\includegraphics[width=0.48\linewidth]{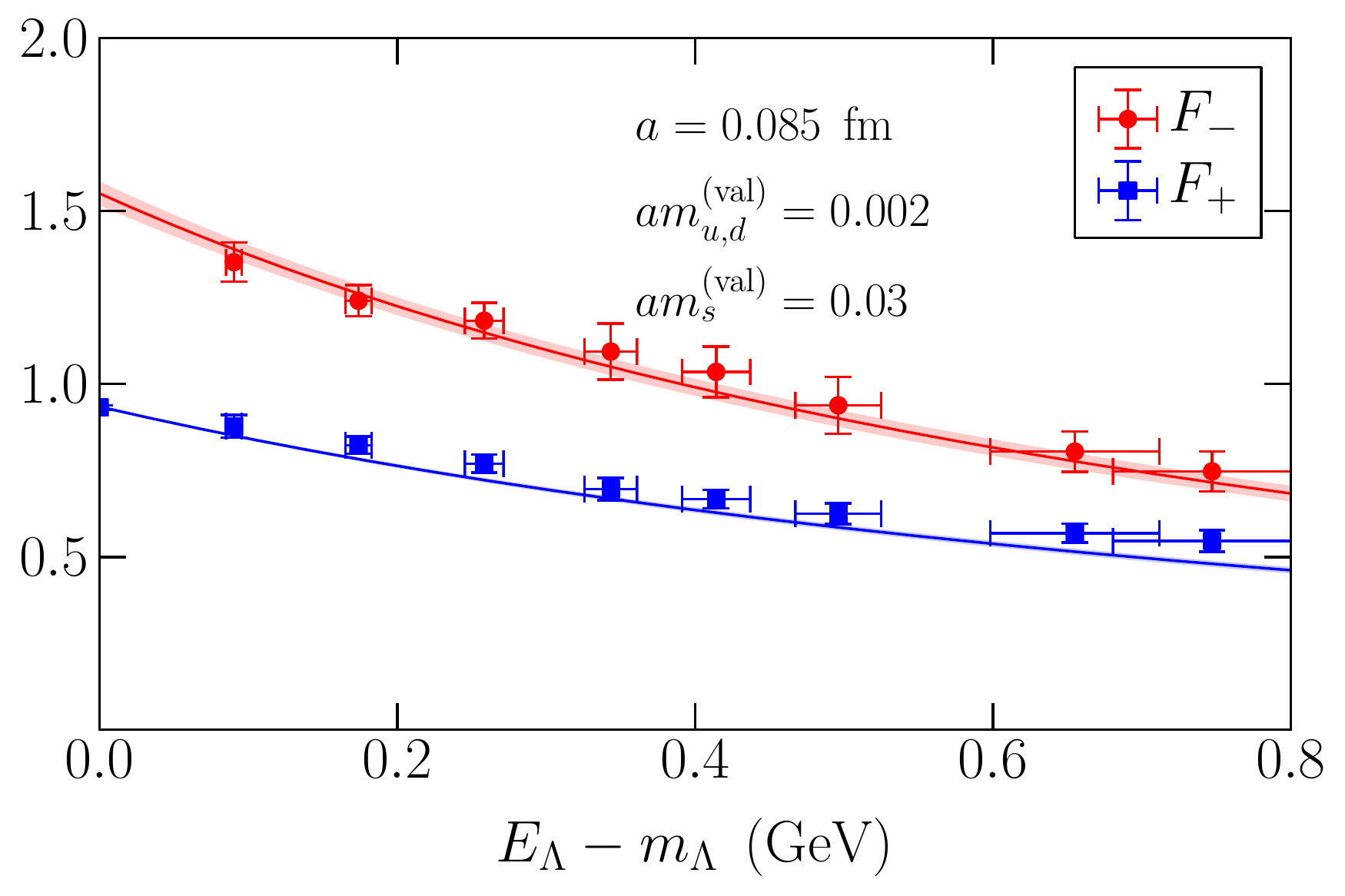}\hfill\includegraphics[width=0.48\linewidth]{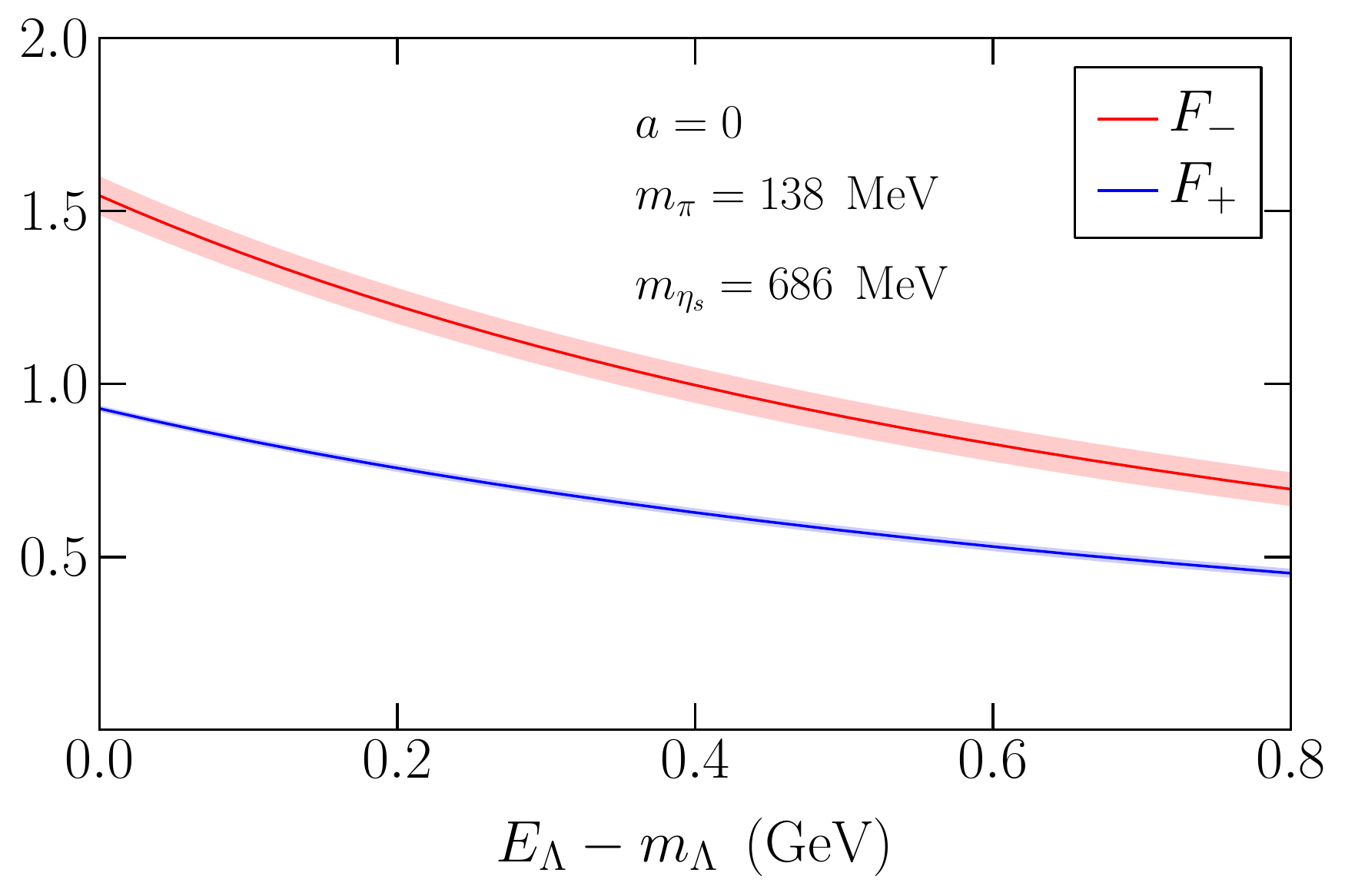} \\
\caption{\label{fig:qsqrasqrextrapall}Fits of the form factor data for $F_+$ and $F_-$ using Eq.~(\protect\ref{eq:dipole}).
The fit of $F_+$ has $\chi^2/{\rm dof}=0.84$, and the fit of $F_-$ has $\chi^2/{\rm dof}=0.72$. The lowest right plot
shows the continuum form factors for the physical light- and strange quark masses.}
\end{figure}

The fits using Eq.~(\ref{eq:dipole}) are visualized as a function of $E_\Lambda-m_\Lambda$ in Fig.~\ref{fig:qsqrasqrextrapall}.
There, we show the results for $F_\pm^{i,n}$ from Tables \ref{tab:Fplus} and \ref{tab:Fminus}, along with the fitted functions
(\ref{eq:dipole}) evaluated at the corresponding lattice spacings $a^i$ and pseudoscalar masses, $m_\pi^i$ and $m_{\eta_s}^i$.
The data are described well by the fitted functions (the \texttt{F63} set fluctuates downward, but the overall values of
$\chi^2/{\rm dof}$ are smaller than 1). The bottom-right plot in Fig.~\ref{fig:qsqrasqrextrapall} shows the fit functions
evaluated in the continuum limit ($a=0$) and for the physical values of the pseudoscalar masses. By construction, in this
physical limit, Eq.~(\ref{eq:dipole}) reduces to
\begin{eqnarray}
 F_\pm &=& \frac{N_\pm}{(X_\pm+E_\Lambda-m_\Lambda)^2}, \label{eq:Fplusminusphysical}
\end{eqnarray}
which only depends on the parameters $N_\pm$ and $X_\pm$. Our results for $N_\pm$ and $X_\pm$ are given in Table
\ref{tab:dipolefitresults}. The results for the parameters describing the dependence on the quark masses and the
lattice-spacing are $c_{l,+}= 0.094(32)\:\:{\rm GeV}^{-1}$, $c_{s,+}=-0.019(27) \:\:{\rm GeV}^{-1}$, $d_+=0.027(27)$,
$c_{l,-}=0.04(20) \:\:{\rm GeV}^{-1}$, $c_{s,-}=-0.14(11) \:\:{\rm GeV}^{-1}$, and $d_-=-0.036(67)$, which are all very
small and mostly consistent with zero.

\begin{figure}
\includegraphics[width=0.48\linewidth]{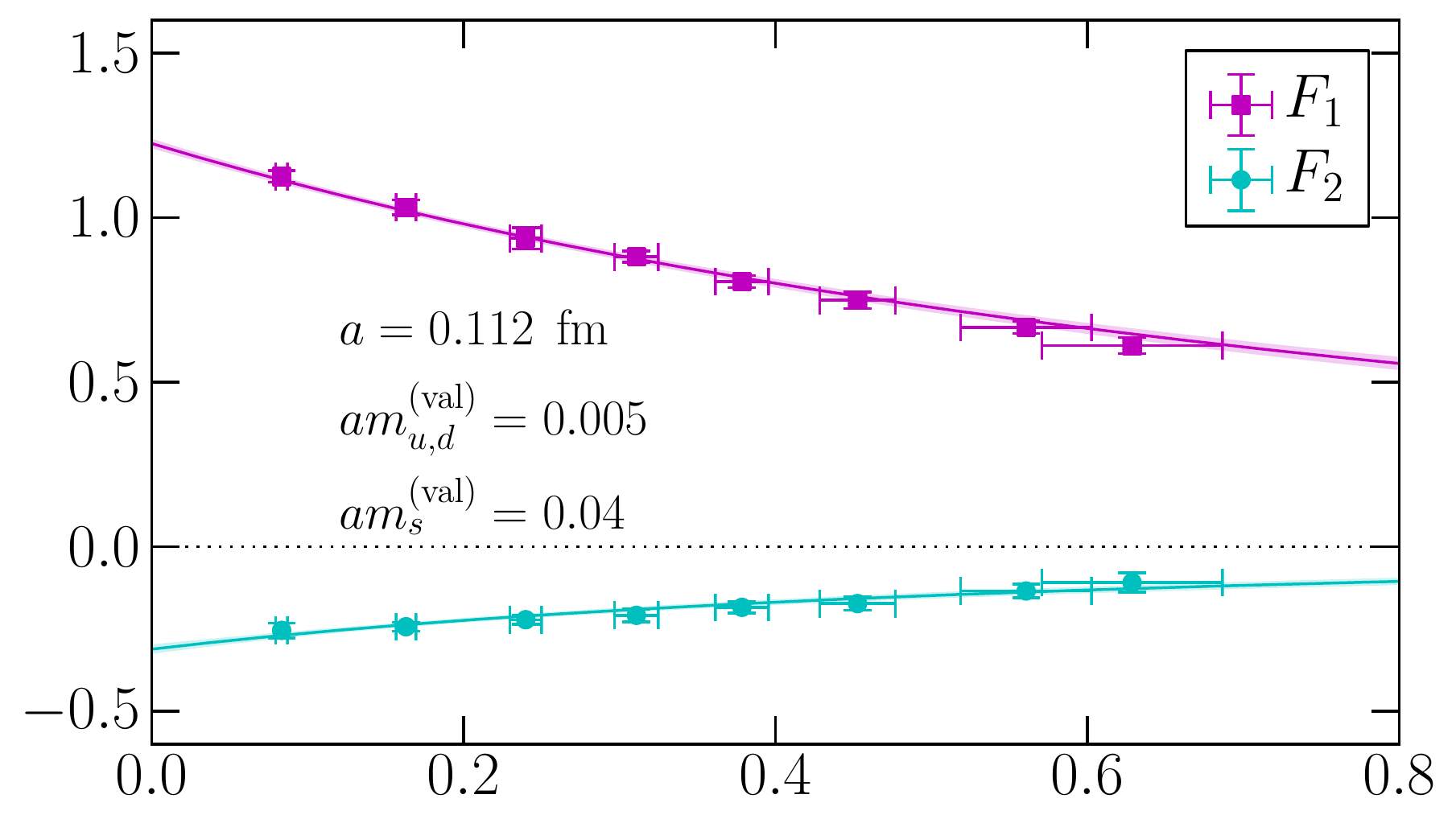}\hfill\includegraphics[width=0.48\linewidth]{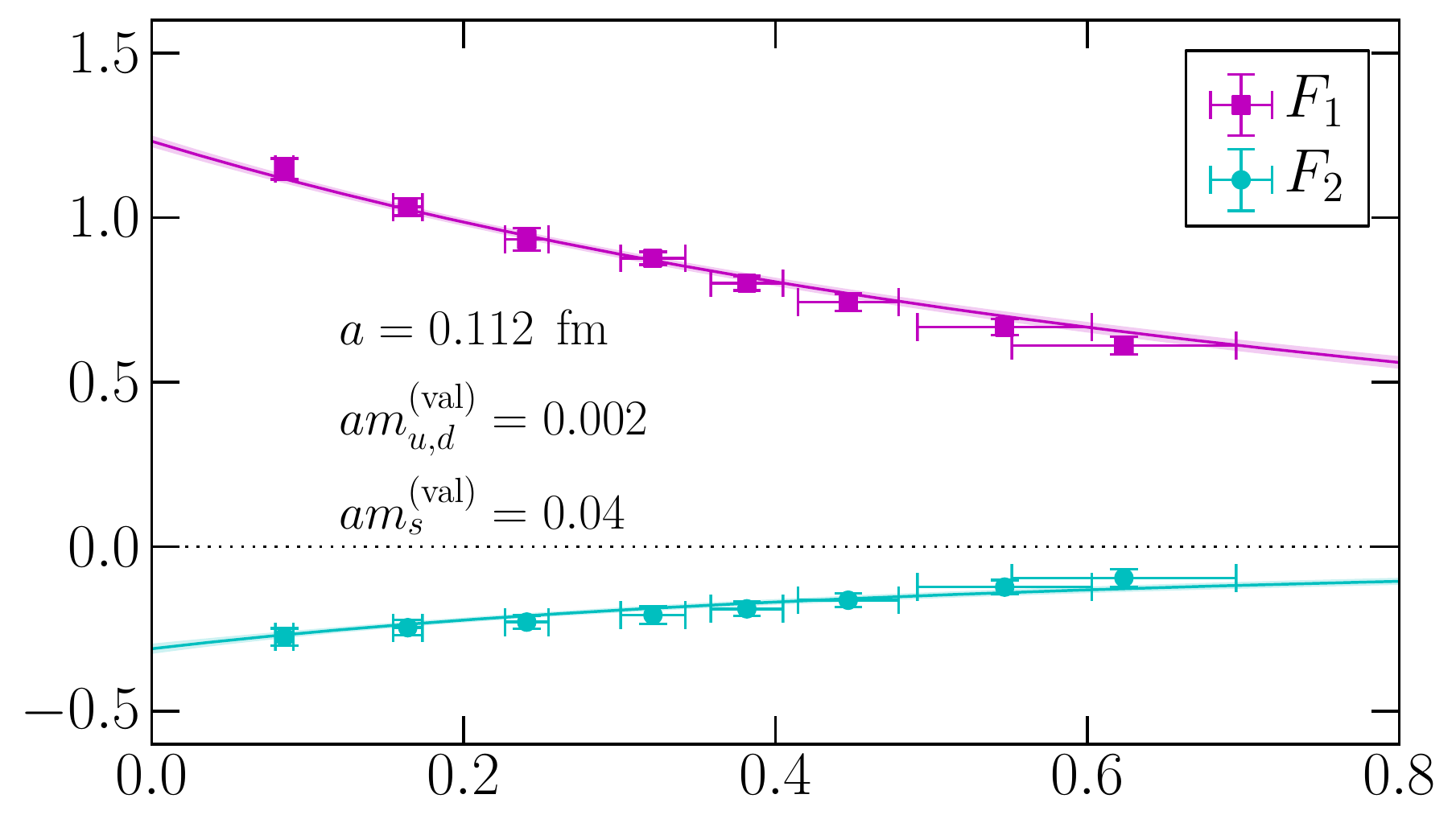} \\
\includegraphics[width=0.48\linewidth]{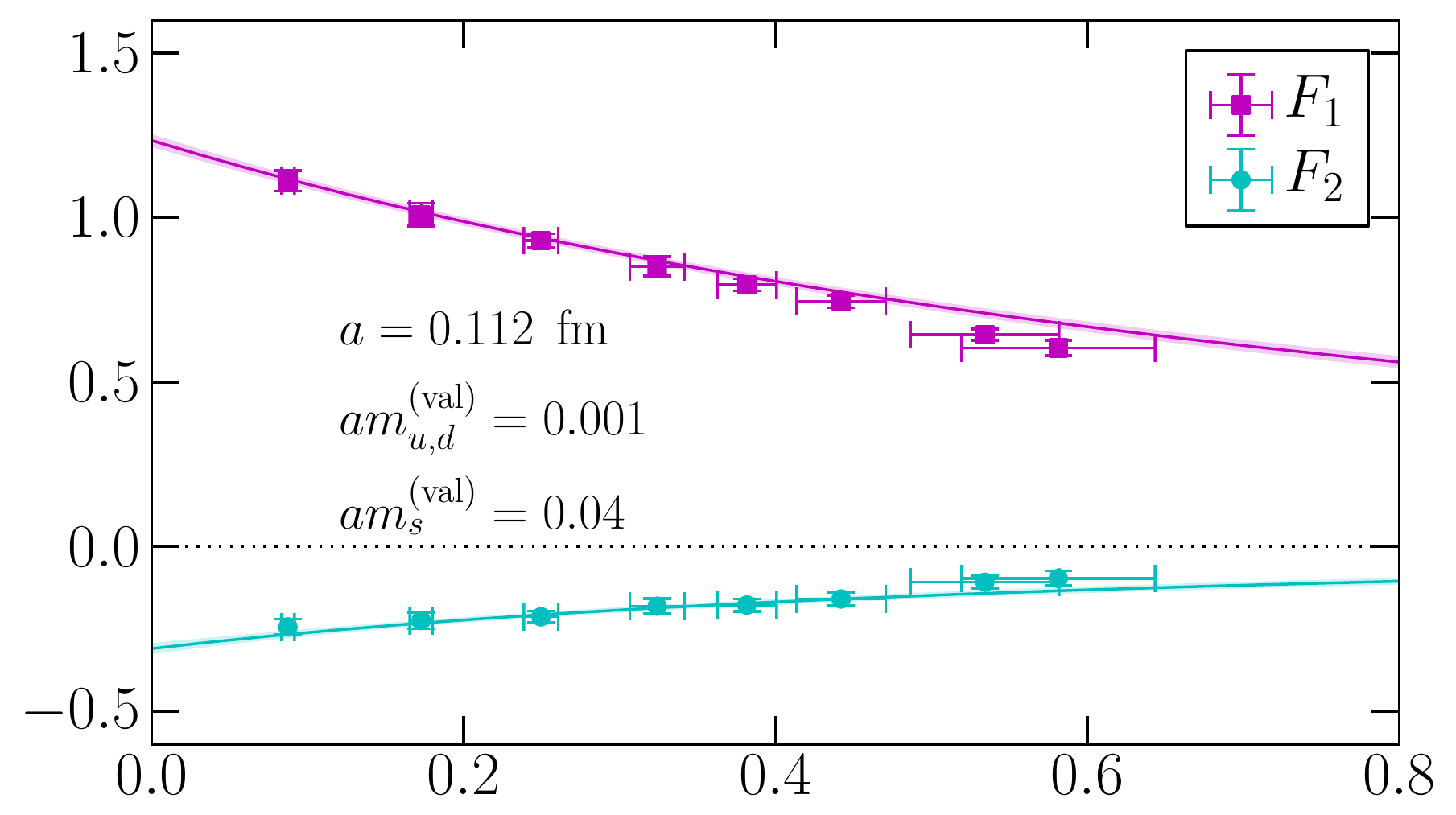}\hfill\includegraphics[width=0.48\linewidth]{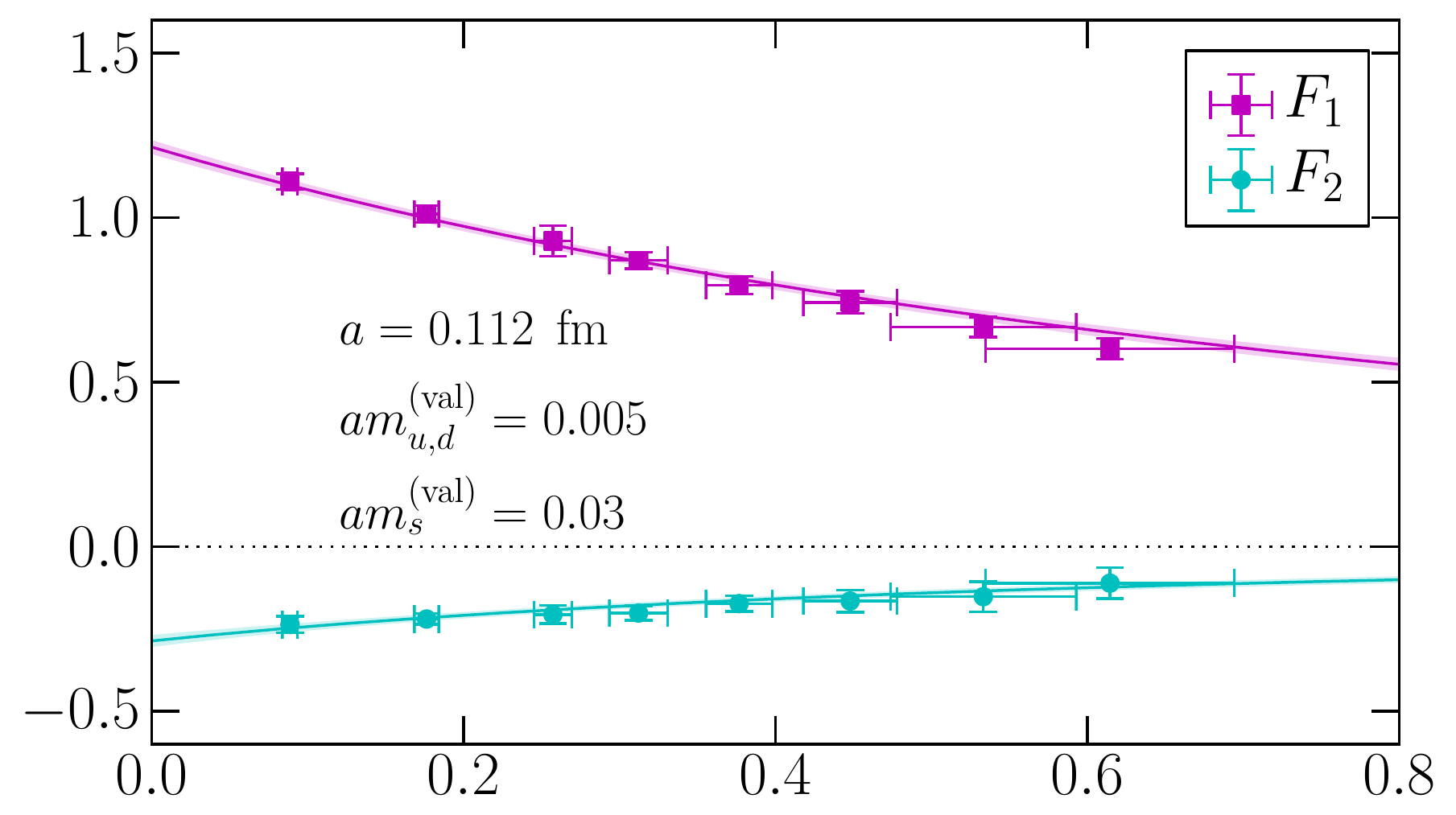} \\
\includegraphics[width=0.48\linewidth]{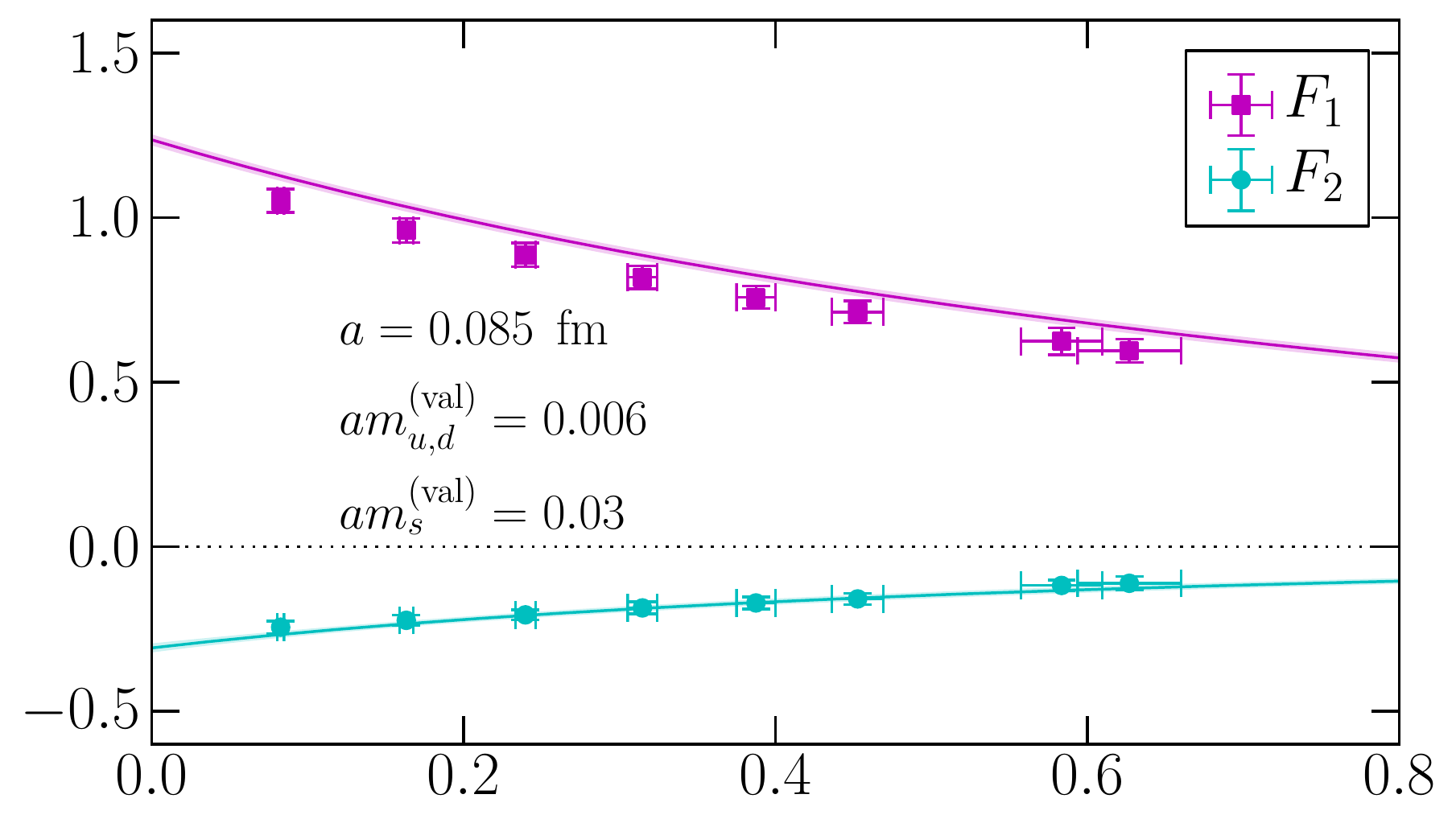}\hfill\includegraphics[width=0.48\linewidth]{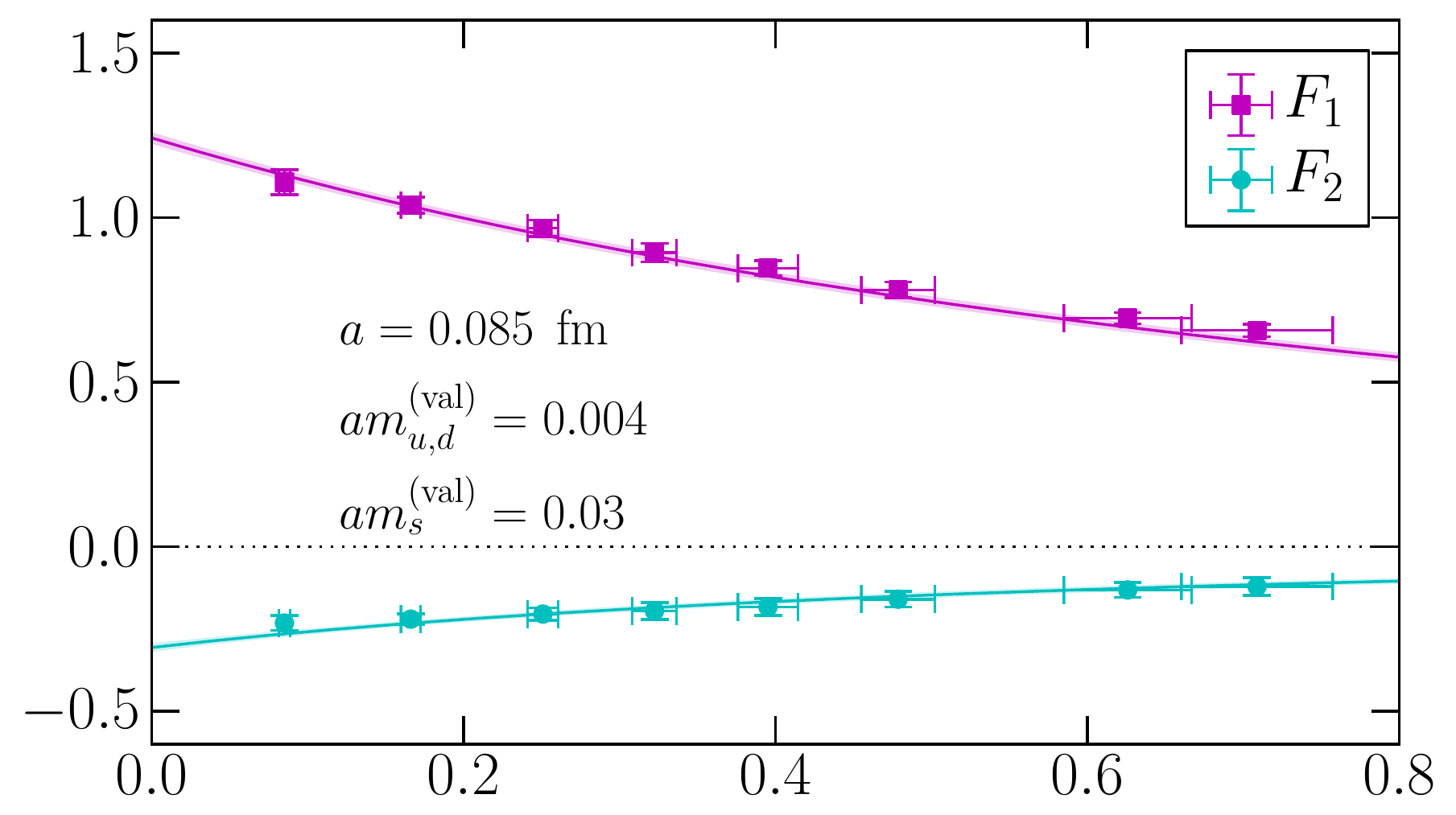} \\
\includegraphics[width=0.48\linewidth]{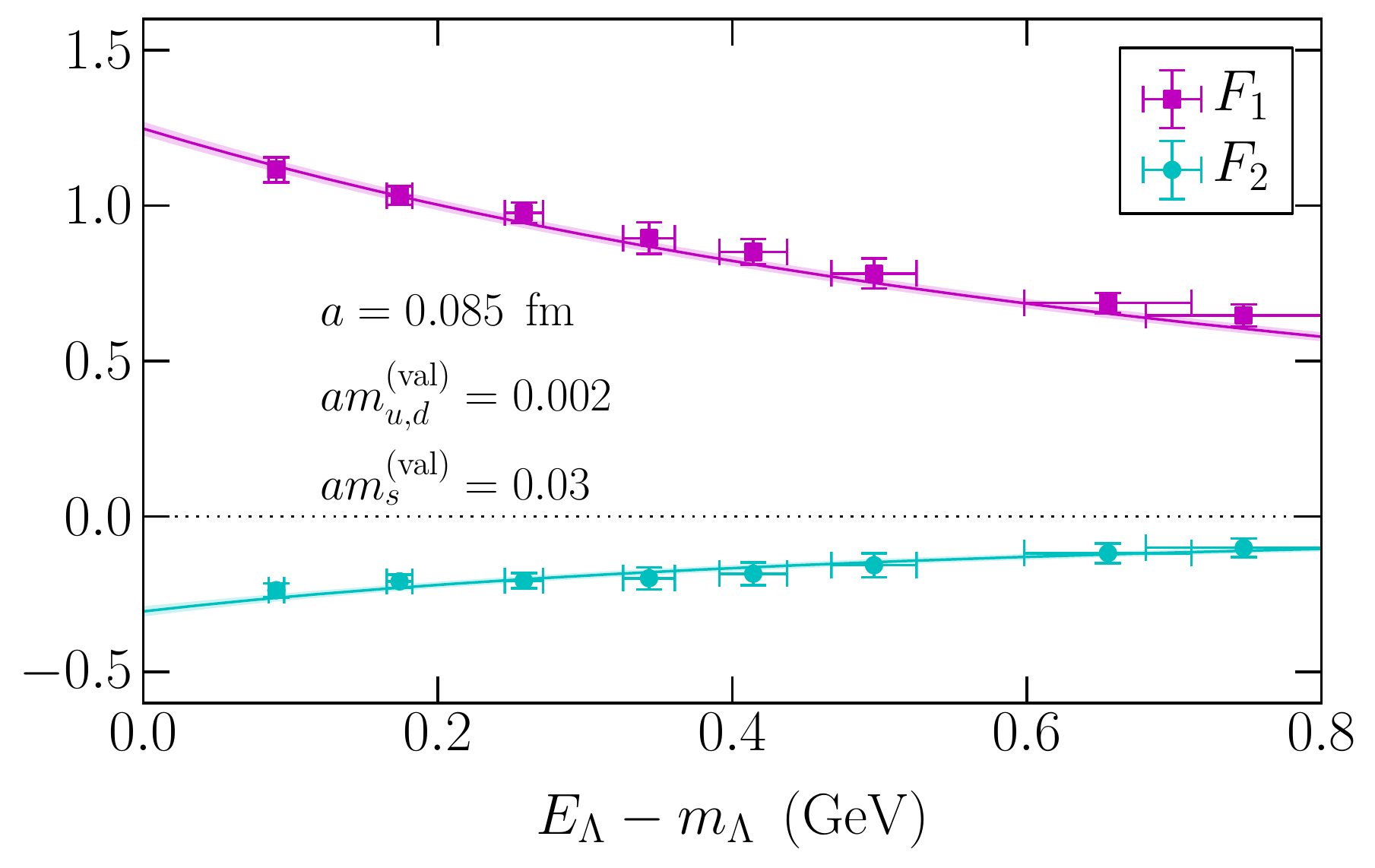}\hfill\includegraphics[width=0.48\linewidth]{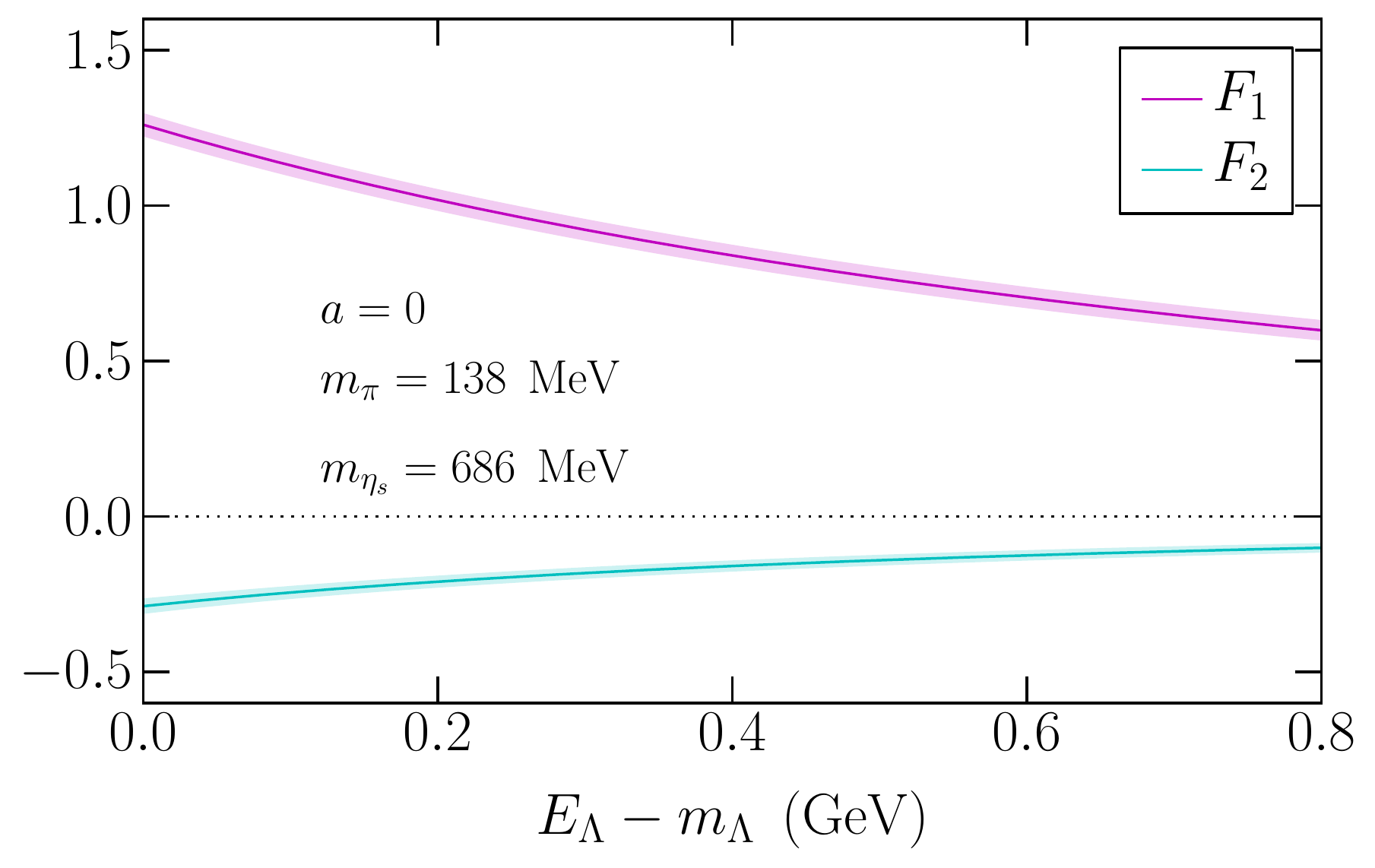} \\
\caption{\label{fig:qsqrasqrextrapallF1F2}Fits of the form factor data for $F_1$ and $F_2$ using Eq.~(\protect\ref{eq:dipoleF1F2}).
The fit of $F_1$ has $\chi^2/{\rm dof}=0.59$, and the fit of $F_2$ has $\chi^2/{\rm dof}=0.85$. The lowest right plot shows
the continuum form factors for the physical light- and strange quark masses.}
\end{figure}

Functions for the form factors $F_1$ and $F_2$ could be obtained from (\ref{eq:Fplusminusphysical}) by taking the linear
combinations $F_1=(F_+ + F_-)/2$ and $F_2=(F_+ - F_-)/2$. However, because we use independent pole parameters $X_+$ and $X_-$,
these linear combinations are no longer of the simple dipole form. Alternatively, we can also perform new fits to the lattice
data $F^{i,n}_1 = (F^{i,n}_+ + F^{i,n}_-)/2$ and $F^{i,n}_2=(F^{i,n}_+ - F^{i,n}_-)/2$ using functions of the same form as
in Eq.~(\ref{eq:dipole}), but with new parameters labeled by the subscripts $1,2$ instead of $+,-$:
\begin{eqnarray}
 F_{1,2}^{i,n} &=& \frac{N_{1,2}}{(X_{1,2}^i+E_\Lambda^{i,n}-m_\Lambda^i)^2}\cdot [1 + d_{1,2} (a^i E_\Lambda^{i,n})^2]. \label{eq:dipoleF1F2}
\end{eqnarray}
These fits are visualized in Fig.~\ref{fig:qsqrasqrextrapallF1F2}, and the resulting parameters $N_{1,2}$, $X_{1,2}$ are
given in Table \ref{tab:dipolefitresultsF1F2}. In this case, the results for the other fit parameters were
$c_{l,1}=0.09(17) \:\:{\rm GeV}^{-1}$, $c_{s,1}=-0.067(94) \:\:{\rm GeV}^{-1}$, $d_1=-0.049(53)$, $c_{l,2}=-0.06(38) \:\:{\rm GeV}^{-1}$,
$c_{s,2}=-0.35(22) \:\:{\rm GeV}^{-1}$, and $d_2=0.00(15)$.

\begin{table}
\begin{tabular}{ccc}
\hline\hline
Parameter & \hspace{1ex} & Result \\
\hline
$N_+$  && $3.188 \pm 0.268$ ${\rm GeV}^2$  \\
$X_+$  && $1.852 \pm 0.074$ ${\rm GeV}^{\phantom{2}}$  \\
$N_-$  && $4.124 \pm 0.750$ ${\rm GeV}^2$ \\
$X_-$  && $1.634 \pm 0.144$ ${\rm GeV}^{\phantom{2}}$  \\
\hline\hline
\end{tabular}
\caption{\label{tab:dipolefitresults} Fit results for $N_\pm$ and $X_\pm$ using Eq.~(\protect\ref{eq:dipole}).
The covariances are ${\rm Cov}(N_+,X_+)=0.0198\:\:{\rm GeV}^3$, ${\rm Cov}(N_-,X_-)=0.106\:\:{\rm GeV}^3$.
The results are renormalized in the $\overline{\rm MS}$ scheme at $\mu=m_b$.}
\end{table}

\begin{table}
\begin{tabular}{ccc}
\hline\hline
Parameter & \hspace{1ex} & Result \\
\hline
$N_1$  && $\wm3.975 \pm 0.576$ ${\rm GeV}^2$  \\
$X_1$  && $\wm1.776 \pm 0.123$ ${\rm GeV}^{\phantom{2}}$  \\
$N_2$  && $-0.385 \pm 0.132$ ${\rm GeV}^2$ \\
$X_2$  && $\wm1.156 \pm 0.200$ ${\rm GeV}^{\phantom{2}}$  \\
\hline\hline
\end{tabular}
\caption{\label{tab:dipolefitresultsF1F2} Fit results for $N_{1,2}$ and $X_{1,2}$ as discussed in the main text.
The covariances are ${\rm Cov}(N_1,X_1)=0.0692\:\:{\rm GeV}^3$, ${\rm Cov}(N_2,X_2)=-0.0256\:\:{\rm GeV}^3$.
The results are renormalized in the $\overline{\rm MS}$ scheme at $\mu=m_b$.}
\end{table}

\subsection{\label{sec:systerrs}Estimates of systematic uncertainties}

\begin{figure}
\includegraphics[height=5.65cm]{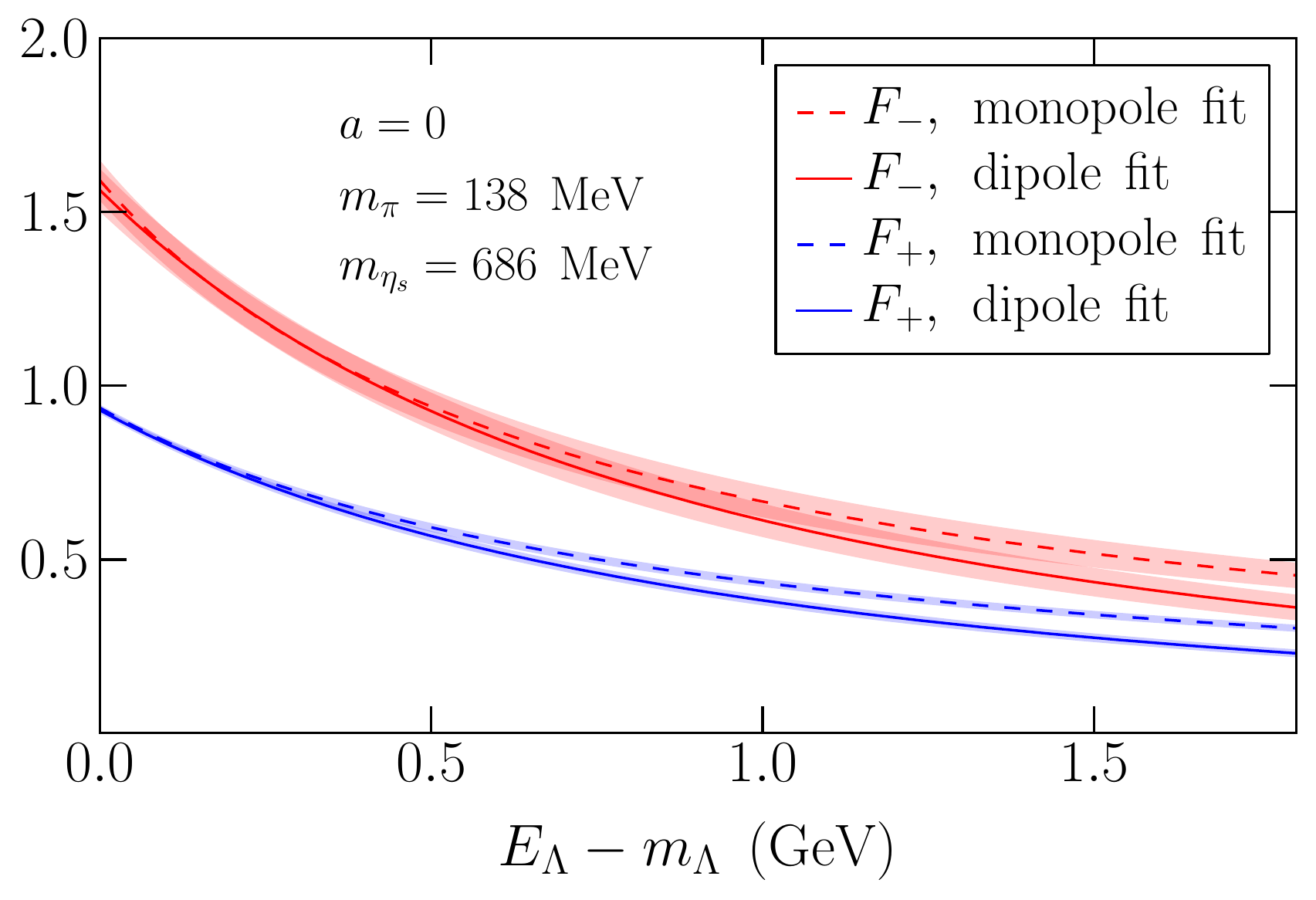}\hfill\includegraphics[height=5.5cm]{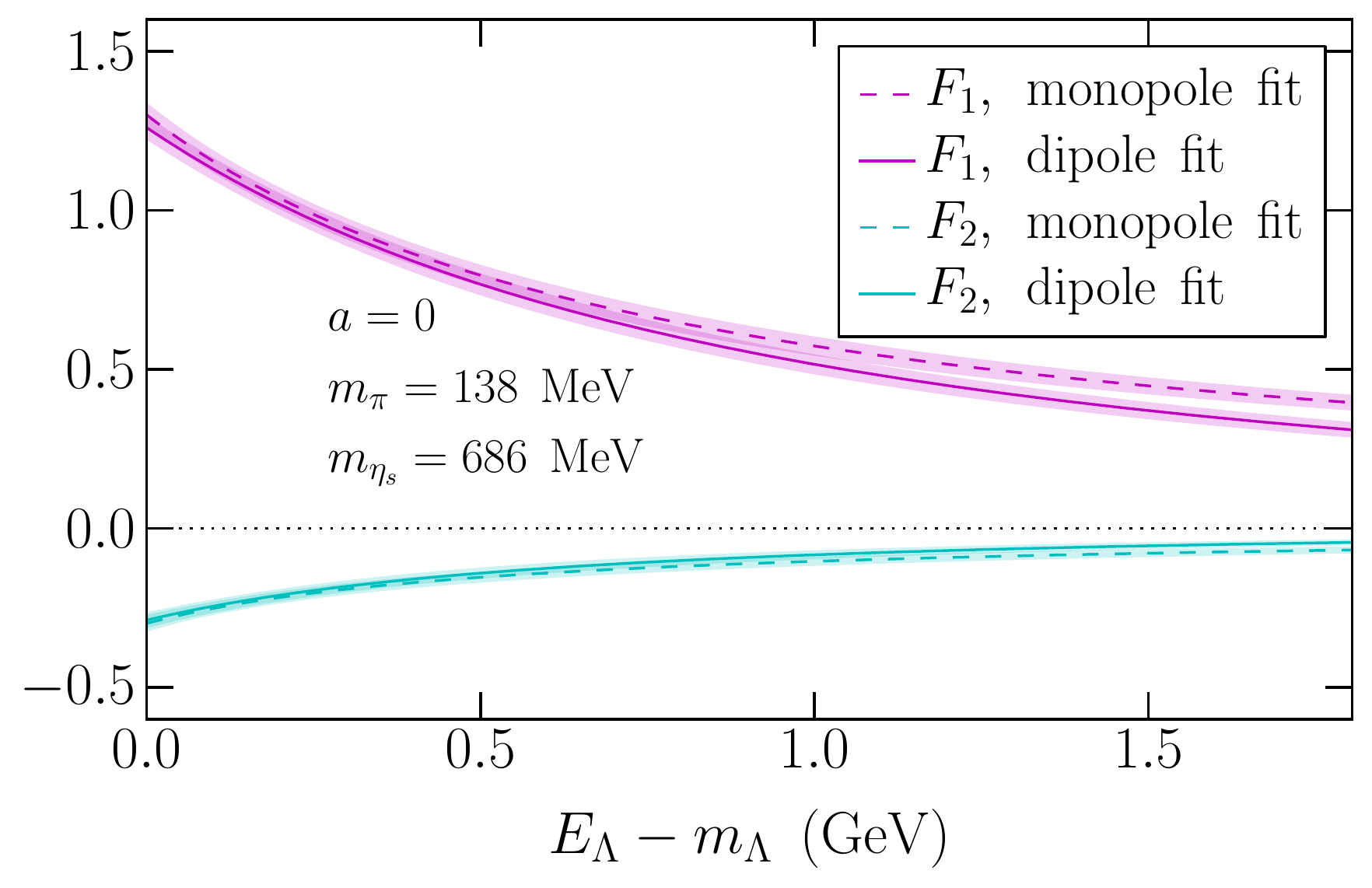} \\
\caption{\label{fig:dipole_vs_monopole}Comparison of our results for the $\Lambda_Q \to \Lambda$ form factors from dipole fits
using Eqs.~(\ref{eq:dipole}) and (\ref{eq:dipoleF1F2}) to results from monopole fits
(the same equations without the power of 2 in the denominator). Shown here is the entire kinematic range needed for the
decay $\Lambda_b\to\Lambda\ell^+\ell^-$ with $m_\ell=0$ (the point $q^2=0$ corresponds to $E_\Lambda-m_\Lambda\approx1.8$ GeV).
In the range where we have lattice data ($E_\Lambda-m_\Lambda \lesssim 0.7$ GeV), the dipole and monopole functions are
consistent with each other, but for large $E_\Lambda-m_\Lambda$, model-dependence can be seen.
In our main analysis we choose the dipole fits as they have slightly lower values of $\chi^2/{\rm dof}$.}
\end{figure}

The remaining systematic uncertainties in our form factor results include missing higher-order renormalization corrections
to the heavy-light current, finite-volume effects, chiral-extrapolation errors, and residual discretization errors. We discuss each of these below.
Furthermore, for large $E_\Lambda-m_\Lambda$, where we do not have lattice data, our assumption of a dipole shape in Eqs.~(\ref{eq:dipole}) and
(\ref{eq:dipoleF1F2}) introduces an unknown model-dependence. This is illustrated in Fig.~\ref{fig:dipole_vs_monopole}), where we compare
the dipole fits to monopole fits. However, we do not have confidence that this difference is a reliable estimate of a fitting
form systematic uncertainty (or indeed that such a systematic uncertainty can be constructed) and so leave this to the judgment of the reader.

To estimate the systematic uncertainty due to missing higher-order renormalization corrections to the heavy-light current (\ref{eq:LHQETcurrent}),
we vary the scale $\mu$ in the matching coefficients $\mathcal{Z}(\mu)$, $c^{(m_s a)}(\mu)$, $c^{(p_s a)}(\mu)$, and in the renormalization-group
running $U(m_b,\mu)$. We then recompute the ratios (\ref{eq:Rplus}) and (\ref{eq:Rminus}) with the modified current.
Changing $\mu$ from $a^{-1}$ to $2 a^{-1}$ results in a 7\% change of both $R_+$ and $R_-$ at the coarse lattice spacing and a 6\%
change of both $R_+$ an $R_-$ at the fine lattice spacing. These relative changes are nearly independent of the source-sink separation,
the momentum, and the quark masses. Thus, we take the renormalization uncertainty in the final form factor results to be 6\%.

Finite-volume effects in the lattice data are unknown (as in the chiral extrapolation, no low-energy effective theory
exists to guide us over the full range of $E_\Lambda$), but are expected to be of order $\exp(-m_\pi L)$. The lowest pion
mass used in our calculation is $m_\pi\approx227$ MeV, corresponding to $m_\pi L\approx 3.1$ and $\exp(-m_\pi L)\approx0.04$.
The average value of $\exp(-m_\pi L)$ for the different data sets (see Table \ref{tab:params}) is about 0.02.
Given these values, we estimate the systematic uncertainty in our final results due to finite-volume effects to be 3\%.

The chiral extrapolations of the form factors were performed quadratically in the valence pseudoscalar masses, i.e.~linearly
in the valence-quark masses, ignoring that some of the data were partially quenched and ignoring possible nonanalytic dependence
on the quark masses. To study the effect of the quark-mass extrapolations, we perform new fits with with either $c_{l,\pm}$ or
$c_{s,\pm}$, or both, set to zero, and consider the changes in the extracted form factors $F_+$ and $F_-$
(analogously also for $F_1$ and $F_2$). This corresponds to replacing the linear fits of the quark-mass dependence by constant fits.
The resulting relative changes in $F_+$, $F_-$, $F_1$, and $F_2$ when setting $c_l=0$ are below 1\% throughout
the kinematic range where we have lattice data; the biggest relative change (5\%) is seen in $F_2$ at zero recoil when
setting $c_s=0$. However, all of the changes are consistent with zero within statistical uncertainties.

\begin{figure}
\includegraphics[height=5.65cm]{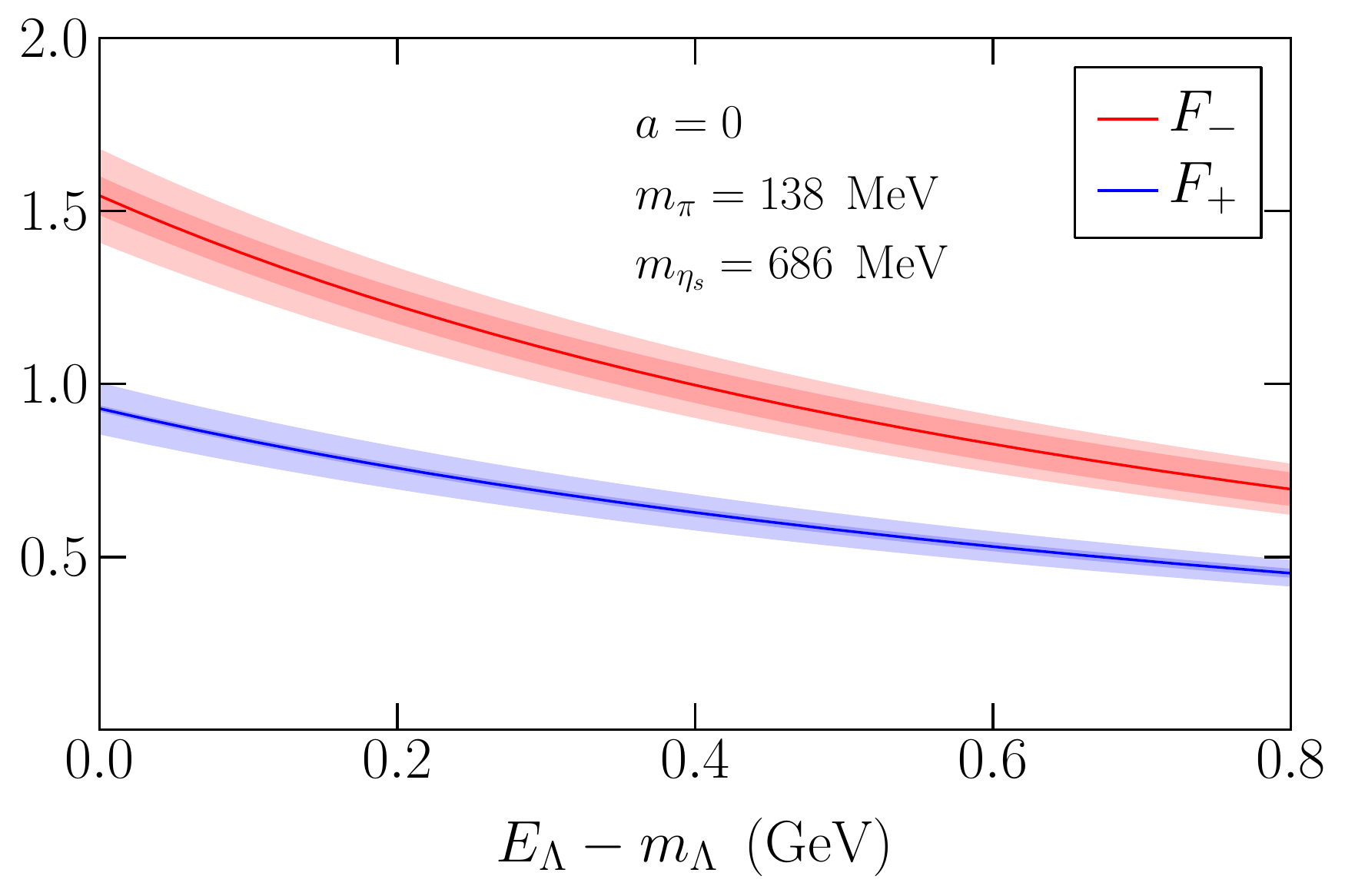}\hfill\includegraphics[height=5.5cm]{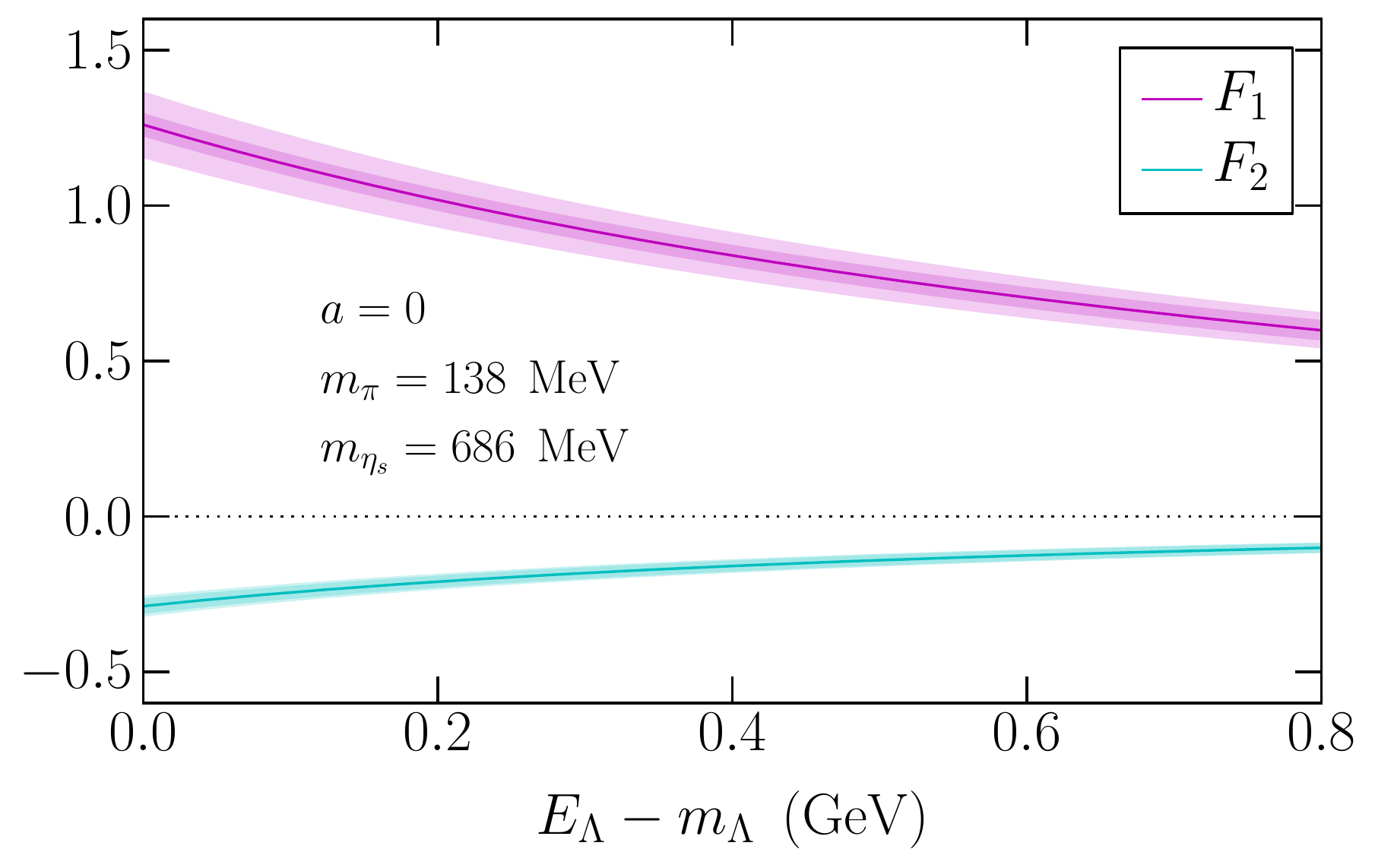} \\
\caption{\label{fig:finalFFs}Final results for the $\Lambda_Q \to \Lambda$ form factors, given by
$F_\pm = N_\pm/(X_\pm+E_\Lambda-m_\Lambda)^2$ and $F_{1,2} = N_{1,2}/(X_{1,2}+E_\Lambda-m_\Lambda)^2$
with parameters as in Tables \protect\ref{tab:dipolefitresults} and \protect\ref{tab:dipolefitresultsF1F2}.
The dark shaded bands show the statistical uncertainty, and the light shaded bands show the total
(including 8\% systematic) uncertainty. The results are renormalized in the $\overline{\rm MS}$ scheme at $\mu=m_b$.}
\end{figure}

\begin{figure}
\includegraphics[width=0.5\linewidth]{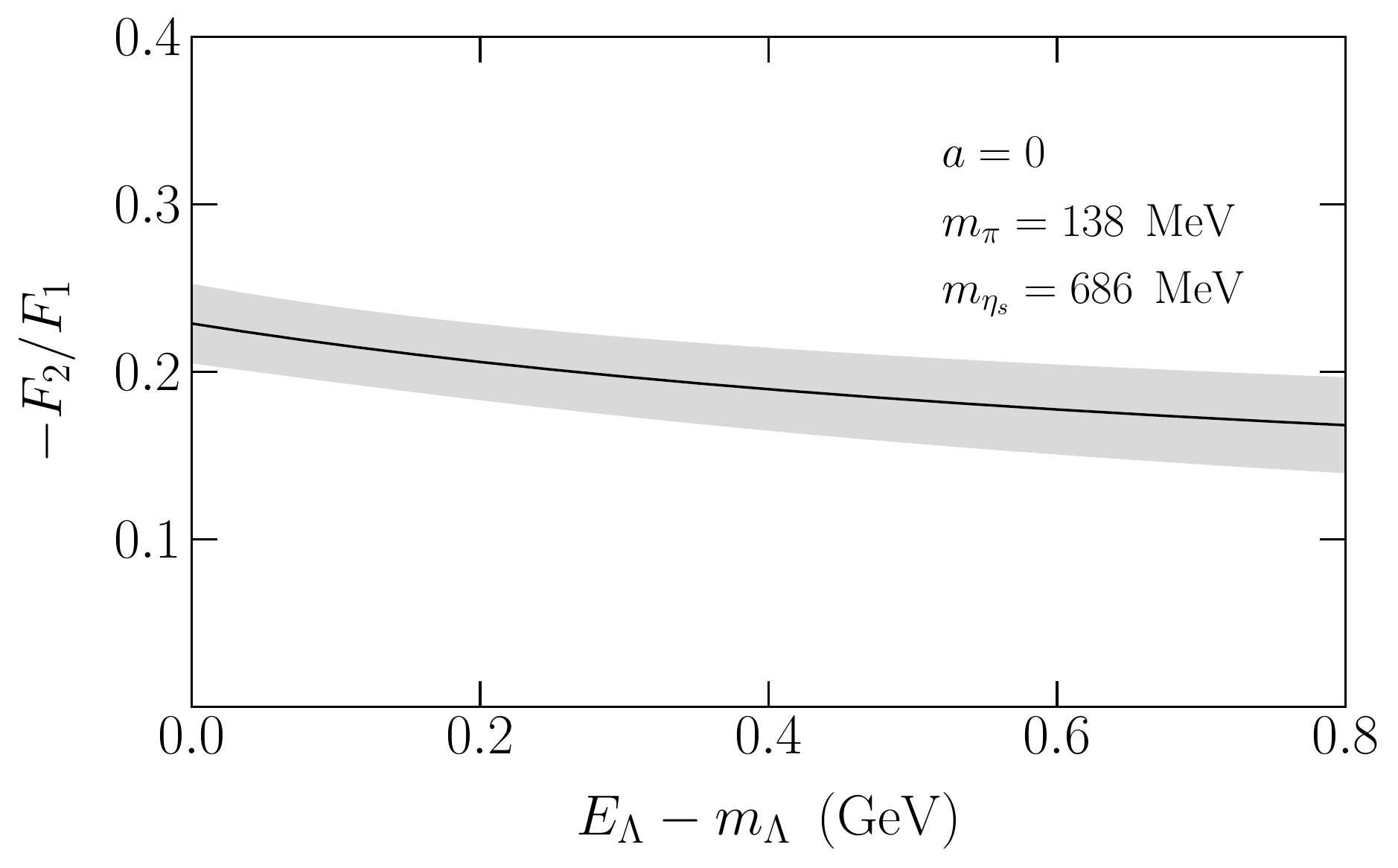}
\caption{\label{fig:F1F2ratio}Final results for the form factor ratio $-F_2/F_1$, given by
$-(N_{2}/N_{1}) (X_{1}+E_\Lambda-m_\Lambda)^2 / (X_{2}+E_\Lambda-m_\Lambda)^2$ with
parameters as in Table \protect\ref{tab:dipolefitresultsF1F2}. The shaded band shows total uncertainty,
which is dominated by the statistical uncertainty.}
\end{figure}

\FloatBarrier

By ``residual discretization errors'' we mean discretization errors that are not eliminated through the continuum
extrapolation using the factors of $[1 + d_\pm (a^i E_\Lambda^{i,n})^2]$ in Eq.~(\ref{eq:dipole})
and $[1 + d_{1,2} (a^i E_\Lambda^{i,n})^2]$ in Eq.~(\ref{eq:dipoleF1F2}). While we know that the leading
discretization orders are quadratic in $a$, we do not know how they depend on $E_\Lambda$.
To study the effect of the factors $[1 + d_\pm (a^i E_\Lambda^{i,n})^2]$ and $[1 + d_{1,2} (a^i E_\Lambda^{i,n})^2]$,
we perform new fits with $d_\pm$ (or $d_{1,2}$) set to zero. In this case, the resulting changes in all form factors
are less than 4\% for $E_\Lambda-m_\Lambda < 0.8$ GeV, and are consistent with zero within the statistical uncertainties.

Combining the uncertainties in the above discussion, we estimate the total systematic uncertainty of our final results for the form factors for
$E_\Lambda-m_\Lambda < 0.8$ GeV to be 8\%. Plots of the form factors including this systematic uncertainty are shown
in Fig.~\ref{fig:finalFFs}. The relatively large systematic uncertainty from the current matching cancels in ratios such as $F_2/F_1$.
This ratio is shown in Fig.~\ref{fig:F1F2ratio}, and we estimate the total systematic uncertainty in $F_2/F_1$ to be 5\%.

\section{\label{sec:Lambdabdecay}The decay $\Lambda_b \to \Lambda\: \ell^+ \ell^-$}
\FloatBarrier

As a first application of our results for the $\Lambda_Q \to \Lambda$ form factors, we calculate the differential
branching fraction for the decays $\Lambda_b \to \Lambda \:\ell^+ \ell^-$ with $\ell=e,\mu,\tau$ in the standard model.
Long-distance contributions (discussed at the end of this section) are not included. In the following, it is convenient to use the notation
\begin{equation}
\mathcal{H}_{\rm eff} = -\frac{2 G_F}{\sqrt{2}}V_{tb}V_{ts}^* \sum_{i=7,9,10} \Big( C_{i,{\rm eff}}^V O_i^V - C_{i,{\rm eff}}^A  O_i^A  \Big),
\end{equation}
with
\begin{align}
\nonumber O_7^V    =& \frac{e}{16\pi^2} m_b \bar{s} \sigma^{\mu\nu}  b \: F_{\mu\nu}^{(\rm e.m.)},
&  O_7^A =& -\frac{e}{16\pi^2} m_b  \bar{s} \sigma^{\mu\nu} \gamma_5 b \: F_{\mu\nu}^{(\rm e.m.)}, \\
\nonumber O_9^V    =& \frac{e^2}{16\pi^2} \bar{s} \gamma^\mu  b\: \bar{l} \gamma_\mu l,
& O_9^A =& \frac{e^2}{16\pi^2} \bar{s} \gamma^\mu \gamma_5 b\: \bar{l} \gamma_\mu l, \\
O_{10}^V =& \frac{e^2}{16\pi^2} \bar{s} \gamma^\mu  b\: \bar{l} \gamma_\mu \gamma_5 l,
& O_{10}^A =& \frac{e^2}{16\pi^2} \bar{s} \gamma^\mu \gamma_5 b\: \bar{l} \gamma_\mu \gamma_5 l, 
\end{align}
and
\begin{eqnarray}
\nonumber C_{i,{\rm eff}}^V &=& C_{i,{\rm eff}} + C_{i,{\rm eff}}', \\
C_{i,{\rm eff}}^A &=& C_{i,{\rm eff}} - C_{i,{\rm eff}}'.
\end{eqnarray}
The ``effective'' Wilson coefficients $C_{i,{\rm eff}}$ and $C_{i,{\rm eff}}'$ ($i=7,9,10$), which are defined
at the scale $\mu=m_b$, contain the one-loop matrix elements of the four-quark operators $O_1,...,O_6$ \cite{Buras:1993xp}.

The invariant matrix element of $\mathcal{H}_{\rm eff}$ is given by
\begin{eqnarray}
\nonumber \mathcal{M}&=&-\langle \Lambda(p',s')\:\ell^+(p_+,s_+)\:\ell^-(p_-,s_-) | \mathcal{H}_{\rm eff} | \Lambda_b(p,s) \rangle \\
\nonumber &=& \frac{G_F\:\alpha_{\rm em}}{2\sqrt{2}\pi}V_{tb}V_{ts}^* \Big[ A_\mu \:\bar{u}(p_+,s_+)\gamma^\mu v(p_-,s_-)
+ B_\mu \: \bar{u}(p_+,s_+)\gamma^\mu\gamma_5 v(p_-,s_-)  \Big],
\end{eqnarray}
with the hadronic matrix elements
\begin{eqnarray}
\nonumber A_\mu&=& \langle \Lambda(p',s')|\Big(C_{9,{\rm eff}}^V(q^2) \:\bar{s} \gamma_\mu b
- C_{9,{\rm eff}}^A(q^2) \:\bar{s} \gamma_\mu\gamma_5 b - C_{7,\rm eff}^V \frac{2m_b}{q^2} q^\nu \bar{s} i\sigma_{\mu\nu} b
- C_{7,\rm eff}^A \frac{2m_b}{q^2} q^\nu \bar{s} i\sigma_{\mu\nu}\gamma_5 b \Big)| \Lambda_b(p,s) \rangle, \\
 B_\mu&=& \langle \Lambda(p',s')| \Big(C_{10,\rm eff}^V \:\bar{s} \gamma_\mu b
- C_{10,\rm eff}^A \:\bar{s} \gamma_\mu\gamma_5 b \Big) | \Lambda_b(p,s) \rangle. \label{eq:hadMEqcd}
\end{eqnarray}
Here we have contracted the electromagnetic field strength tensor in $O_7^V$ and $O_7^A$
with a perturbative insertion of the leptonic QED interaction term.

In order to use the HQET relation (\ref{eq:FFdef}) to express the hadronic matrix elements in terms of the form factors
$F_1$ and $F_2$, we first need to match the QCD $b \to s$ currents in Eq.~(\ref{eq:hadMEqcd}) to HQET
currents. We use the one-loop perturbative results in naive dimensional regularization from Ref.~\cite{Eichten:1989zv}:
\begin{eqnarray}
\nonumber \bar{s} \gamma_\mu b &=& c_\gamma\: \bar{s} \gamma_\mu Q + c_v\: \bar{s} v_\mu Q, \\
\nonumber \bar{s} \gamma_\mu\gamma_5 b &=& c_\gamma\: \bar{s} \gamma_\mu\gamma_5 Q - c_v\: \bar{s} v_\mu \gamma_5 Q, \\
\nonumber \bar{s} \sigma_{\mu\nu} b &=& c_\sigma\: \bar{s} \sigma_{\mu\nu} Q, \\
 \bar{s} \sigma_{\mu\nu}\gamma_5 b &=& c_\sigma\: \bar{s} \sigma_{\mu\nu}\gamma_5 Q,
\end{eqnarray}
with
\begin{eqnarray}
\nonumber  c_\gamma &=& 1-\frac{\alpha_s(\mu)}{\pi}\left[ \frac{4}{3} + \ln\left( \frac{\mu}{m_b} \right) \right], \\
\nonumber  c_v      &=& \frac{2}{3} \frac{\alpha_s(\mu)}{\pi}, \\
 c_\sigma      &=& 1-\frac{\alpha_s(\mu)}{\pi}\left[ \frac{4}{3} + \frac{5}{3}\ln\left( \frac{\mu}{m_b} \right) \right].
\end{eqnarray}
This gives
\begin{eqnarray}
\nonumber A_\mu &=& \bar{u}(p',s')\Big(F_1+\slashed{v} F_2\Big)\Bigg(C_{9,{\rm eff}}^V(q^2) c_\gamma \gamma_\mu
+ C_{9,{\rm eff}}^V(q^2) c_v \: v_\mu   - C_{9,{\rm eff}}^A(q^2) c_\gamma \: \gamma_\mu\gamma_5 + C_{9,{\rm eff}}^A(q^2) c_v\: v_\mu\gamma_5 \\
\nonumber && \hspace{30ex}  - C_{7,\rm eff}^V c_\sigma \frac{2m_b}{q^2} q^\nu  i\sigma_{\mu\nu}
- C_{7,\rm eff}^A c_\sigma \frac{2m_b}{q^2} q^\nu  i\sigma_{\mu\nu}\gamma_5  \Bigg) u(p,s),  \\
 B_\mu &=& \bar{u}(p',s')\Big(F_1+\slashed{v} F_2\Big) \Big(C_{10,{\rm eff}}^V c_\gamma \: \gamma_\mu
+ C_{10,{\rm eff}}^Vc_v \: v_\mu   - C_{10,{\rm eff}}^A c_\gamma \: \gamma_\mu\gamma_5
+ C_{10,{\rm eff}}^A c_v \: v_\mu\gamma_5  \Big) u(p,s). \label{eq:hadMEHQET}
\end{eqnarray}
Note that here we use spinors with the standard relativistic normalization for all particles, including the $\Lambda_b$. In terms of
the HQET spinors (\ref{eq:HQETspinors}), we have $u(p,s)=\sqrt{m_{\Lambda_b}}\:\mathcal{U}(v, s)$, with $p=m_{\Lambda_b} v$.
For a given value of $q^2$, the form factors $F_1$ and $F_2$ in Eq.~(\ref{eq:hadMEHQET}) are evaluated at
\begin{equation}
 E_{\Lambda} = p'\cdot v = \frac{m_{\Lambda_b}^2+m_\Lambda^2-q^2}{2m_{\Lambda_b}}, \label{eq:ELambda}
\end{equation}
where the masses take their physical values. The fully differential decay rate with polarized particles is given by
\begin{equation}
 \mathrm{d}\Gamma = \frac{1}{2m_{\Lambda_b}}  \frac{\mathrm{d}^3p'}{(2\pi)^3 2E_\Lambda} \frac{\mathrm{d}^3p_-}{(2\pi)^3 2E_{\ell^-}}
 \frac{\mathrm{d}^3p_+}{(2\pi)^3 2 E_{\ell^+}} (2\pi)^4 \delta^4(p-p'-p_--p_+)|\mathcal{M}|^2. \label{eq:dGamma}
\end{equation}
For the standard model calculation, we set the right-handed couplings to zero ($C_{7,\rm eff}'=C_{9,\rm eff}'=C_{10,\rm eff}'=0$)
and use the following Wilson coefficients (at $\mu=4.8$ GeV), which are of next-to-next-to-leading-logarithm accuracy \cite{Altmannshofer:2008dz}:
\begin{eqnarray}
 \nonumber C_{7,\rm eff} &=& -0.304, \\
 \nonumber C_{9,\rm eff}(q^2) &=& 4.211 + Y(q^2), \\
C_{10,\rm eff} &=& -4.103.
\end{eqnarray}
The function $Y(q^2)$ is defined as in Ref.~\cite{Altmannshofer:2008dz}. Furthermore, we use $|V_{ts}|=0.04002$ and $|V_{tb}|=0.999142$ from Ref.~\cite{ckm}.

\begin{figure}
\includegraphics[width=0.49\linewidth]{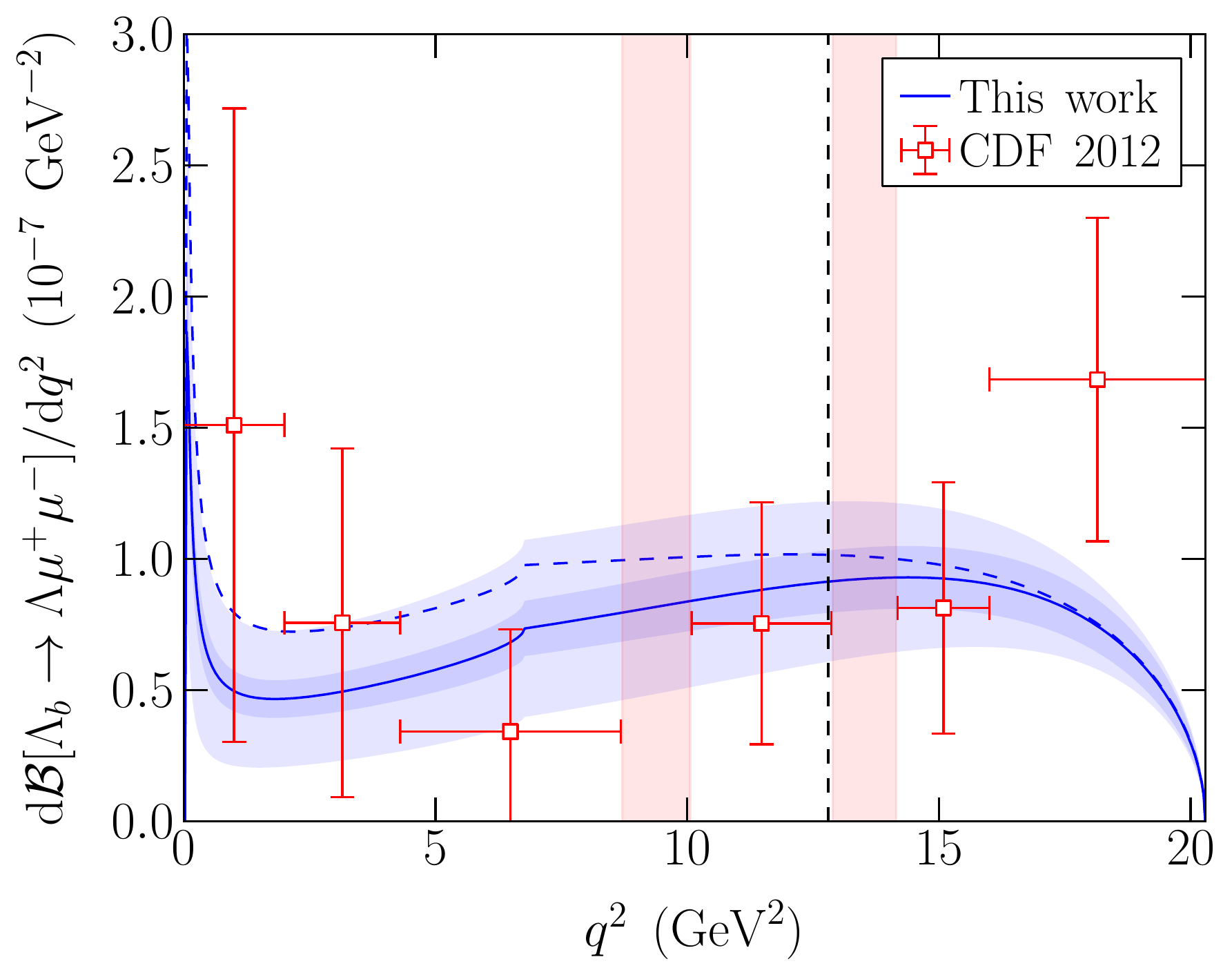} \hfill \includegraphics[width=0.49\linewidth]{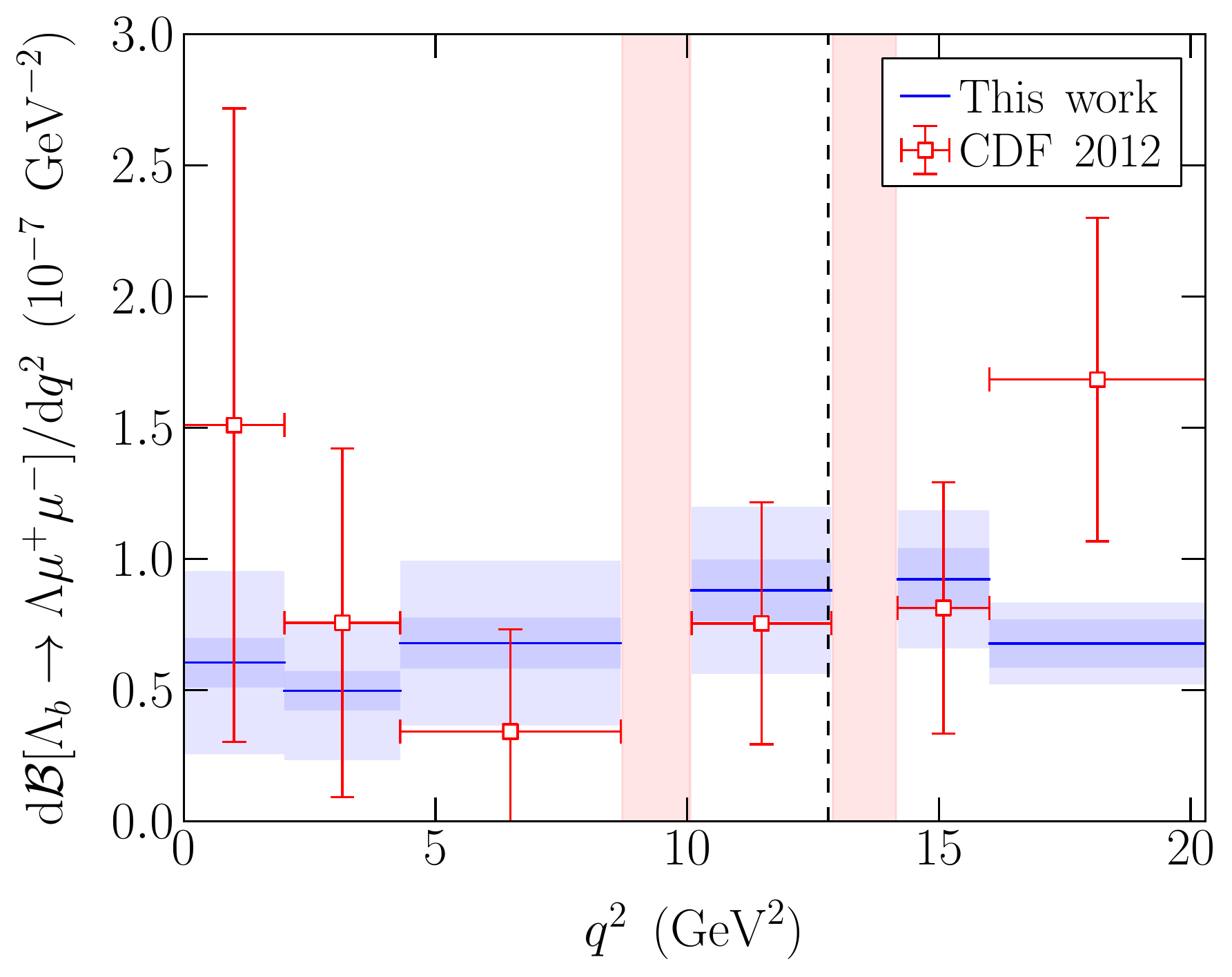}
\caption{\label{fig:dGamma}Left panel: Differential branching fraction for $\Lambda_b\to\Lambda \mu^+\mu^-$. The solid curve
is our prediction using the form factors from lattice QCD. Long-distance effects are not included in the calculation.
The inner, dark shaded band around the curve indicates the uncertainty in $\mathrm{d}\mathcal{B}/\mathrm{d}q^2$ that
results from the statistical plus systematic uncertainty in the form factors $F_\pm$. The outer, light shaded band
additionally includes an estimate of the systematic uncertainty in $\mathrm{d}\mathcal{B}/\mathrm{d}q^2$ that results
from our use of the static approximation for the $b$ quark. The vertical dashed line indicates the lowest value of $q^2$
where we have lattice data; to the left of that line the form factors are extrapolated. To illustrate the model-dependence
resulting from the extrapolation of the form factors to low $q^2$, the dashed curve shows $\mathrm{d}\mathcal{B}/\mathrm{d}q^2$
computed with form factors extrapolated using a different ansatz (monopole instead of dipole, see Fig.~\ref{fig:dipole_vs_monopole};
the uncertainty for the dashed curve is not shown for clarity). The experimental data are from Ref.~\protect\cite{CDF2012},
which is an update of Ref.~\cite{Aaltonen:2011qs}. The error bars shown for the experimental data include systematic uncertainties.
The vertical shaded bands indicate the charmonium veto regions, where long-distance effects are large.
Right panel: with binning applied to the theory prediction.}
\end{figure}

\begin{figure}
\includegraphics[width=0.49\linewidth]{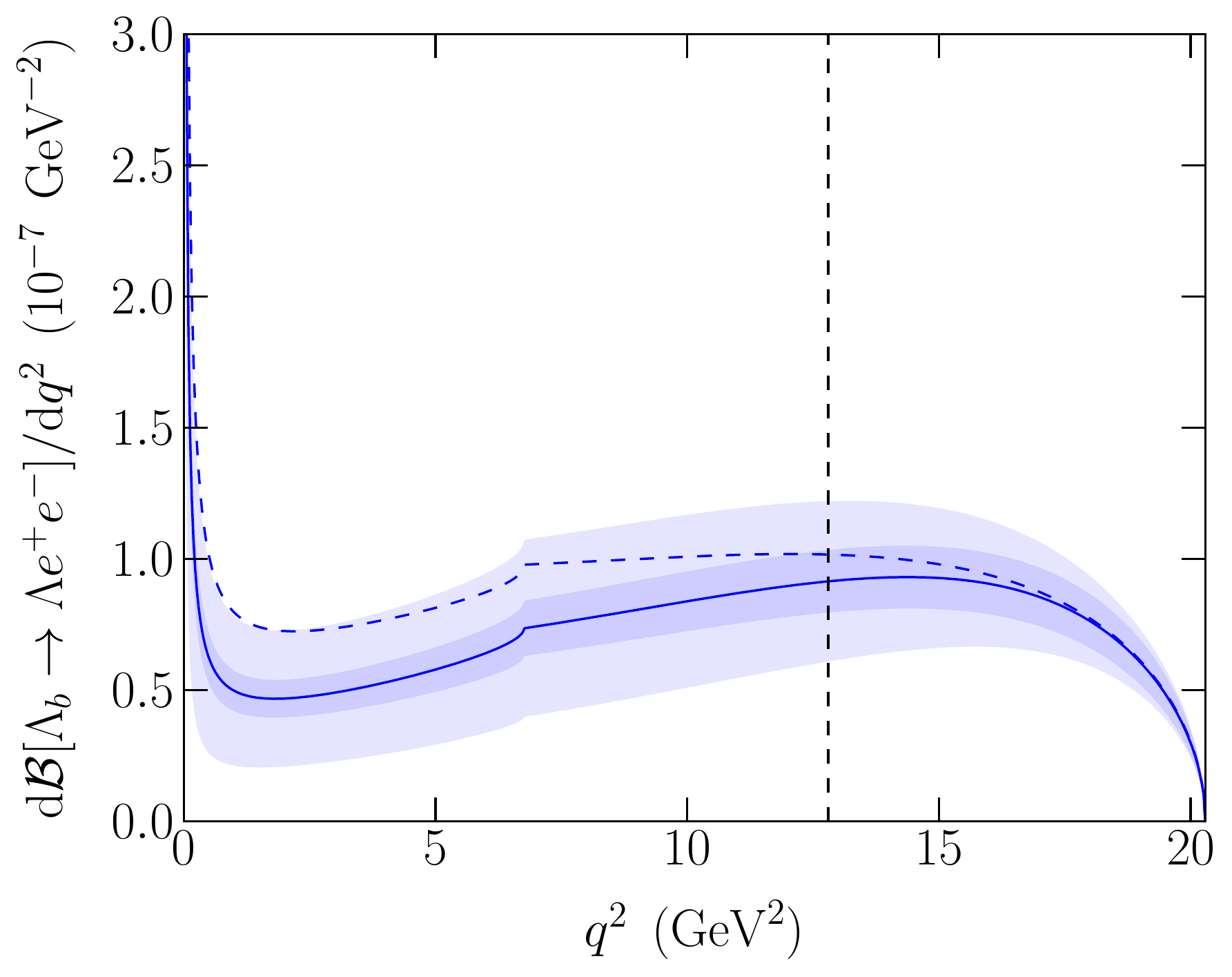} \hfill \includegraphics[width=0.49\linewidth]{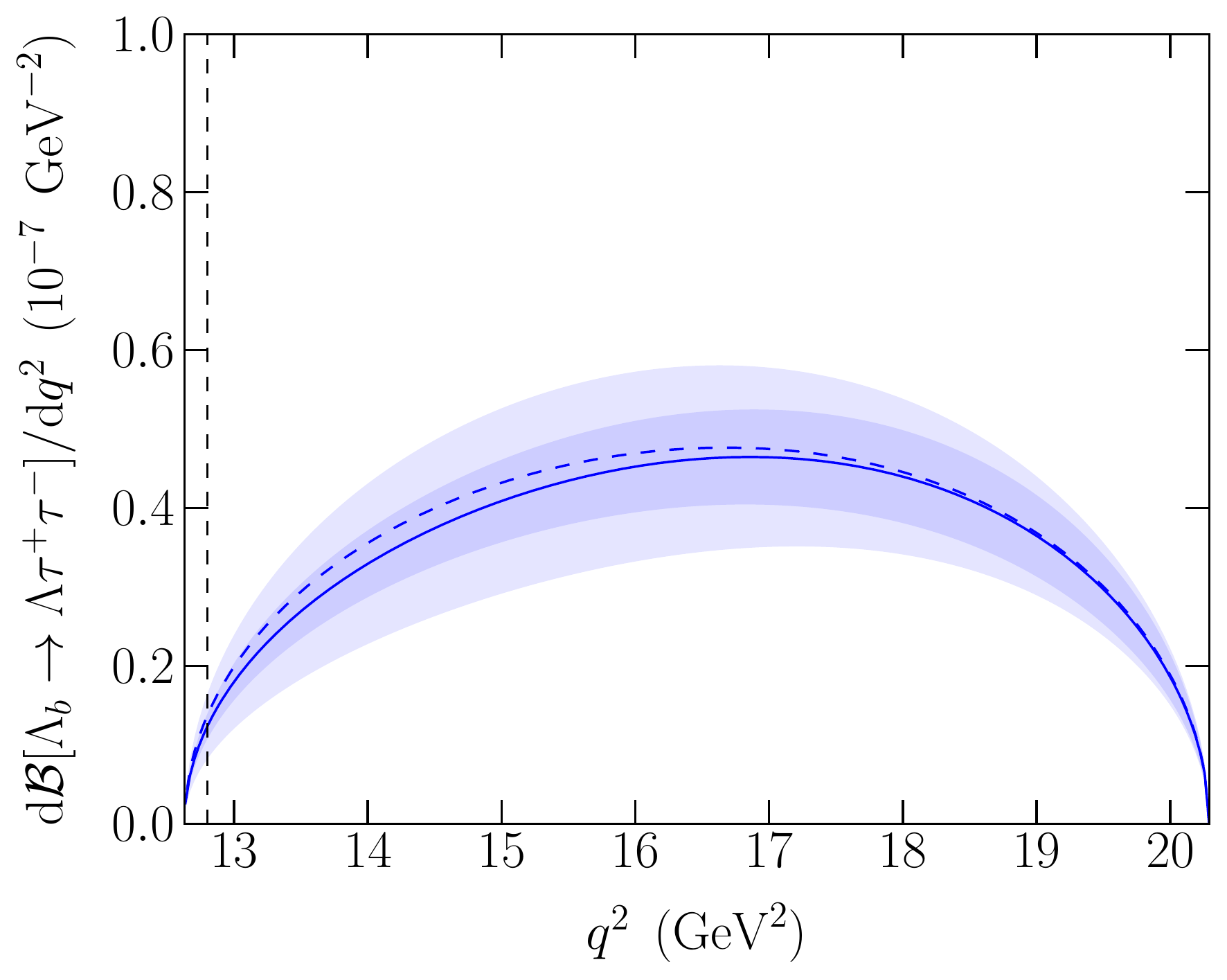}
\caption{\label{fig:dGammaetau}Differential branching fractions for $\Lambda_b\to\Lambda e^+e^-$ (left)
and $\Lambda_b\to\Lambda \tau^+\tau^-$ (right). See the caption of Fig.~\ref{fig:dGamma} for explanations.}
\end{figure}

To calculate $\mathrm{d}\Gamma/\mathrm{d}q^2$, we integrate (\ref{eq:dGamma}) over the lepton momenta and the direction of the $\Lambda$,
sum over the spins of the $\Lambda$, $\ell^+$, $\ell^-$, and average over the $\Lambda_b$ spin (because $\mathrm{d}\Gamma/\mathrm{d}q^2$ is rotationally
symmetric, it has to be independent of the $\Lambda_b$ polarization, and therefore we can treat the $\Lambda_b$ as unpolarized here). The result is given by
\begin{eqnarray}
\nonumber \frac{\mathrm{d}\Gamma}{\mathrm{d}q^2} &=& \frac{\alpha_{\rm em} ^2 G_F^2 |V_{tb} V_{ts}^*|^2}{6144\:\pi^5\: q^4\: m_{\Lambda_b}^5}
\sqrt{1-\frac{4 m_l^2}{q^2}} \sqrt{((m_{\Lambda_b}-m_{\Lambda})^2-q^2) ((m_{\Lambda_b}+m_\Lambda)^2-q^2)}\\
\nonumber &&\times \Big[ q^2 |C_{10,{\rm eff}}|^2 \mathcal{A}_{10,10}
+ 16 c_\sigma^2 m_b^2 (q^2+2 m_l^2) |C_{7,{\rm eff}}|^2 \mathcal{A}_{7,7} + q^2(q^2+2 m_l^2) |C_{9,{\rm eff}}(q^2)|^2 \mathcal{A}_{9,9}\\
&&\:\:\:\:+ 8 q^2 c_\sigma m_b (q^2+2 m_l^2) m_{\Lambda_b} \Re[C_{7,{\rm eff}}\: C_{9,{\rm eff}}(q^2) ] \mathcal{A}_{7,9}  \Big], \label{eq:dGammadqsqr}
\end{eqnarray}
with
\begin{eqnarray}
\nonumber \mathcal{A}_{10,10}&=& \Big[
   \left(2 c_\gamma^2+2 c_\gamma c_v+c_v^2\right) \left(2 m_l^2+q^2\right)\left(m_{\Lambda_b}^4 -2 m_{\Lambda_b}^2
   m_\Lambda^2 + \left(q^2-m_\Lambda^2\right)^2 \right)\\
   &&\hspace{3ex}+2 m_{\Lambda_b}^2 q^2 \left(4 c_\gamma^2 \left(q^2-4 m_l^2\right)-(2 c_\gamma c_v+c_v^2)
   \left(q^2-10 m_l^2\right)\right)\Big] \mathcal{F} +4  c_\gamma \left(c_\gamma+c_v\right) \left(2
   m_l^2+q^2\right) \mathcal{G} F_+ F_-, \hspace{4ex}\\
 \mathcal{A}_{7,7}&=&\left(m_{\Lambda
   _b}^4+m_{\Lambda_b}^2 \left(q^2-2 m_\Lambda^2\right)+\left(q^2-m_{\Lambda
   }^2\right)^2\right)\mathcal{F}+2 \mathcal{G} F_+ F_-,\\
\nonumber \mathcal{A}_{9,9}&=& \Big[\left(2 c_\gamma^2+2 c_\gamma
   c_v+c_v^2\right) \left(m_{\Lambda_b}^4 + \left(q^2-m_\Lambda^2\right)^2 \right) -2 m_{\Lambda_b}^2
   \left(2 c_\gamma^2 \left(m_\Lambda^2-2 q^2\right)+(2 c_\gamma c_v+c_v^2) \left(m_{\Lambda
   }^2+q^2\right)\right) \Big] \mathcal{F} \\
&&+4 c_\gamma
   \left(c_\gamma+c_v\right) \mathcal{G} F_+ F_-,\\
 \mathcal{A}_{7,9}&=& 3 c_\gamma \left(m_{\Lambda_b}^2-m_\Lambda^2+q^2\right)\mathcal{F}\: +2 \left(3 c_\gamma+c_v\right) \left(m_\Lambda^4-2 m_\Lambda^2
   \left(m_{\Lambda_b}^2+q^2\right)+\left(q^2-m_{\Lambda_b}^2\right)^2\right)F_+ F_-,
\end{eqnarray}
where
\begin{eqnarray}
 \mathcal{F} &=& ((m_{\Lambda_b}-m_\Lambda)^2-q^2) F_-^2  + ((m_{\Lambda_b}+m_\Lambda)^2-q^2) F_+^2, \\
 \mathcal{G} &=& m_{\Lambda_b}^6-m_{\Lambda_b}^4 \left(3 m_{\Lambda
   }^2+q^2\right)-m_{\Lambda_b}^2 \left(q^2-m_\Lambda^2\right) \left(3 m_{\Lambda
   }^2+q^2\right)+\left(q^2-m_\Lambda^2\right)^3.
\end{eqnarray}
To obtain the differential branching fraction $\mathrm{d}\mathcal{B}/\mathrm{d}q^2 = \tau_{\Lambda_b}\mathrm{d}\Gamma/\mathrm{d}q^2$,
we use the experimental value of the $\Lambda_b$ lifetime, $\tau_{\Lambda_b}=1.425\cdot10^{-12}$ s \cite{PDG2012}. The
form factors $F_+$ and $F_-$ are given by the functions (\ref{eq:Fplusminusphysical}) with parameters $N_\pm$ and $X_\pm$ as in Table \ref{tab:dipolefitresults},
and with additional systematic uncertainties of 8\% included (see Fig.~\ref{fig:finalFFs}). The
resulting differential branching fraction for $\Lambda_b \to \Lambda \mu^+ \mu^-$ is shown in Fig.~\ref{fig:dGamma}, along with
recent experimental results from CDF \cite{CDF2012}. The agreement of the standard model with the experimental data is clear, with no evidence
for physics beyond the standard model. Further predictions for $\Lambda_b \to \Lambda \ell^+ \ell^-$ with $\ell=e,\tau$ are shown
in Fig.~\ref{fig:dGammaetau}.

In Figs.~\ref{fig:dGamma} and \ref{fig:dGammaetau}, the inner shaded bands around the curves correspond to the statistical
plus systematic uncertainty in the form factors $F_\pm$. However, note that we have lattice data only in the region
$q^2 \gtrsim 13\:\:{\rm GeV}^2$, as indicated by the vertical dashed lines in Figs.~\ref{fig:dGamma} and \ref{fig:dGammaetau}.
Below that region, we rely on extrapolations of the form factors, which are model-dependent. This was shown in
Fig.~\ref{fig:dipole_vs_monopole}, where we compared the form factors from dipole and monopole fits.
Our main results for the differential branching fractions are based on the dipole form factors. To illustrate the model-dependence,
the dashed curves in Figs.~\ref{fig:dGamma} and \ref{fig:dGammaetau} give the differential branching fractions calculated
with the monopole form factors (the uncertainties of the dashed curves are not shown for clarity, but are of similar size
as with the dipole form factors). In the large-$q^2$ region, both curves are consistent with each other. At low $q^2$,
model-dependence can be seen, but as already discussed in Sec.~\ref{sec:systerrs}, a comparison between any two fit models
can only give a qualitative picture of the model-dependence.

The outer shaded bands in Figs.~\ref{fig:dGamma} and \ref{fig:dGammaetau} include an estimate of the systematic uncertainty
in $\mathrm{d}\mathcal{B}/\mathrm{d}q^2$ which arises from the use of the static approximation (i.e., leading-order HQET)
for the $b$ quark. In general, the uncertainty associated with this approximation is of order $\Lambda_{\rm QCD}/m_b$.
However, the non-zero momentum $\mathbf{p}'$ of the $\Lambda$ baryon in the $\Lambda_b$ rest frame is an additional relevant
scale, which may lead to errors of order $|\mathbf{p'}|/m_b$. Thus, we add these two errors in quadrature and
estimate the relative systematic uncertainty in $\mathrm{d}\mathcal{B}/\mathrm{d}q^2$ due to the use of HQET to be
\begin{equation}
 \sqrt{\frac{\Lambda_{\rm QCD}^2}{m_b^2}+\frac{|\mathbf{p'}|^2}{m_b^2}},
\end{equation}
where we take $\Lambda_{\rm QCD}=500$ MeV.

Another major cause of systematic uncertainty in the branching fraction is that we have neglected long-distance effects.
The most important type of long-distance effects is associated with photon exchange between the lepton and quark
electromagnetic currents, where the quark electromagnetic current combines with the four-quark operators $O_1$ to $O_6$
in the effective Hamiltonian (\ref{eq:Heff}). The resulting contribution to the decay amplitude is described by nonlocal
hadronic matrix elements of the form
\begin{equation}
 \frac{1}{q^2} \int \mathrm{d}^4 x \:\: e^{iq\cdot x}
 \langle \Lambda(p',s')| \mathsf{T}\: O_i (0)\: j^\mu_{\rm e.m.}(x)\: | \Lambda_b(p,s) \rangle,
\end{equation}
where $j^\mu_{\rm e.m.}(x)$ is the quark electromagnetic current (see for example \cite{Grinstein:2004vb,Beylich:2011aq}).
These are challenging matrix elements to compute from lattice QCD, but may give large contributions when $\sqrt{q^2}$
coincides with the mass of a hadronic resonance with $J^{PC}=1^{--}$. Of the four-quark operators,
$O_1=(\bar{c}^b\gamma^\mu P_L b^a)(\bar{s}^a\gamma_\mu P_L c^b) $ and
$O_2=(\bar{c}^a\gamma^\mu P_L b^a)(\bar{s}^b\gamma_\mu P_L c^b)$ (where the superscripts $a$ and $b$ are color indices)
have the largest Wilson coefficients, and at $q^2=m_{J/\psi}^2, m_{\psi'}^2$ the decay amplitude is dominated
by the non-local matrix elements of $O_1$ and $O_2$. The experimental analysis of Ref.~\cite{CDF2012} excludes these
$q^2$ regions, as shown by the vertical shaded bands in Fig.~\ref{fig:dGamma}. Away from these resonances,
however, we expect the short distance contributions from $O_{7,9,10}$ to be the most important.
Several approaches show that for large $q^2$, these and other long-distance effects can be treated as small corrections
to the leading order behavior given by matrix elements of $O_{7,9,10}$ \cite{Beneke:2001at, Grinstein:2004vb,Beylich:2011aq}.
Although these papers discuss $B\to K^{(*)} \ell^+\ell^-$, the same principles apply to $\Lambda_b \to \Lambda \ell^+\ell^-$.

\FloatBarrier
\section{Conclusions}
\FloatBarrier

Theoretical studies of the rare baryon decays $\Lambda_b \to \Lambda \ell^+\ell^-$ and $\Lambda_b \to \Lambda \gamma$
require knowledge of the hadronic matrix elements $\langle \Lambda | \:\bar{s} \Gamma b\: | \Lambda_b \rangle$ in
nonperturbative QCD. At leading order in  heavy-quark effective theory, these matrix elements are given by two independent
form factors, $F_1$ and $F_2$ (or, equivalently, $F_+=F_1+F_2$ and $F_-=F_1-F_2$), which are functions of the energy of
the $\Lambda$ baryon in the $\Lambda_b$ rest frame \cite{Mannel:1990vg, Hussain:1990uu, Hussain:1992rb}. Here, we have
performed the first lattice QCD calculation of these form factors. Our final results for $F_\pm$ and $F_{1,2}$, in the
continuum limit and for the physical values of the up, down, and strange-quark masses, are shown in Fig.~\ref{fig:finalFFs}.
High precision determinations were achieved by analyzing the ratios $R_+(|\mathbf{p'}|^2, t)$ and $R_-(|\mathbf{p'}|^2, t)$,
defined in Eqs.~(\ref{eq:Rplus}) and (\ref{eq:Rminus}), for a wide range of source-sink separations $t$, and by using
multiple light-quark masses as well as two different lattice spacings. Systematic uncertainties in the form factors are
estimated to be $\sim$8\%. A further reduction in systematic uncertainties would require two-loop or nonperturbative
current matching, finer lattice spacings, and light-quark masses at or very close to the physical values (which would
also require increased lattice sizes). However, already at our current level of uncertainty in the form factors,
the precision in phenomenological applications is primarily limited by the use of leading-order HQET for the heavy quark.

To compare our form factor results to the literature, let us first consider the zero recoil point
($E_\Lambda=m_\Lambda$, or $q^2=q^2_{\rm max}$). There we obtain (matched to the $\overline{\rm MS}$ scheme at $\mu=m_b$)
\begin{eqnarray}
\nonumber F_1(q^2_{\rm max})&=&1.26(4)(10),\\
\nonumber F_2(q^2_{\rm max})&=&-0.288(25)(23),\\
F_2(q^2_{\rm max})/F_1(q^2_{\rm max})&=&-0.229(21)(11), \label{eq:ratioqsqrmax}
\end{eqnarray}
where the first uncertainty is statistical/fitting, and the second uncertainty is systematic. The ratio $F_2/F_1$ has
previously been estimated by the CLEO collaboration using experimental data for the semileptonic
$\Lambda_c \to \Lambda\: e^+ \nu_e$ decay, assuming the same shape for $F_1$ and $F_2$ and ignoring $\Lambda_{\rm QCD}/m_c$
corrections, to be $F_2/F_1=-0.31(5)(4)$ \cite{Hinson:2004pj}. This is consistent with our results, given the expected
size of $\Lambda_{\rm QCD}/m_c$ corrections, but significantly less precise. In Ref.~\cite{Cheng:1994kp}, an earlier CLEO
extraction of $F_2/F_1$ \cite{Crawford:1995wz} was combined with the MIT bag model to obtain
$F_1(q^2_{\rm max})=1.02$, $F_2(q^2_{\rm max})=-0.34$. The authors of Ref.~\cite{Mannel:1997xy} combined the CLEO data
from Ref.~\cite{Crawford:1995wz} with the measured $\Lambda_c$ lifetime to get $F_1(q^2_{\rm max})=1.21$,
$F_2(q^2_{\rm max})=-0.30$, which happen to be quite close to our results. The sum rule calculation of
Ref.~\cite{Huang:1998ek} gave $F_1(q^2_{\rm max})\approx 0.81$, $F_2(q^2_{\rm max})\approx -0.34$, and
$F_2(q^2_{\rm max})/F_1(q^2_{\rm max})\approx-0.42$, in marked disagreement with (\ref{eq:ratioqsqrmax}).

At large recoil, leading-order soft-collinear effective theory predicts that $F_1$ becomes equal to a soft form factor $\xi_
\Lambda$, while $F_2$ vanishes \cite{Feldmann:2011xf, Mannel:2011xg, Wang:2011uv}. Using light-cone sum rules, the authors of
Ref.~\cite{Feldmann:2011xf} obtained $\xi_\Lambda(q^2=0)\approx 0.38$. This is rather close to our results for $F_1$ at that
point ($q^2=0$ corresponds to $E_\Lambda-m_\Lambda\approx 1.8$ GeV), as can be seen in Fig.~(\ref{fig:dipole_vs_monopole}).
However, we stress that our results are not reliable at such large values of $E_\Lambda-m_\Lambda$, where we do not have
lattice data and rely on extrapolation.

Our form factor results can be used to make theoretical predictions for several observables in $\Lambda_b \to \Lambda\: \ell^+\ell^-$
decays. As a first example, here we have calculated the differential branching fractions of these decays in the standard model.
For $\ell=\mu$, experimental data are already available from CDF \cite{CDF2012}, and our calculation agrees with the data
within uncertainties (see Fig.~\ref{fig:dGamma}). The theoretical uncertainties in the branching fraction are dominated by
errors of order $\Lambda_{\rm QCD}/m_b$ and $|\mathbf{p'}|/m_b$ associated with the use of leading-order HQET for the heavy quark,
and by missing long-distance contributions to the decay amplitudes. At high recoil, there is also an unknown uncertainty
associated with the extrapolation of the form factors. The long-distance contributions have not been calculated in lattice QCD,
and need to be estimated using other approaches. The HQET uncertainties could be reduced by using higher-order lattice HQET,
lattice nonrelativistic QCD, or a relativistic action for the $b$ quarks. In such calculations one would have to deal with
more complicated current matching and the full set of ten $\Lambda_b \to \Lambda$ form factors.

Another possible application of our form factor results to the phenomenology of $\Lambda_b \to \Lambda\: \ell^+\ell^-$
decays is to study various angular distributions which depend on the baryon polarization and probe the helicity structure
of the effective weak Hamiltonian. In particular, one should compute the angular distribution of the two-stage decay
$\Lambda_b \to (\Lambda \to p \pi^-)\: \ell^+\ell^-$ for partially polarized $\Lambda_b$ baryons. To make numerical
predictions of these angular distributions for the LHC, the polarization of the $\Lambda_b$ baryons produced in proton-proton
collisions needs to be determined, for example using the method discussed in Ref.~\cite{Hrivnac:1994jx}. It remains to be
seen whether the decay $\Lambda_b \to (\Lambda \to p \pi^-)\: \ell^+\ell^-$ will be competitive with
$B\to K^*(\to K\pi) \ell^+\ell^-$ \cite{Altmannshofer:2008dz} in constraining new-physics models.

\begin{acknowledgments}
We thank the RBC and UKQCD collaborations for access to their gauge field configurations. The domain-wall propagators
used in this work were computed with the Chroma software system \cite{Edwards:2004sx}. This work is supported by the
U.S.~Department of Energy under cooperative research agreement Contract Number DE-FG02-94ER40818. Numerical calculations
were performed using resources at the National Energy Research Scientific Computing Center (U.S.~Department of Energy
Grant Number DE-AC02-05CH11231), and XSEDE resources at the National Institute for Computational Sciences (National Science
Foundation Grant Number OCI-1053575). The work of WD was supported in part by the Jeffress Memorial Trust, J-968. WD and
SM were also supported by DOE Outstanding Junior Investigator Award DE-S{C0}0{0-17}84. CJDL is supported by Taiwanese
NSC Grant Number 99-2112-M-009-004-MY3. MW is supported by the STFC and received additional travel support from an
IPPP Associateship.
\end{acknowledgments}


\newpage

\begin{thebibliography}{10}

\bibitem{Grinstein:1988me} 
  B.~Grinstein, M.~J.~Savage, and M.~B.~Wise,
  Nucl.\ Phys.\ B {\bf 319}, 271 (1989).

\bibitem{Grinstein:1990tj} 
  B.~Grinstein, R.~P.~Springer, and M.~B.~Wise,
  Nucl.\ Phys.\ B {\bf 339}, 269 (1990).

\bibitem{Misiak:1992bc} 
  M.~Misiak,
  Nucl.\ Phys.\ B {\bf 393}, 23 (1993)
  [Erratum-ibid.\ B {\bf 439}, 461 (1995)].

\bibitem{Buras:1993xp} 
  A.~J.~Buras, M.~Misiak, M.~M\"unz, and S.~Pokorski,
  Nucl.\ Phys.\ B {\bf 424}, 374 (1994)
  [arXiv:hep-ph/9311345].

\bibitem{Buras:1994dj} 
  A.~J.~Buras and M.~M\"unz,
  Phys.\ Rev.\ D {\bf 52}, 186 (1995)
  [arXiv:hep-ph/9501281].

\bibitem{Buchalla:1995vs} 
  G.~Buchalla, A.~J.~Buras, and M.~E.~Lautenbacher,
  Rev.\ Mod.\ Phys.\  {\bf 68}, 1125 (1996)
  [arXiv:hep-ph/9512380].

\bibitem{Chetyrkin:1996vx} 
  K.~G.~Chetyrkin, M.~Misiak, and M.~M\"unz,
  Phys.\ Lett.\ B {\bf 400}, 206 (1997)
  [Erratum-ibid.\ B {\bf 425}, 414 (1998)]
  [arXiv:hep-ph/9612313].

\bibitem{Bobeth:1999mk} 
  C.~Bobeth, M.~Misiak, and J.~Urban,
  Nucl.\ Phys.\ B {\bf 574}, 291 (2000)
  [arXiv:hep-ph/9910220].

\bibitem{Altmannshofer:2012az} 
  W.~Altmannshofer and D.~M.~Straub,
  JHEP {\bf 1208}, 121 (2012)
  [arXiv:1206.0273].

\bibitem{Aaltonen:2011qs} 
  T.~Aaltonen {\it et al.}  (CDF Collaboration),
  Phys.\ Rev.\ Lett.\  {\bf 107}, 201802 (2011)
  [arXiv:1107.3753].

\bibitem{Falk:1993rf} 
  A.~F.~Falk and M.~E.~Peskin,
  Phys.\ Rev.\ D {\bf 49}, 3320 (1994)
  [arXiv:hep-ph/9308241].

\bibitem{Bonvicini:1994mr} 
  G.~Bonvicini and L.~Randall,
  Phys.\ Rev.\ Lett.\  {\bf 73}, 392 (1994)
  [arXiv:hep-ph/9401299].

\bibitem{Diaconu:1995mp} 
  C.~Diaconu, M.~Talby, J.~G.~K\"orner, and D.~Pirjol,
  Phys.\ Rev.\ D {\bf 53}, 6186 (1996)
  [arXiv:hep-ph/9512330].

\bibitem{Buskulic:1995mf} 
  D.~Buskulic {\it et al.}  (ALEPH Collaboration),
  Phys.\ Lett.\ B {\bf 365}, 437 (1996).

\bibitem{Abbiendi:1998uz} 
  G.~Abbiendi {\it et al.}  (OPAL Collaboration),
  Phys.\ Lett.\ B {\bf 444}, 539 (1998)
  [arXiv:hep-ex/9808006].

\bibitem{Abreu:1999gf} 
  P.~Abreu {\it et al.}  (DELPHI Collaboration),
  Phys.\ Lett.\ B {\bf 474}, 205 (2000).

\bibitem{Ajaltouni:2004zu} 
  Z.~J.~Ajaltouni, E.~Conte, and O.~Leitner,
  Phys.\ Lett.\ B {\bf 614}, 165 (2005)
  [arXiv:hep-ph/0412116].

\bibitem{Dharmaratna:1996xd} 
  W.~G.~D.~Dharmaratna and G.~R.~Goldstein,
  Phys.\ Rev.\ D {\bf 53}, 1073 (1996).

\bibitem{Hiller:2007ur} 
  G.~Hiller, M.~Knecht, F.~Legger, and T.~Schietinger,
  Phys.\ Lett.\ B {\bf 649}, 152 (2007)
  [arXiv:hep-ph/0702191].

\bibitem{Hrivnac:1994jx} 
  J.~Hrivnac, R.~Lednicky, and M.~Smizanska,
  J.\ Phys.\ G {\bf 21}, 629 (1995)
  [arXiv:hep-ph/9405231].

\bibitem{Gremm:1995nx} 
  M.~Gremm, F.~Kr\"uger, and L.~M.~Sehgal,
  Phys.\ Lett.\ B {\bf 355}, 579 (1995)
  [arXiv:hep-ph/9505354].

\bibitem{Hiller:2001zj} 
  G.~Hiller and A.~Kagan,
  Phys.\ Rev.\ D {\bf 65}, 074038 (2002)
  [arXiv:hep-ph/0108074].

\bibitem{Mannel:1997xy} 
  T.~Mannel and S.~Recksiegel,
  J.\ Phys.\ G {\bf 24}, 979 (1998)
  [arXiv:hep-ph/9701399].

\bibitem{Chua:1998dx} 
  C.-K.~Chua, X.-G.~He, and W.-S.~Hou,
  Phys.\ Rev.\ D {\bf 60}, 014003 (1999)
  [arXiv:hep-ph/9808431].

\bibitem{Huang:1998ek} 
  C.-S.~Huang and H.-G.~Yan,
  Phys.\ Rev.\ D {\bf 59}, 114022 (1999)
  [Erratum-ibid.\ D {\bf 61}, 039901 (2000)]
  [arXiv:hep-ph/9811303].

\bibitem{Wang:2008sm} 
  Y.-M.~Wang, Y.~Li, and C.-D.~Lu,
  Eur.\ Phys.\ J.\ C {\bf 59}, 861 (2009)
  [arXiv:0804.0648].

\bibitem{Mannel:2011xg} 
  T.~Mannel and Y.-M.~Wang,
  JHEP {\bf 1112}, 067 (2011)
  [arXiv:1111.1849].

\bibitem{Chen:2001ki} 
  C.-H.~Chen and C.~Q.~Geng,
  Phys.\ Rev.\ D {\bf 63}, 114024 (2001)
  [arXiv:hep-ph/0101171].

\bibitem{Chen:2002rg} 
  C.-H.~Chen, C.~Q.~Geng, and J.~N.~Ng,
  Phys.\ Rev.\ D {\bf 65}, 091502 (2002)
  [arXiv:hep-ph/0202103].

\bibitem{Aslam:2008hp} 
  M.~J.~Aslam, Y.-M.~Wang, and C.-D.~Lu,
  Phys.\ Rev.\ D {\bf 78}, 114032 (2008)
  [arXiv:0808.2113].

\bibitem{Chen:2001sj} 
  C.-H.~Chen and C.~Q.~Geng,
  Phys.\ Lett.\ B {\bf 516}, 327 (2001)
  [arXiv:hep-ph/0101201].

\bibitem{Chen:2001zc} 
  C.-H.~Chen and C.~Q.~Geng,
  Phys.\ Rev.\ D {\bf 64}, 074001 (2001)
  [arXiv:hep-ph/0106193].

\bibitem{Mannel:1990vg} 
  T.~Mannel, W.~Roberts, and Z.~Ryzak,
  Nucl.\ Phys.\ B {\bf 355}, 38 (1991).

\bibitem{Hussain:1990uu} 
  F.~Hussain, J.~G.~K\"orner, M.~Kramer, and G.~Thompson,
  Z.\ Phys.\ C {\bf 51}, 321 (1991).

\bibitem{Hussain:1992rb} 
  F.~Hussain, D.-S.~Liu, M.~Kramer, J.~G.~K\"orner, and S.~Tawfiq,
  Nucl.\ Phys.\ B {\bf 370}, 259 (1992).

\bibitem{Feldmann:2011xf} 
  T.~Feldmann and M.~W.~Y.~Yip,
  Phys.\ Rev.\ D {\bf 85}, 014035 (2012)
  [Erratum-ibid.\ D {\bf 86}, 079901 (2012)]
  [arXiv:1111.1844].

\bibitem{Wang:2011uv} 
  W.~Wang,
  Phys.\ Lett.\ B {\bf 708}, 119 (2012)
  [arXiv:1112.0237].

\bibitem{Cheng:1994kp} 
  H.-Y.~Cheng, C.-Y.~Cheung, G.-L.~Lin, Y.~C.~Lin, T.-M.~Yan, and H.-L.~Yu,
  Phys.\ Rev.\ D {\bf 51}, 1199 (1995)
  [arXiv:hep-ph/9407303].

\bibitem{Cheng:1995fe} 
  H.-Y.~Cheng and B.~Tseng,
  Phys.\ Rev.\ D {\bf 53}, 1457 (1996)
  [Erratum-ibid.\ D {\bf 55}, 1697 (1997)]
  [arXiv:hep-ph/9502391].

\bibitem{Mohanta:1999id} 
  R.~Mohanta, A.~K.~Giri, M.~P.~Khanna, M.~Ishida, and S.~Ishida,
  Prog.\ Theor.\ Phys.\  {\bf 102}, 645 (1999)
  [arXiv:hep-ph/9908291].

\bibitem{Mott:2011cx} 
  L.~Mott and W.~Roberts,
  Int.\ J.\ Mod.\ Phys.\ A {\bf 27}, 1250016 (2012)
  [arXiv:1108.6129].

\bibitem{He:2006ud} 
  X.-G.~He, T.~Li, X.-Q.~Li, and Y.-M.~Wang,
  Phys.\ Rev.\ D {\bf 74}, 034026 (2006)
  [arXiv:hep-ph/0606025].

\bibitem{Wang:2009hra} 
  Y.-M.~Wang, Y.-L.~Shen, and C.-D.~Lu,
  Phys.\ Rev.\ D {\bf 80}, 074012 (2009)
  [arXiv:0907.4008].

\bibitem{Crawford:1995wz} 
  G.~D.~Crawford {\it et al.}  (CLEO Collaboration),
  Phys.\ Rev.\ Lett.\  {\bf 75}, 624 (1995).

\bibitem{Hinson:2004pj} 
  J.~W.~Hinson {\it et al.}  (CLEO Collaboration),
  Phys.\ Rev.\ Lett.\  {\bf 94}, 191801 (2005)
  [arXiv:hep-ex/0501002].

\bibitem{Detmold:2012ug} 
  W.~Detmold, C.~J.~D.~Lin, S.~Meinel, and M.~Wingate,
  PoS LATTICE {\bf 2012}, 123 (2012)
  [arXiv:1211.5127].

\bibitem{Eichten:1989kb}
  E.~Eichten and B.~R.~Hill,
  Phys.\ Lett.\  B {\bf 240}, 193 (1990).

\bibitem{Kaplan:1992bt}
  D.~B.~Kaplan,
  Phys.\ Lett.\  B {\bf 288}, 342 (1992)
  [arXiv:hep-lat/9206013].

\bibitem{Shamir:1993zy}
  Y.~Shamir,
  Nucl.\ Phys.\  B {\bf 406}, 90 (1993)
  [arXiv:hep-lat/9303005].

\bibitem{Furman:1994ky}
  V.~Furman and Y.~Shamir,
  Nucl.\ Phys.\  B {\bf 439}, 54 (1995)
  [arXiv:hep-lat/9405004].

\bibitem{Aoki:2010dy}
  Y.~Aoki {\it et al.}  (RBC/UKQCD Collaboration),
  Phys.\ Rev.\  D {\bf 83}, 074508 (2011)
  [arXiv:1011.0892].

\bibitem{Hasenfratz:2001hp}
  A.~Hasenfratz and F.~Knechtli,
  Phys.\ Rev.\  D {\bf 64}, 034504 (2001)
  [arXiv:hep-lat/0103029].

\bibitem{DellaMorte:2003mn}
  M.~Della Morte, S.~D\"urr, J.~Heitger, H.~Molke, J.~Rolf, A.~Shindler, and R.~Sommer
                  (ALPHA Collaboration),
  Phys.\ Lett.\  B {\bf 581}, 93 (2004)
  [Erratum-ibid.\  B {\bf 612}, 313 (2005)]
  [arXiv:hep-lat/0307021].

\bibitem{Iwasaki:1983ck}
  Y.~Iwasaki,
  Report No. UTHEP-118 (1983).

\bibitem{Iwasaki:1984cj}
  Y.~Iwasaki and T.~Yoshie,
  Phys.\ Lett.\  B {\bf 143}, 449 (1984).
  
\bibitem{Ishikawa:2011dd} 
  T.~Ishikawa, Y.~Aoki, J.~M.~Flynn, T.~Izubuchi, and O.~Loktik,
  JHEP {\bf 1105}, 040 (2011)
  [arXiv:1101.1072].

\bibitem{Ji:1991pr} 
  X.-D.~Ji and M.~J.~Musolf,
  Phys.\ Lett.\ B {\bf 257}, 409 (1991).

\bibitem{Broadhurst:1991fz} 
  D.~J.~Broadhurst and A.~G.~Grozin,
  Phys.\ Lett.\ B {\bf 267}, 105 (1991)
  [arXiv:hep-ph/9908362].

\bibitem{Bowler:1997ej} 
  K.~C.~Bowler {\it et al.}  (UKQCD Collaboration),
  Phys.\ Rev.\ D {\bf 57}, 6948 (1998)
  [arXiv:hep-lat/9709028].

\bibitem{Meinel:2010pv} 
  S.~Meinel,
  Phys.\ Rev.\ D {\bf 82}, 114502 (2010)
  [arXiv:1007.3966].

\bibitem{Davies:2009tsa} 
  C.~T.~H.~Davies {\it et al.}  (HPQCD Collaboration),
  Phys.\ Rev.\ D {\bf 81}, 034506 (2010)
  [arXiv:0910.1229].

\bibitem{Della Morte:2005yc} 
  M.~Della Morte, A.~Shindler, and R.~Sommer,
  JHEP {\bf 0508}, 051 (2005)
  [arXiv:hep-lat/0506008].

\bibitem{Eichten:1989zv} 
  E.~Eichten and B.~R.~Hill,
  Phys.\ Lett.\ B {\bf 234}, 511 (1990).

\bibitem{Altmannshofer:2008dz} 
  W.~Altmannshofer, P.~Ball, A.~Bharucha, A.~J.~Buras, D.~M.~Straub, and M.~Wick,
  JHEP {\bf 0901}, 019 (2009)
  [arXiv:0811.1214].

\bibitem{ckm}
  J.~Charles {\it et al.} (CKMfitter Group),
  ``Updated results on the CKM matrix, including results presented up to ICHEP 2012''
  (\url{http://ckmfitter.in2p3.fr/www/results/plots_ichep12/num/ckmEval_results_ICHEP12.pdf}).

\bibitem{PDG2012} 
  J.~Beringer {\it et al.}  (Particle Data Group Collaboration),
  Phys.\ Rev.\ D {\bf 86}, 010001 (2012).

\bibitem{CDF2012}
  CDF Collaboration,
  ``Precise measurements of exclusive $b \to s \mu^+ \mu^-$ decay amplitudes using the full CDF data set'',
  Public Note 108xx, Version 0.1 (\url{http://www-cdf.fnal.gov/physics/new/bottom/bottom.html}).

\bibitem{Grinstein:2004vb} 
  B.~Grinstein and D.~Pirjol,
  Phys.\ Rev.\ D {\bf 70}, 114005 (2004)
  [arXiv:hep-ph/0404250].

\bibitem{Beylich:2011aq} 
  M.~Beylich, G.~Buchalla, and T.~Feldmann,
  Eur.\ Phys.\ J.\ C {\bf 71}, 1635 (2011)
  [arXiv:1101.5118].

\bibitem{Beneke:2001at} 
  M.~Beneke, T.~Feldmann, and D.~Seidel,
  Nucl.\ Phys.\ B {\bf 612}, 25 (2001)
  [arXiv:hep-ph/0106067].

\bibitem{Edwards:2004sx}
  R.~G.~Edwards and B.~Jo\'o,
  Nucl.\ Phys.\ Proc.\ Suppl.\  {\bf 140}, 832 (2005)
  [arXiv:hep-lat/0409003].

\end{thebibliography}
\end{document}